\documentclass[twocolumn,iop]{openjournal}
\usepackage{graphicx}
\usepackage{txfonts}
\usepackage[colorlinks,allcolors=blue]{hyperref}

\usepackage{standalone}
\usepackage{tikz}
\usepackage{ifthen}
\usepackage{float}
\usetikzlibrary{matrix,calc}

\begin{document} 

\journalinfo{The Open Journal of Astrophysics}
\submitted{submitted X, accepted X}

   \title{New dwarf galaxy candidates in the M\,106, NGC\,3521, and UGCA127 groups with the Hyper Suprime Camera}
    \shorttitle{Dwarf candidates around M\,106, NGC\,3521, and UGCA127}

   \author{Oliver M\"uller$^{\star1,2,3}$}
\author{Helmut Jerjen$^{4}$}
\author{Salvatore Taibi$^{1}$}
\author{Nick Heesters$^{1}$}
\author{Ethan Crosby$^{4}$}
\author{Marcel S. Pawlowski$^{5}$}

\affiliation{$^{1}$Institute of Physics, Laboratory of Astrophysics, Ecole Polytechnique Fédérale de Lausanne (EPFL), 1290 Sauverny, Switzerland}
\affiliation{$^{2}$Visiting Fellow, Clare Hall, University of Cambridge, Cambridge, UK}
\affiliation{$^{3}$Institute of Astronomy, Madingley Rd, Cambridge CB3 0HA, UK}
\affiliation{$^{4}$Research School of Astronomy and Astrophysics, Australian National University, Canberra, ACT 2611, Australia}
\affiliation{$^{5}$Leibniz-Institut f\"ur Astrophysik Potsdam, An der Sternwarte 16, D-14482 Potsdam, Germany}

\thanks{$^\star$ E-mail: \nolinkurl{om420@cam.ac.uk}}


 
  \begin{abstract}
  The  local universe is still full of hidden dwarf galaxies to be discovered using deep imaging campaigns. Here we present the third paper in a series to search for low-surface brightness dwarf galaxies around nearby isolated luminous host galaxies with the Subaru Hyper Suprime Camera. Based on  visual inspection, we found 11, 0, 4, and 6 dwarf galaxy candidates around M\,106, NGC\,2903, NGC\,3521, and UGCA127, respectively. This adds to the 40 candidates around M\,104 and 4 candidates around NGC\,2683 found in the previous papers. Artificial galaxy experiments show that we are complete down to a mean effective surface brightness of 26 mag/arcsec$^2$. The new dwarf galaxy candidates follow known scaling relation in size, surface brightness and luminosity, making them good candidates based on their morphology and photometric properties. We trace the luminosity function of these galaxies down to magnitude of $\approx-$9 in the V band for all galaxies targeted in our survey footprint so far. While the most massive galaxy (M\,104) has a significant higher abundance of dwarfs, NGC\,3521, NGC\,2903, and NGC\,2683 have a similar luminosity function as the Milky Way. These latter three galaxies also have a similar stellar mass and might be considered Milky Way analogs. UGCA127 is a low-mass galaxy but almost reaches the same number of dwarfs as the Milky Way at our limiting magnitude. We have searched for hints of lopsidedness in the satellite distributions, but found none to be significant. The next step will be to confirm these members through either distance or velocity measurements.
  \end{abstract}


   \maketitle
%

\section{Introduction}

Inspections of the Milky Way's and Andromeda's gravitational spheres of influence have shown that their dwarf satellite galaxy systems are co-rotating in highly anisotropic planar structures \citep{1976MNRAS.174..695L,1976RGOB..182..241K,2005A&A...431..517K,2006AJ....131..895K,2006MNRAS.365..902M,2007MNRAS.374.1125M,2012MNRAS.423.1109P,2015MNRAS.453.1047P,2013Natur.493...62I,2013ApJ...766..120C,2020MNRAS.499.3755S,2020MNRAS.491.3042P,2020ApJ...901...43S,2024A&A...681A..73T}. Planes of satellites are also known to exist around the nearby galaxies Centaurus A
\citep{2015ApJ...802L..25T,Muller2018,Muller2019,Muller2021b,2023MNRAS.519.6184K}, M\,81 \citep{2013AJ....146..126C,2024A&A...683A.250M}, NGC\,253 (\citealt{2021A&A...652A..48M}, however, see also \citealt{2024ApJ...966..188M}), NGC\,4490 \citep{2024MNRAS.528.2805K,2024A&A...688A.153P} and other more distant galaxy groups \citep{2021ApJ...917L..18P,2021A&A...654A.161H}. However, not all discovered flattened structures feature a co-rotational signal (see e.g. \citealt{2024arXiv241106795M} for NGC\,474 and \citealt{2024A&A...683A.250M} for M\,81).

High-resolution  cosmological simulations such as the Millennium-II simulation \citep{2009MNRAS.398.1150B} or hydrodynamical cosmological simulations such as TNG from the Illustris suite \citep{2018MNRAS.475..676S,2018MNRAS.475..624N} predict that most satellite galaxy systems are close to isotropic with random motions \citep[e.g., ][]{2015ApJ...815...19P,Muller2021b} leading to a still ongoing discussion about the significance of the observational results from those well studied host galaxy environments in the context of near-field cosmology \citep{2009MNRAS.399..550L,2014ApJ...784L...6I,2014MNRAS.442.2362P,2015ApJ...800...34G,2015MNRAS.453.3839P,2021MNRAS.504.1379S}. The results could be interpreted as rare statistical outliers \citep{2005ApJ...629..219Z,2015MNRAS.452.3838C,2020ApJ...897...71S,2023NatAs...7..481S} or as a serious challenge to cosmology \citep{2010A&A...523A..32K,2014ApJ...784L...6I,2021NatAs...5.1185P,2022NatAs...6..897S}. 

Another type of anisotropy  detected in the distribution of satellites around pairs of large galaxies is to find the satellites preferentially occupying the space between the host galaxies rather than on opposing sides, this has been found around the Andromeda galaxy \citep{2013ApJ...766..120C}, as well as in a statistical sample from the Sloan Digital Sky Survey (SDSS) at the 5$\sigma$ level \citep{2016ApJ...830..121L}. The same features are found in cosmological simulations \citep{2017ApJ...850..132P}. \citet{2019MNRAS.488.3100G} suggested that the signal seen in the observations is dominated by dynamically young active environments. Such studies have now been extended to isolated systems \citep{2021ApJ...914...78W,2024MNRAS.529.1405L,2024A&A...690A.110H}, with the Andromeda galaxy satellite system being the most extreme \citep{2025arXiv250408047K}. \citet{2024A&A...690A.110H} studied different metrics to detect lopsidedness in a system. They found that the so-called maximum wedge population metric (or short wedge metric) best captures any anisotropy in a satellite system. 
With this, \citet{2024A&A...690A.110H} found that 16 percent of their MATLAS and ELVES sample showed a statistically significant lopsidedness.

Motivated by the use of dwarf galaxies to study cosmology, we have conducted a survey to search for dwarf galaxies around nearby giant galaxies within the Local Volume. First results were presented in \citet{2023MNRAS.521.4009C} for the NGC\,2683 group and in \citet{2024MNRAS.527.9118C,2025MNRAS.536.2072C} for the M\,104 group. This resulted in the discovery of four dwarf galaxy candidates for the former and 40 candidates for the latter group.

In this paper we present our search for dwarf satellite galaxies around the Local Volume galaxies M\,106, NGC\,3521, and UGCA127 based on our deep Subaru Hyper Suprime-Cam data. We also surveyed the region of NGC\,2903, however, we did not find any new candidates. All our detections around NGC\,2903 were already listed in  \citet{2022ApJ...933...47C}, confirming the completeness of this environment to the quoted luminosity limit. Therefore, we leave the discussion of detection of dwarf candidates for this galaxy out of the paper.
In the sphere of influence of M\,106 18 dwarf galaxies have been reported that are considered companions and are confirmed through TRGB or surface brightness fluctuation (SBF) measurements \citep{2022ApJ...933...47C}. For NGC\,3521, there are 11 known dwarf galaxies based on SBF \citep{2022ApJ...933...47C} and redshift measurements \citep{2007AJ....134.1849W} and one candidate, and for UGCA127 there are two known dwarfs based on redshift measurements \citep{2007AJ....134.1849W}.

\section{Observations and data reduction}
We surveyed the M\,106 (NGC\,4257), NGC\,3521, and UGCA127 regions in the $g$-band using the Hyper Suprime Camera (HSC) at the 8.2m Subaru telescope at the Mauna Kea Observatories. The data acquisition was completed as part of the observing proposal S18B0118QN (PI: H. Jerjen) on 2019 January 30-31, targeting eight Local Volume host galaxy environments for near-field cosmology studies with dwarf galaxies. The average seeing for this set of observations was $1\farcs11\pm 0\farcs29$. The HSC is equipped with an array of 104 4k$\times$2k science CCD detectors covering a 1.5 degree diameter FoV with a $0\farcs168$/pixel resolution \citep{Miyazaki_2018}. We used 7 pointings per galaxy. At the distance of M\,106 ($D=7.66$\,Mpc, \citealt{2009AJ....138..323T}) that FoV corresponds to a physical scale of 201\,kpc, at that of NGC\,3521 ($D=10.70$\,Mpc, \citealt{2013AJ....145..101K}) to 280\,kpc, and at that of UGCA127 ($D=8.50$\,Mpc, \citealt{2013AJ....145..101K}) to 223\,kpc. 
We processed the CCD data using the open-source HSC Pipeline ({\it hscPipe}) software version 7.9.1\footnote{\url{https://hscdata.mtk.nao.ac.jp/hsc_bin_dist}}, which is based on the pipeline in development for the Rubin Legacy Survey of Space and Time (LSST) Data Management System \citep{Bosch_2019}. The {\it hscPipe} is described in algorithmic detail by the overview paper \citep{Bosch_2018}, and in practical use by the Hyper Suprime-Cam Subaru Strategic Program (HSC-SSP) data release papers \citep{Aihara_2018, Aihara_2019}.

Our data reduction workflow is based on the tutorial in the HSC Pipeline {\it hscPipe}  user manual\footnote{\url{https://hsc.mtk.nao.ac.jp/pipedoc/pipedoc_7_e}} and the LSST Science Pipelines user manual\footnote{\url{https://pipelines.lsst.io/getting-started}}. The main data reduction steps were: instrumental corrections, single-visit processing, mosaicking, stacking, and multi-visit processing. We adopt specific hscpipe/LSST terminology to describe our data during these stages. A \texttt{visit} is a single exposure with HSC, represented by a unique integer. Each visit contains 112 images, one from each CCD in the imaging array. During the early reduction stages (instrumental corrections and single-visit processing), our data is comprised of 42 visits (7 pointings times 3 exposures, repeated twice). In the later stages (mosaicking, stacking, and multi-visit processing), our data is subdivided into individual patch sub-areas of $4000\times4000$ pixels, contained within a single tract. Adjacent patches have an overlap of 1000 pixels on their edges. For our data set, this amounts to 676 individual patches (in a $26\times 26$ grid), of which 428 patches are non-empty and contain the observation data.

The calibration data provided by the Subaru telescope included 10 bias exposures, 5 dark exposures, and 10 dome flat exposures. Using {\it hscpipe}, we generated calibration images to remove instrumental signatures from the raw images. The calibration steps included bias removal, dark current correction, flat-fielding, masking of bad pixels, and background subtraction. These instrumental corrections were applied to the raw CCD images to produce calibrated images.

When searching for dwarf galaxies, it is important to consider the quality of any sky background subtraction, as the dwarf galaxies are objects with surface brightness levels lower than the night sky. We are using a new and improved sky subtraction algorithm, first included with {\it hscpipe} version 7, which operates in two stages. One stage involves global sky subtraction, while another involves local sky subtraction. The global sky subtraction empirically models the background over the entire focal plane, suppressing the effects that arise from the gaps between the individual CCDs. The local sky subtraction generates a scaled sky frame from a 
$128\times 128$ pixel mesh of the combined observations. This scaled sky frame represents the mean response of the instrument to the sky for a particular filter, allowing static features to be subtracted at smaller scales than the empirical model.

Astrometric and photometric calibrations were conducted relative to the Pan-STARRS1 Survey (PS1) reference catalogue \citep{2020ApJS..251....7F}. The astrometric and photometric fits were initially performed on a per-CCD level during single-visit processing, and then repeated and normalised during mosaicking and multi-visit processing. A successful multi-visit calibration required consistent positions and fluxes for sources that appeared on different parts of the focal plane in different visits.

The mosaicking routine generates a photometric quality assurance diagnostic, which examines the internal and external consistency of the photometric calibration. This diagnostic calculates the deviation in magnitudes between measured sources and the reference catalogue. The internal scatter of our photometry is measured to be $\sigma_{int}=0.015$\,mag, which is comparable to the results presented by \cite{Aihara_2019} for the HSC-SSP.

\section{Search for satellite candidates}

\begin{figure*}[ht]
	\includegraphics[width=0.49\linewidth]{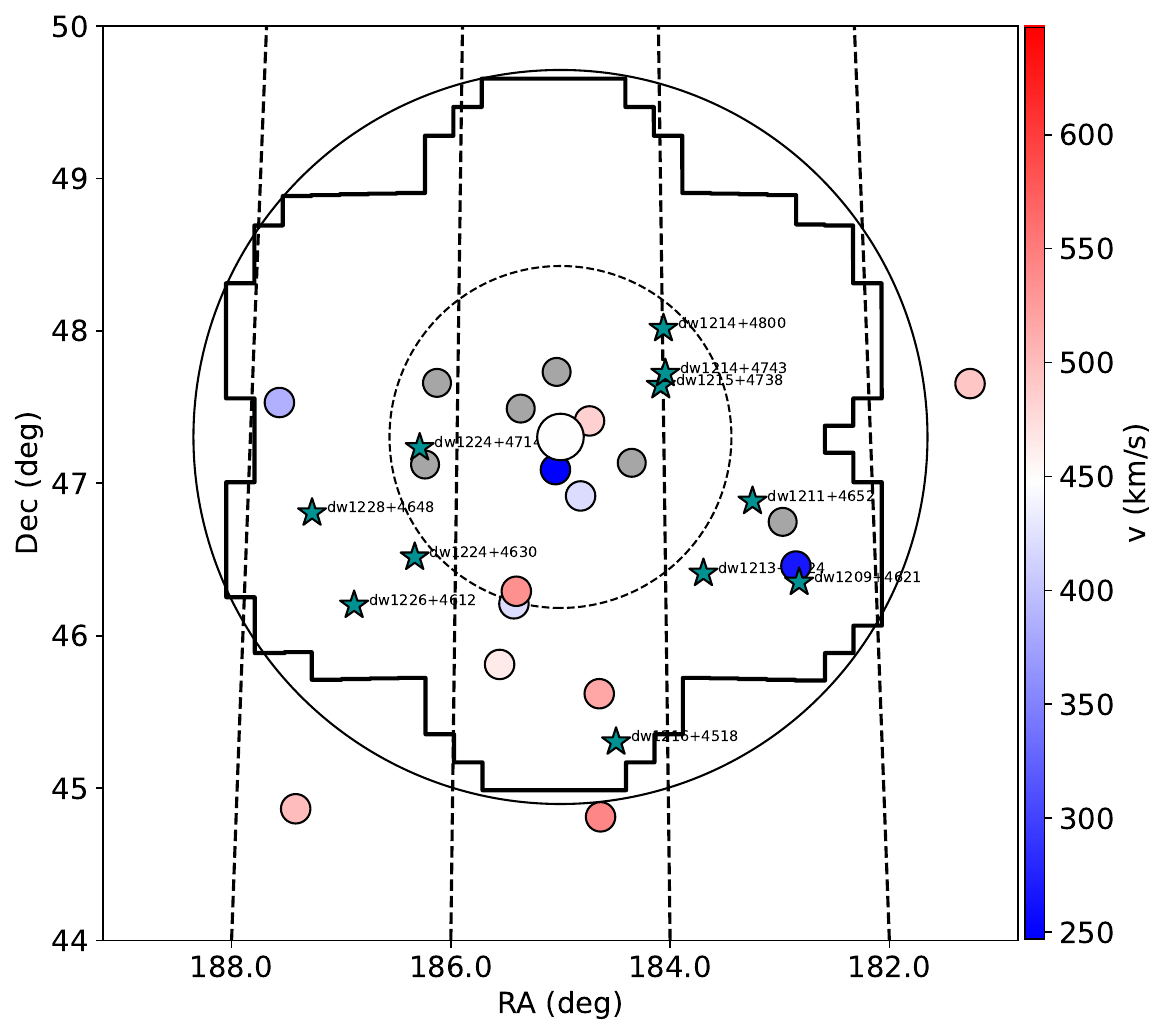}
    	\includegraphics[width=0.49\linewidth]{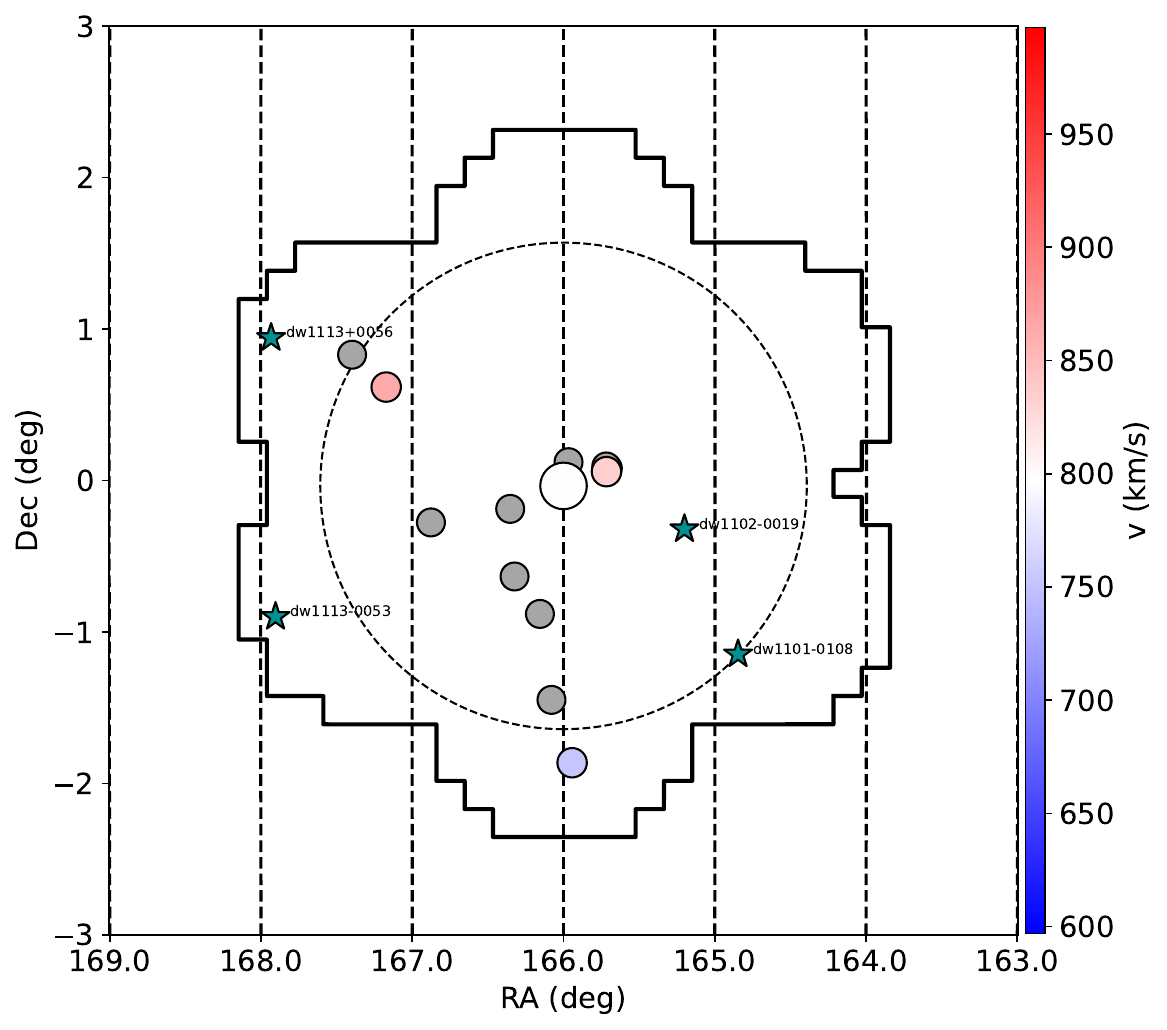}\\
        	\includegraphics[width=0.49\linewidth]{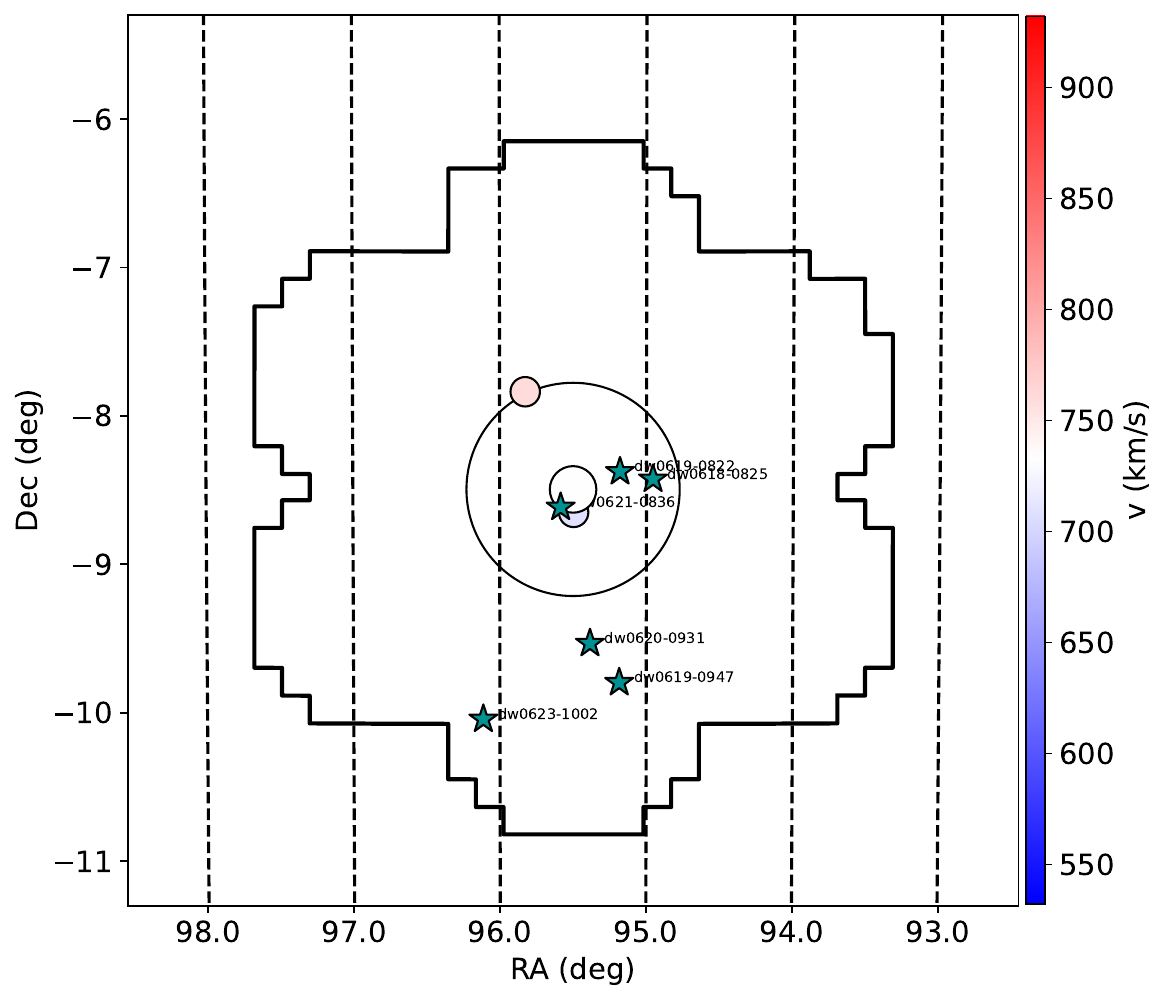}
	\caption{The surveyed field of M\,106 (top left) NGC\,3521 ( top right) and UGCA127 (bottom left). The circles are known galaxies, with the color representing their velocity, and gray indicating that no velocity information is available. The stars are the newly discovered dwarf galaxy candidates. The large solid circles represents the virial radius of the hosts, and the dashed circles the search footprint of ELVES.}
	\label{fig:fields}
\end{figure*}

In order to find new satellite galaxy candidates we searched for unresolved, low surface brightness objects in the stacked HSC images through visual inspection. This is a time-tested approach for the identification of low-surface brightness objects \citep[e.g., ][]{1985AJ.....90.1681B,2000A&A...362..544K,2000AJ....119..593J,2007AJ....133.1756S,2015A&A...583A..79M,2016A&A...588A..89J,2024MNRAS.527.9118C}. We employ this approach through a meticulous scanning of the entire survey area. The entire data set was reviewed multiple times and possible galaxy group members were logged and categorised based on their appearance. Features include qualitative characteristics such as surface brightness profile, morphology and angular size. The presence of faint spiral arms, a disk, or overall high surface brightness indicate that the object is a background galaxy. Some low surface brightness ghosts and artefacts were caused by scattered light from bright foreground stars. Each object was also checked in the SIMBAD \citep{SIMBAD} and NED databases for possible optical or 21cm-based redshift measurements. 

We found a total of 11, 4, and 6 dwarf galaxy candidates around M\,106, NGC\,3531, and UGCA127 satellites, respectively. Their locations relative to their hosts are shown in Fig.\ref{fig:fields} and individual $g$-band images are shown in Figs.\,\ref{fig:M106_known}, \ref{fig:NGC3521_known}, and \ref{fig:UGCA127_known} in the appendix.  Their photometric properties are listed in Table\,\ref{tab:group_M106}, \ref{tab:group_NGC3521}, and \ref{tab:group_UGCA127}, also in the appendix.

To validate and quantify the performance of our detection process, we injected model dwarf galaxies into our HSC images and attempt to recover them visually. To represent a realistic sample of dwarf galaxies we generated 5000 random 2D light profiles uniformly distributed in the parameter space of total apparent magnitude $15<m_g<23$ mag, effective radius $2<r_{\text{eff}}<35$ pixels, position angle $0<PA<180$ deg, ellipticity $0<\epsilon<0.5$, and S\'ersic index $0.5<n<2.0$. To estimate the completeness limits of our visual detection process as a function of surface brightness and angular size we fitted the 50 percent ridge line in the histogram with equation 11 from \cite{2005PASA...22..118G}, for the mean surface brightness within the effective radius:
$$ 
m_{g}=\langle\mu\rangle_{\text{eff,50\%}}-2.5\log(2\pi r_{\text{eff}}^2) 
$$
We find a completeness of  $\left<\mu\right>_{\text{eff,50\%}}=26.01\pm0.06$\,mag/arcsec$^{2}$. Intrinsically difficult to identify as satellite candidates independently of surface brightness are objects smaller than $r_{\text{eff}}<3$\,arcsec. They fall into two categories, the ones that have low surface brightness and blend into the sky background and the others that have relative high surface brightness. Although clearly visible the latter closely resemble background ellipticals or spirals. In our selection we excluded any object smaller than $r_{\text{eff}}=3$\,arcsec to avoid introducing a large fraction of background contamination. At 10\,Mpc, this corresponds to dwarf galaxies with an effective radius less than 145\,pc.

\section{Photometry}
{ 
To calculate the photometric properties of the candidate satellites and all known dwarf galaxies in the field, we re-binned the data by a factor of four (i.e., mapping $4\times4$ pixels onto 1 pixel using the average value) to increase the S/N. We then masked the possible sources of contamination close to the satellites using the \textit{detect\_sources} function of \texttt{photutils.segmentation}, assuming a 3-$\sigma$ threshold. 
We also used the masked image to refine the local background estimate, subtracting the median value.
We proceeded by using the \textit{Ellipse} function of \texttt{photutils.isophote} to obtain the structural parameters and compute the radial surface brightness profiles. We ran \textit{Ellipse} with initial guess values for the center, semi-major axis radius, ellipticity and position angle, allowing the code to find the most appropriate parameters within a predefined area, with steps between 1.35\arcsec and 4.00\arcsec (i.e. different steps depending on whether the systems were fainter and smaller or brighter and larger). The fitting area was defined according to where the cumulative intensity asymptotically reached a plateau (generally between 30\arcsec and 150\arcsec).
Finally, we fit the extracted surface brightness profiles with a Sersic function (Sersic 1968):
$$
\mu_{\rm sersic}(r) = \mu_0 + 1.0857 \times (r/r_0)^n
$$
with free parameters the central surface brightness $\mu_0$, the scale radius $r_0$, and the Sersic index $n$.
We then used these parameters to derive the total magnitude $m_g$, the effective radius $R_e$, and the effective surface brightness $\mu_e$ (see Graham \& Driver 2005).
}

\begin{figure}[ht]
	\includegraphics[width=\linewidth]{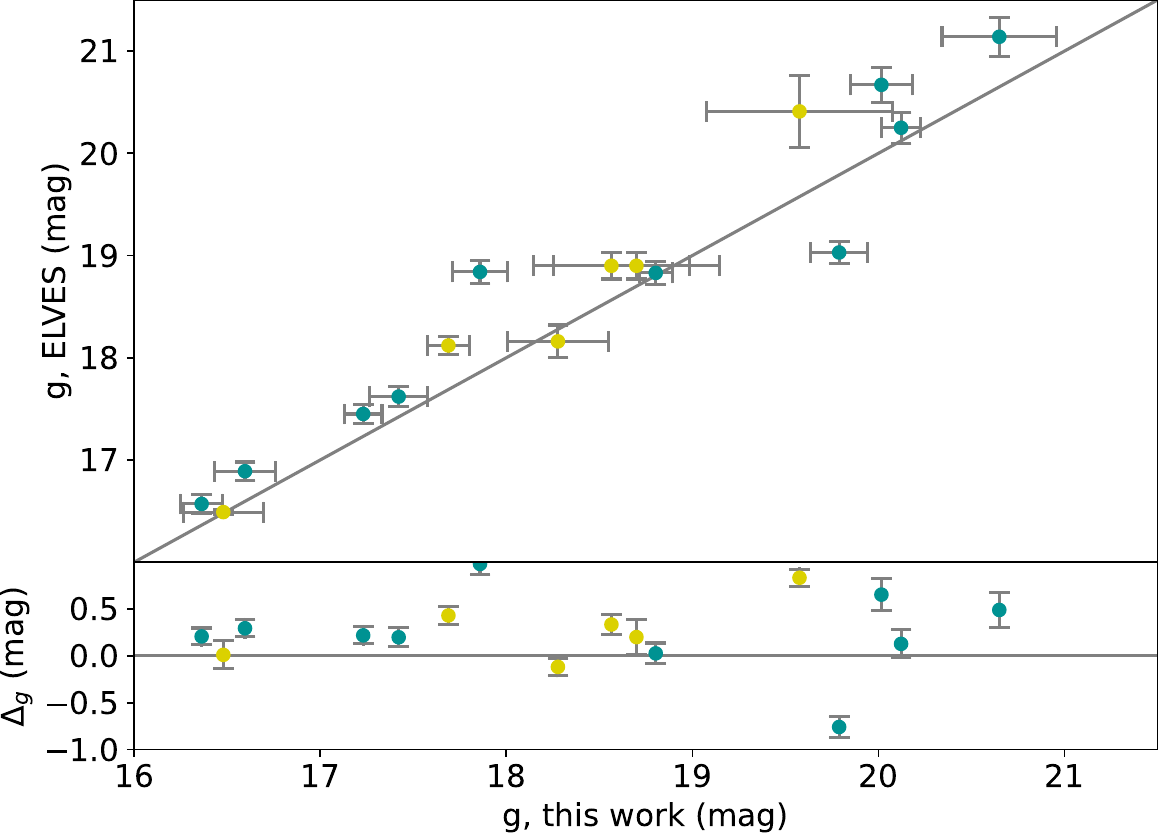}
	\caption{Comparison of the photometry performed here and dwarf galaxies in ELVES. The thin line corresponds to unity. Green dots correspond to NGC\,3521 dwarfs and yellow dots to M\,106 dwarfs.}
	\label{fig:phot}
\end{figure}

To test our photometry, we compare the apparent magnitudes with the ones derived in ELVES \citep{2022ApJ...933...47C}, where 17 dwarf galaxies overlap with the dwarfs studied here. In Fig.\,\ref{fig:phot} we show the difference between our and their photometry. We find that the ELVES photometry yields fainter apparent magnitudes for the same dwarfs with a mean of 0.22\,mag. The standard deviation of the differences is 0.28\,mag. 

\section{Discussion}

\subsection{Scaling relations}

\begin{figure*}[ht]
	\includegraphics[width=0.49\linewidth]{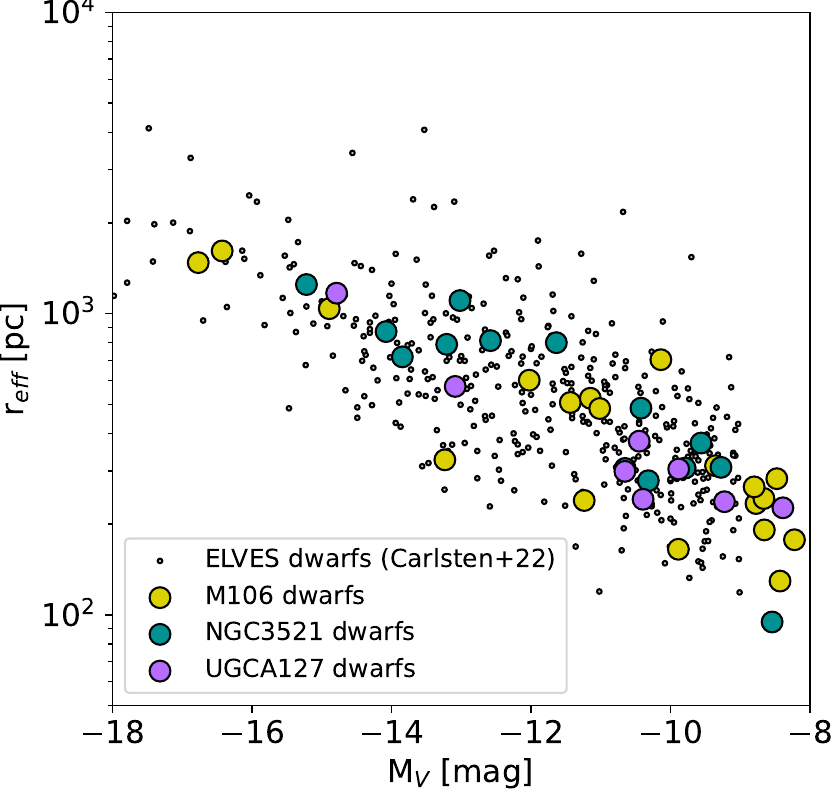}
    	\includegraphics[width=0.49\linewidth]{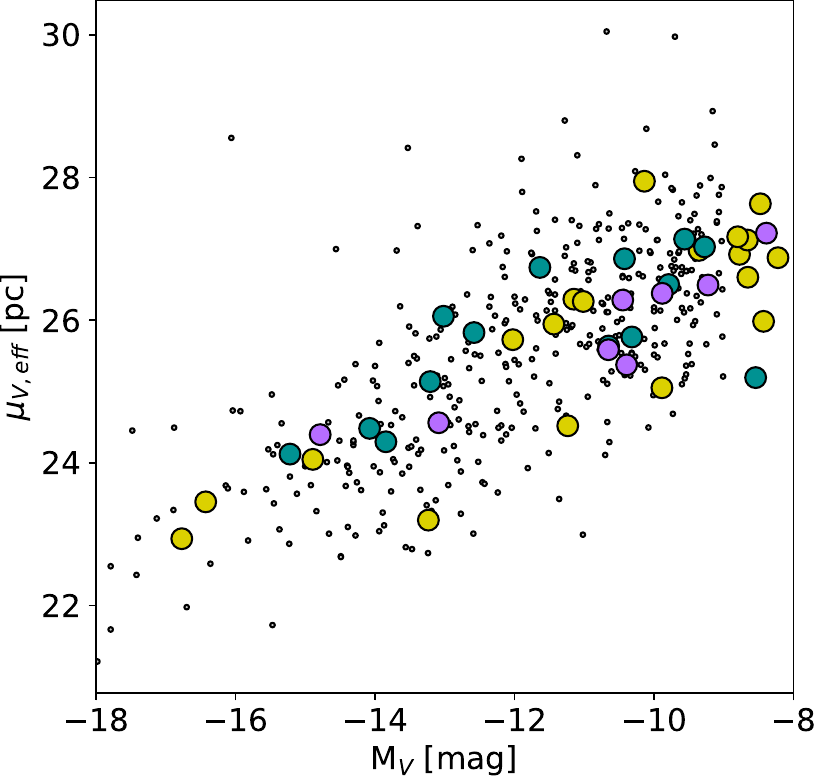}
	\caption{Luminosity - effective radius (left) and luminosity - effective surface brightness (ref) relations for a reference sample from ELVES (small gray dots, \citealt{2022ApJ...933...47C}) and the dwarf galaxies and candidates in our survey footprint (large colored dots).}
	\label{fig:scaling}
\end{figure*}

Dwarf galaxies follow several scaling relations, such as the luminosity - size or luminosity - surface brightness relations \citep{2012AJ....144....4M,2021MNRAS.506.5494P}. Especially interesting is the latter, because the surface brightness is independent of the distance, while the luminosity is not. This means that depending on the assumed distance, a dwarf galaxy candidate may move away from the relation, indicating a non-dwarfish nature \citep{2017A&A...597A...7M}.
In Fig\,\ref{fig:scaling} we compare the structural parameters of dwarf galaxies and dwarf galaxy candidates in our survey footprint with those of the ELVES survey. We plot the luminosity (assuming the distance to the respective host) as a function of the effective radius and surface brightness. All our measured properties are consistent with the scaling relations defined by the ELVES dwarfs, which makes them good dwarf candidates. Our survey extends the limit by roughly 0.5\,mag. 

\subsection{Luminosity function}

\begin{figure}[ht]
	\includegraphics[width=\linewidth]{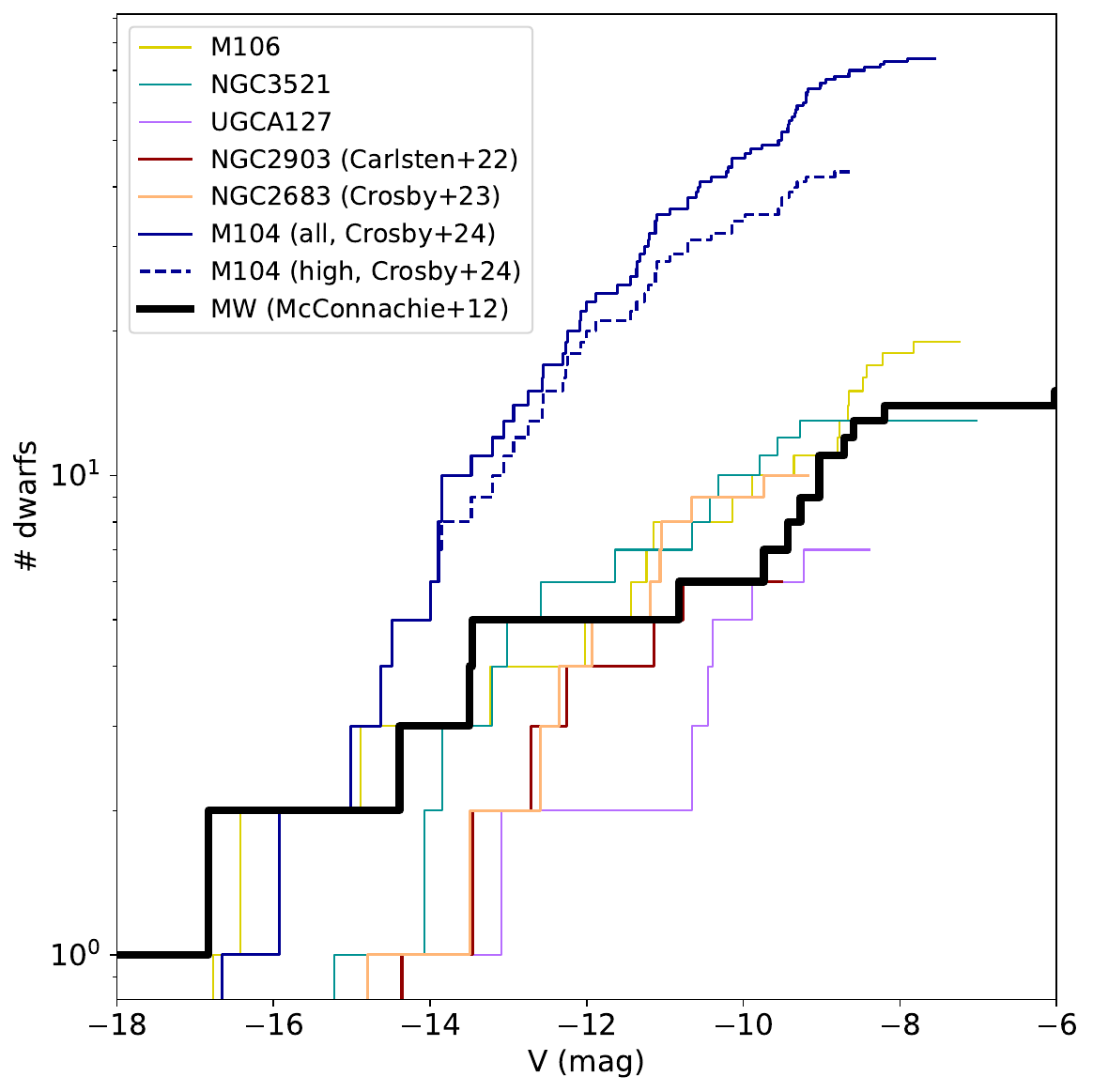}
	\caption{Cumulative luminosity functions for the galaxy systems studied in our Subaru survey (this work, \citealt{2023MNRAS.521.4009C,2024MNRAS.527.9118C}), NGC2903 \citep{2022ApJ...933...47C}, which was in our sample but for which we did not discover additional dwarf galaxies, and the Milky Way \citep{2012AJ....144....4M}. For M\,104 we show both the luminosity function for all dwarfs, including the low-probability candidates (line), and when only considering the high-probability candidates (dashed line).}
	\label{fig:lf}
\end{figure}

A key measurement of galaxy halos is the cumulative luminosity function. It describes the summed number of subhaloes/dwarf galaxies as a function of luminosity. In $\Lambda$CDM, we expect a higher count for more massive haloes \citep{2019ApJ...870...50J}, with ample of scatter though \citep{2019ApJ...885..153B}. In our survey we have a range of masses, from the low-mass galaxy UGCA127\footnote{We calculate the halo mass of UGCA127 based on abundance matching \citep{2010ApJ...717..379B} with a stellar mass of 10$^{9.2}$\,M$\odot$ \citep{2007AJ....134.1849W}.} (M$_{halo}=1.6\times 10^{11}$\,M$_\odot$) to the massive M\,104 (M$_{halo}=1.5\times 10^{13}$\,M$_\odot$, \citealt{2020A&A...643A.124K}), i.e., two orders of magnitudes. This is reflected in the luminosity function shown in Fig.\,\ref{fig:lf}. 

NGC\,3521, NGC\,2683, and NGC\,2903 have a similar luminosity function as the Milky Way. These three galaxies have similar stellar masses (M$_{\rm NGC3521}=12\times10^{10}$\,M$_\odot$, M$_{\rm NGC2683}=6\times10^{10}$\,M$_\odot$, M$_{\rm NGC2903}=7\times10^{10}$\,M$_\odot$, \citealt{2013AJ....145..101K})  as the Milky Way (M$_{\rm MW}=5.0^{+0.4}_{-0.5}\times10^{10}$\,M$_\odot$), so this expected (assuming a similar stellar-to-halo mass ratio). Interestingly enough, UGCA127 almost reaches the same number of dwarf galaxies as the Milky Way at the cut-off. With a stellar mass of only $1.5\times10^{9}$\,M$_\odot$ it has the same mass as the LMC. \citet{2020A&A...644A..91M} investigated the abundance of dwarf galaxies around low-mass galaxies with similar masses and found the observed range to be within 0 to 3 at $<-8$\,mag. Expectations from $\Lambda$CDM for such galaxies range between 0 to 8. In that sense, UGCA127 is within the range of expected dwarf galaxy satellites.

\subsection{Lopsidedness}
Visually, the observed system look lopsided, for example, for M\,106, most satellites are scewed towards South. However, we need to quantify this using statistical means.
Following the implementation of \citet{2024A&A...690A.110H}, we measured the lopsidedness of the satellite systems using the wedge metric. The wedge metric works as following: A wedge is centered on the host galaxy with an opening angle. For each opening angle (ranging from one to 360 degrees), this wedge is rotated by one degree and the number of satellites enclosed within the wedge is estimated. For each size of the opening angle the most populated wedge is recorded. Then the frequency and statistical significance (i.e. the p-value) of such a populated wedge is estimated through random sampling. Because some of the randomly sampled systems might show equal or more extreme wedge
populations at different opening angles, we needed an additional quantification of the significance of the lopsidedness. For that, 100'000 isotropic systems with the same number of satellites as the observed system are generated. Then, on each such system, the wedge metric is applied and a p-value measured. The p-value distribution from these isotropic systems is then compared to the measured p-value of the observed system, which gives the true significance (or meta p-value as \citealt{2024A&A...690A.110H} calls it). 

We have applied the wedge metric to the systems from our Subaru survey (M\,106, M\,104, NGC\,2683, NGC\,3521, and UGCA\,127). We have not applied it for NGC\,2903, because \citet{2024A&A...690A.110H} already applied it and no new dwarf galaxies have been found since their tests. For NGC\,2903, \citet{2024A&A...690A.110H} reports no lopsidedness. Here, we find that none of the systems exhibits a statistically significant lopsided satellite distribution (with an alpha level of 5\%).

\section{Conclusions}

In this paper, we have inspected the  fields around M\,106, NGC\,2903, NGC\,3521, and UGCA127 from our Hyper Suprime Camera survey conducted with the Subaru telescope. The other fields previously investigated in our Subaru survey were around NGC\,2683 and M\,104, where 4 and 40 new dwarf galaxy candidates were discovered. We did not find any new dwarf galaxies around NGC\,2903. The NGC2903 group has 7 confirmed satellites out to 300\,kpc \citep{2022ApJ...933...47C}.   For all these galaxies, we have covered the full virial radius to get a complete census of the dwarf galaxy systems.  To assess our completeness, we have injected artificial dwarf galaxies into the images. We find a limiting surface brightness magnitude of 26.0 mag arcsec$^{-2}$ for the effective surface brightness. Fainter than this value our recovery rate dropped below 50 percent, however, we detect dwarf galaxies as faint as 28 mag arcsec$^{-2}$ (in effective surface brightness). We have further rejected objects with a  effective radius smaller than 3\,arcsec, because confusion with background galaxies becomes severe for such small objects. 
In the present work, we have found 21 new dwarf galaxy candidates. Together with the results from \citet{2023MNRAS.521.4009C,2024MNRAS.527.9118C}, this gives a grand total of 65 newly discovered dwarf galaxy candidates around Local Volume galaxies based on our survey.

We have performed surface photometry for all known dwarf galaxies and candidates in the survey footprint. To assess our photometry, we have compared  our derived apparent magnitudes with the ones from the ELVES survey. We find a general agreement between our numbers and theirs, however, ELVES finds $\approx$0.2 fainter magnitudes than we do. We deem this within the expected scatter from the uncertainties. It could also arise from different masking schemes of foreground stars and background galaxies. 

The dwarf galaxy candidates follow the scaling relations as expected from other dwarfs in the nearby universe, making them good dwarf candidates based on their morphology and structural parameters. Assuming all of them are indeed dwarf galaxies associated with their respective host, we can construct the luminosity function of each target galaxy environment. Unsurprisingly, the most massive galaxy hosts the highest number of dwarf galaxies, while the lowest mass galaxy the least. This trend is expected in a $\Lambda$CDM universe where more massive haloes host a higher number of subhaloes \citep[e.g., ][]{2019ApJ...870...50J}. The three galaxies with similar masses as the Milky Way (NGC\,3521, NGC\,2903, and NGC\,2683) also exhibit a similar luminosity function as the Milky Way. 

We  studied the satellites distributions in terms of their overall lopsidedness using the wedge metric, which was found to be the best metric to find overall lopsidedness. Applying it to the five host galaxies from our survey we find no significant lopsidedness with an alpha level of 5\%. 

To make more detailed investigations, the dwarf galaxies need to be confirmed either through distance or velocity measurements. We have taken up the effort to follow up the candidates with the new IFUM instrument mounted at the Magellan/Clay telescope. All six dwarf galaxy candidates we selected from the dwarf galaxy candidates around M\,104 \citep{2024MNRAS.527.9118C} were confirmed to be members \citep{2025MNRAS.536.2072C} . This is an encouraging result, as it shows the effectiveness of our dwarf galaxy detection strategy by visual inspection. More observations are scheduled in 2025.

\section*{Acknowledgements}
O.M. and N.H. are grateful to the Swiss National Science Foundation for financial support under the grant number PZ00P2\_202104. Marcel S. Pawlowski
acknowledges funding via a Leibniz-Junior Research Group (project number
J94/2020).

\bibliographystyle{aa}
\bibliography{ojap}

\appendix

\newpage

\setcounter{table}{0}
\renewcommand{\thetable}{A\arabic{table}}

\setcounter{figure}{0}
\renewcommand{\thefigure}{A\arabic{figure}}


\begin{table*}[ht]
\vspace{-0.5cm}
\caption{The known dwarfs ($^*$) and new dwarf galaxy candidates around M\,106.}             
\centering                          
\begin{tabular}{lcccrcc}        
\hline\hline                 
       Name &          RA  &          Dec &  $g$ &   $r_{eff}$ &  $\mu_0$ &  
        $\mu_{eff}$ \\ 
        &         (J2000.0) &          (J2000.0) &  (mag) &   (arcsec) &  (mag/arcsec$^2$) &  
       (mag/arcsec$^2$) \\ 
\hline      \\[-2mm]                  
  dw1209+4621 & 12:09:53.30 & +46:21:03.98 &  21.9 $\pm$     0.5 &  6.0 $\pm$   1.5 &  27.5 $\pm$     0.1 &  27.8 $\pm$     0.1 \\
     N4258dwC$^*$ & 12:10:26.76 & +46:44:48.90 &  18.5 $\pm$     0.2 &  6.4 $\pm$   0.6 &  23.0 $\pm$     0.1 &  24.5 $\pm$     0.1 \\
  dw1211+4652 & 12:11:35.61 & +46:52:57.64 &  20.4 $\pm$     0.2 &  8.4 $\pm$   0.9 &  26.4 $\pm$     0.0 &  27.0 $\pm$     0.1 \\
  dw1213+4624 & 12:13:31.85 & +46:24:37.56 &  21.1 $\pm$     0.3 &  5.1 $\pm$   0.8 &  25.9 $\pm$     0.1 &  26.6 $\pm$     0.2 \\
  dw1214+4743 & 12:14:52.87 & +47:43:17.29 &  20.9 $\pm$     0.5 &  6.3 $\pm$   1.6 &  25.7 $\pm$     0.2 &  26.9 $\pm$     0.3 \\
  dw1214+4800 & 12:14:55.55 & +48:00:57.45 &  22.5 $\pm$     1.1 &  6.1 $\pm$   3.7 &  27.2 $\pm$     0.4 &  28.3 $\pm$     0.7 \\
  dw1215+4738 & 12:15:03.88 & +47:38:32.83 &  21.2 $\pm$     0.3 &  7.6 $\pm$   1.5 &  26.9 $\pm$     0.1 &  27.6 $\pm$     0.2 \\
  dw1216+4518 & 12:16:53.40 & +45:18:10.47 &  19.8 $\pm$     0.2 &  4.4 $\pm$   0.6 &  24.6 $\pm$     0.1 &  25.1 $\pm$     0.1 \\
LV-J1218+4655$^*$ & 12:18:11.12 & +46:55:01.59 &  16.5 $\pm$     0.2 &  8.8 $\pm$   0.9 &  21.8 $\pm$     0.1 &  23.2 $\pm$     0.1 \\
        KK132$^*$ & 12:19:06.08 & +47:43:49.24 &  18.3 $\pm$     0.3 & 13.6 $\pm$   1.9 &  24.6 $\pm$     0.1 &  25.9 $\pm$     0.2 \\
       KDG101$^*$ & 12:19:09.09 & +47:05:23.33 &  14.8 $\pm$     0.1 & 27.9 $\pm$   1.3 &  23.4 $\pm$     0.0 &  24.0 $\pm$     0.0 \\
   KKH2011S11$^*$ & 12:20:30.25 & +47:29:26.19 &  19.6 $\pm$     0.5 & 18.9 $\pm$   4.8 &  26.6 $\pm$     0.2 &  27.9 $\pm$     0.3 \\
     N4258dwB$^*$ & 12:20:54.92 & +46:49:47.99 &  18.6 $\pm$     0.4 & 14.1 $\pm$   2.8 &  24.8 $\pm$     0.2 &  26.3 $\pm$     0.2 \\
       DDO120$^*$ & 12:21:15.11 & +45:48:54.34 &  12.9 $\pm$     0.1 & 39.7 $\pm$   2.6 &  22.1 $\pm$     0.1 &  22.9 $\pm$     0.1 \\
       BTS132$^*$ & 12:23:46.16 & +47:39:31.92 &  17.7 $\pm$     0.1 & 16.2 $\pm$   1.0 &  25.1 $\pm$     0.0 &  25.7 $\pm$     0.1 \\
  dw1224+4714 & 12:24:24.13 & +47:14:00.38 &  21.1 $\pm$     0.3 &  6.5 $\pm$   1.0 &  26.8 $\pm$     0.1 &  27.1 $\pm$     0.1 \\
  dw1224+4630 & 12:24:31.60 & +46:30:56.97 &  20.9 $\pm$     0.3 &  7.2 $\pm$   1.4 &  26.4 $\pm$     0.1 &  27.2 $\pm$     0.2 \\
  dw1226+4612 & 12:26:47.88 & +46:12:00.95 &  21.3 $\pm$     0.8 &  3.5 $\pm$   1.3 &  24.5 $\pm$     0.3 &  26.0 $\pm$     0.5 \\
  dw1228+4648 & 12:28:29.61 & +46:48:20.73 &  21.5 $\pm$     0.2 &  4.8 $\pm$   0.7 &  26.3 $\pm$     0.1 &  26.9 $\pm$     0.1 \\
     UGC07639$^*$ & 12:29:53.46 & +47:31:50.50 &  13.3 $\pm$     0.4 & 43.4 $\pm$   9.4 &  22.8 $\pm$     0.2 &  23.5 $\pm$     0.2 \\
\hline
\end{tabular}
\label{tab:group_M106}
\end{table*}

\begin{figure*}[h!]
\raggedright
	\includegraphics[width=0.163\linewidth]{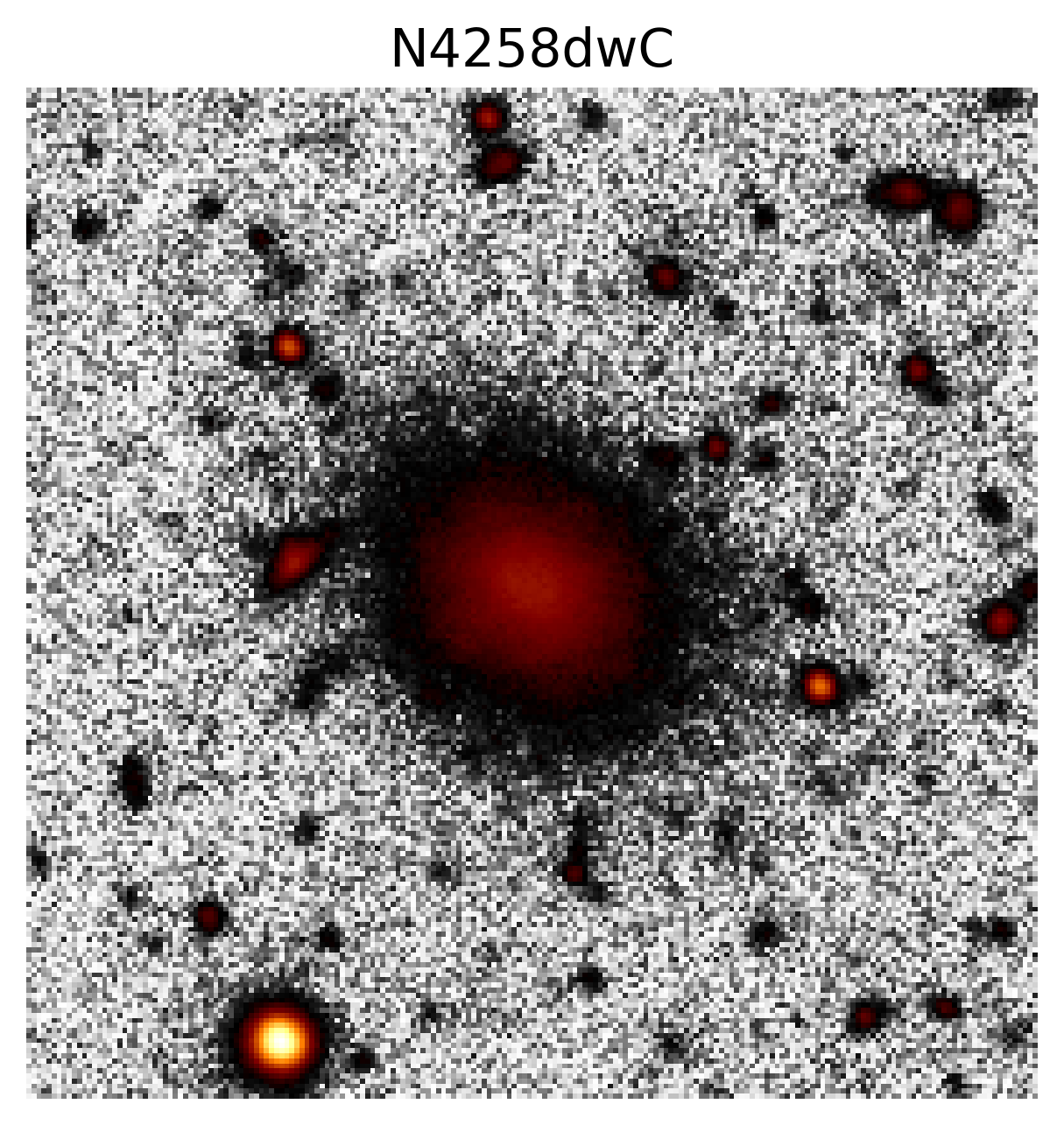}
    \includegraphics[width=0.163\linewidth]{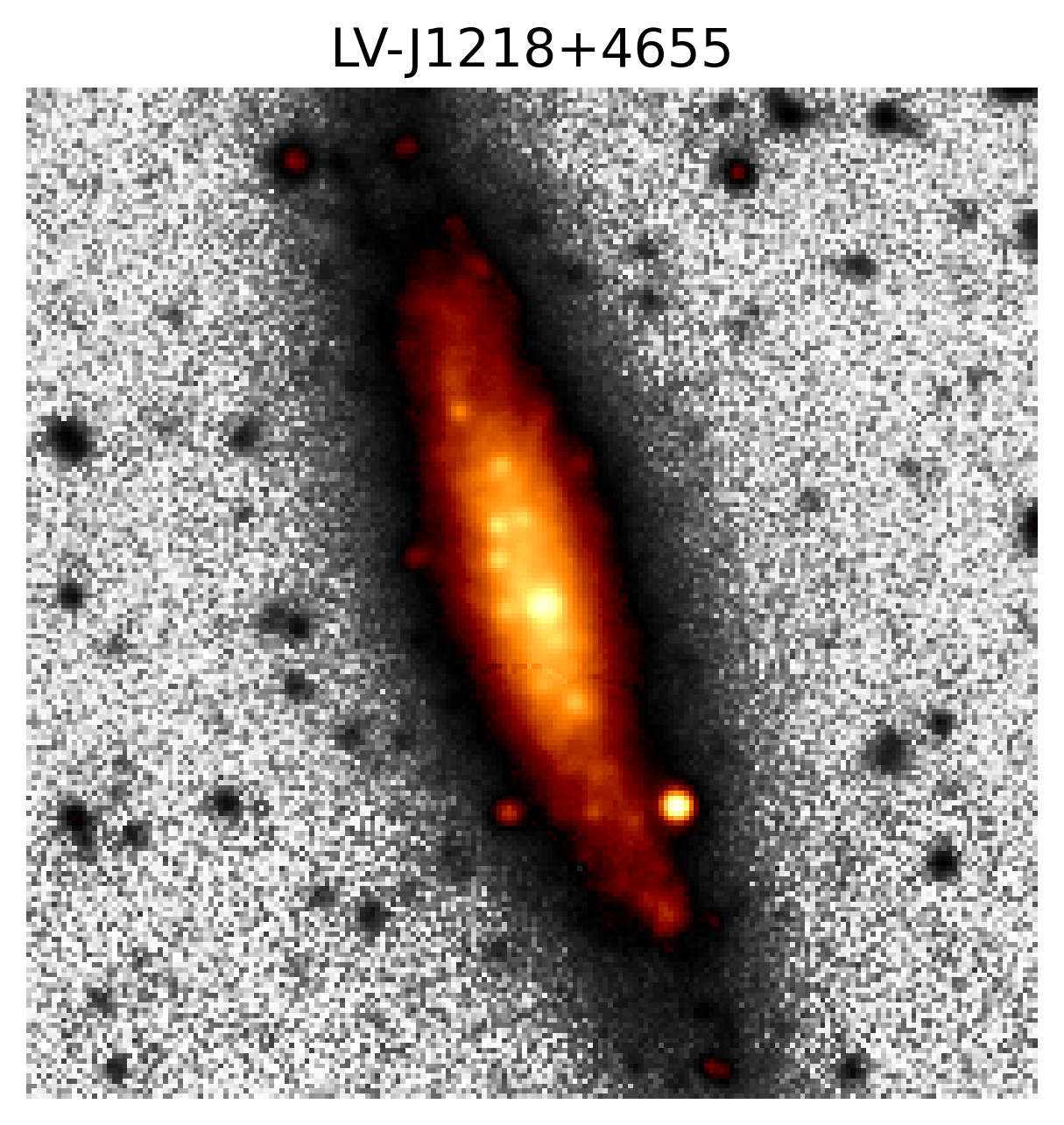}
	\includegraphics[width=0.163\linewidth]{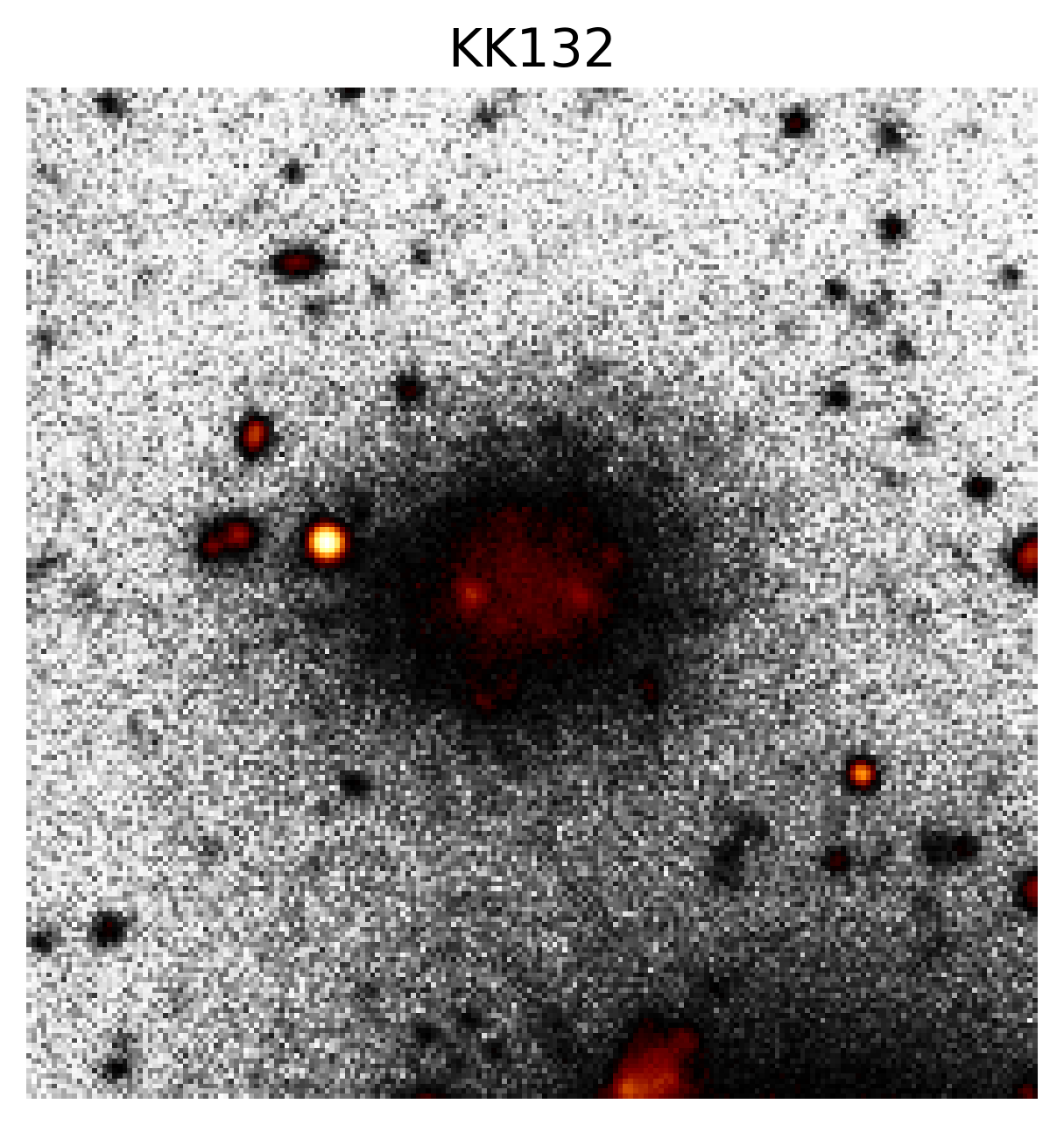}
	\includegraphics[width=0.163\linewidth]{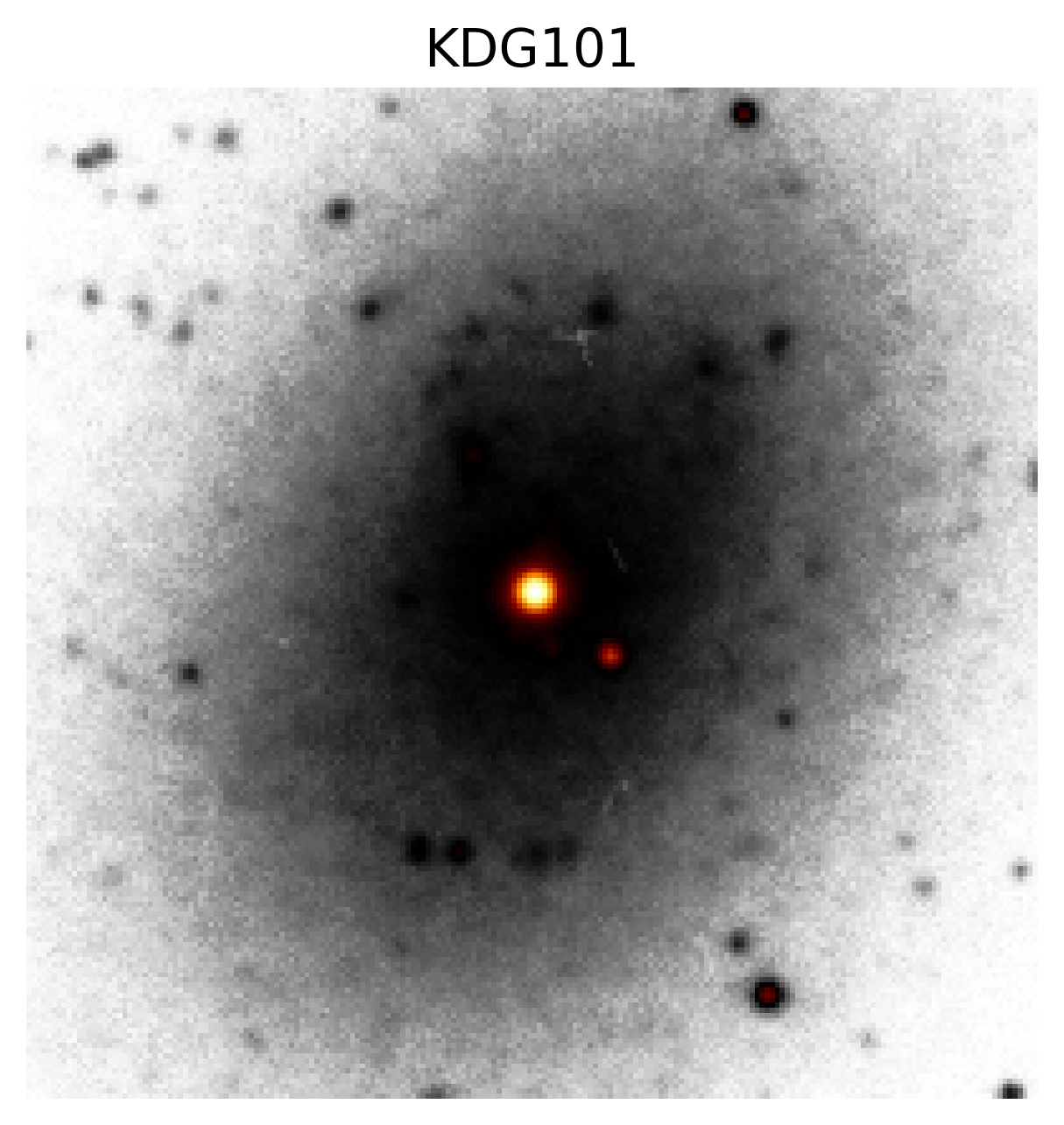}
	\includegraphics[width=0.163\linewidth]{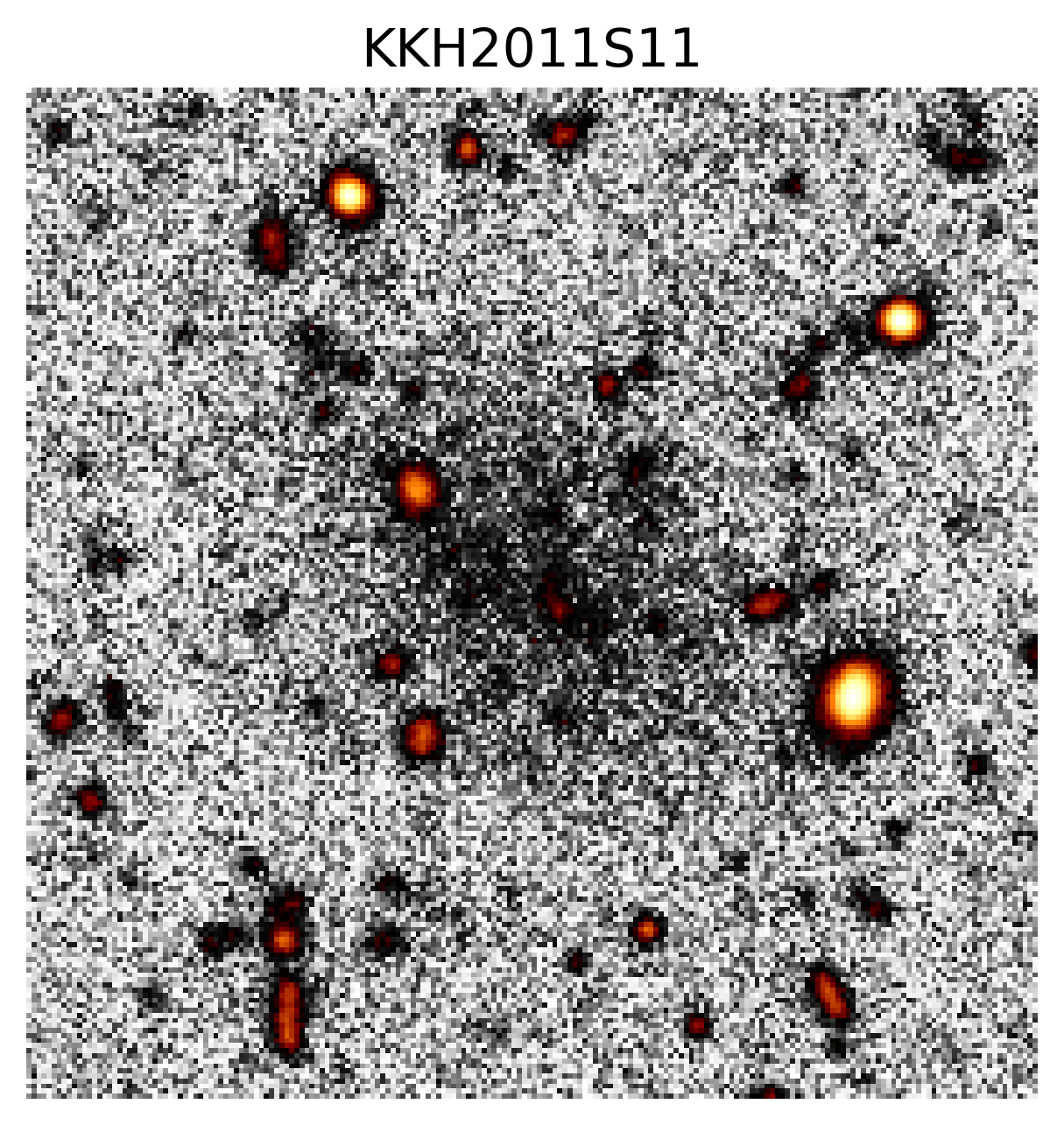}
	\includegraphics[width=0.163\linewidth]{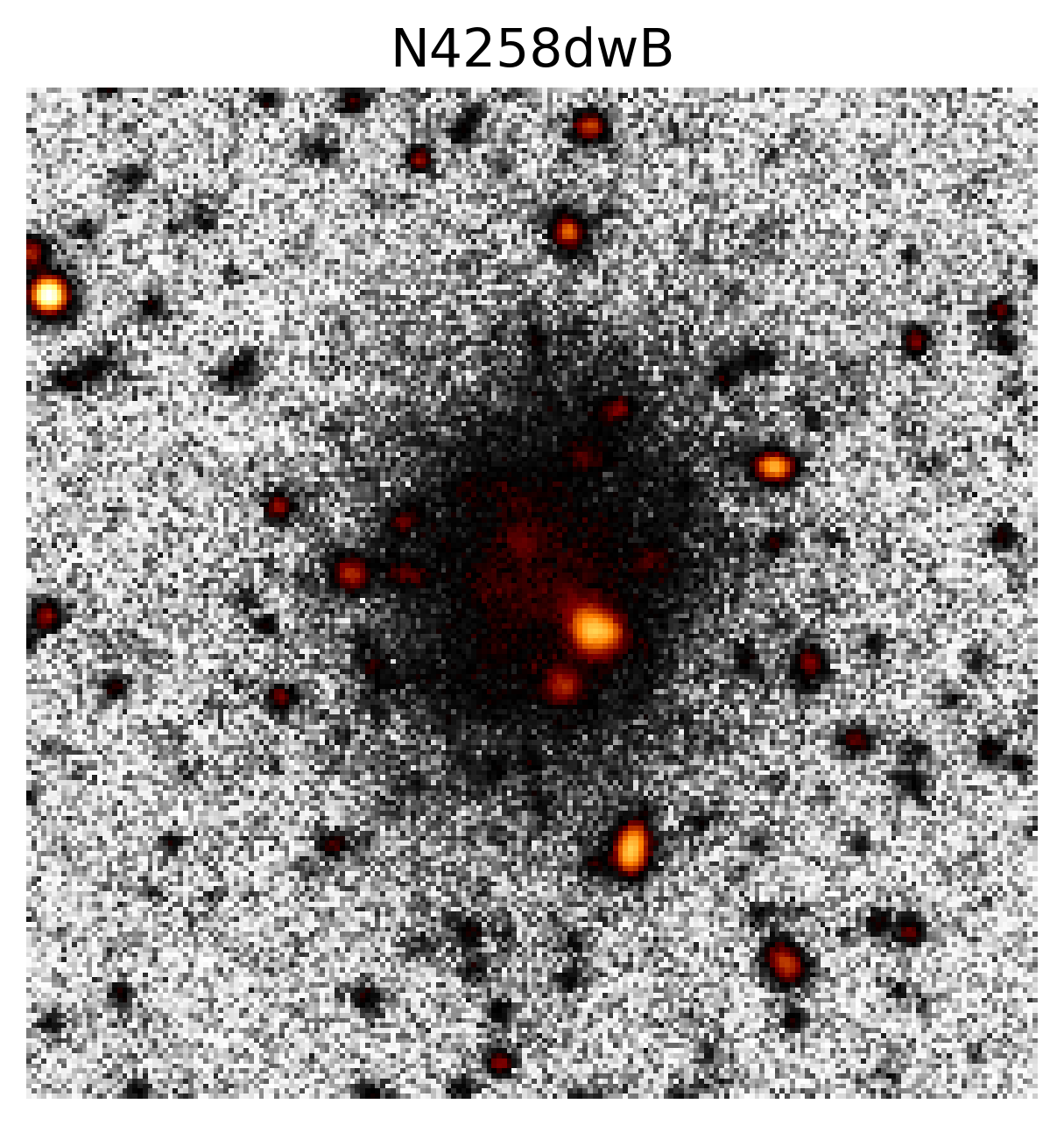}\\
	\includegraphics[width=0.163\linewidth]{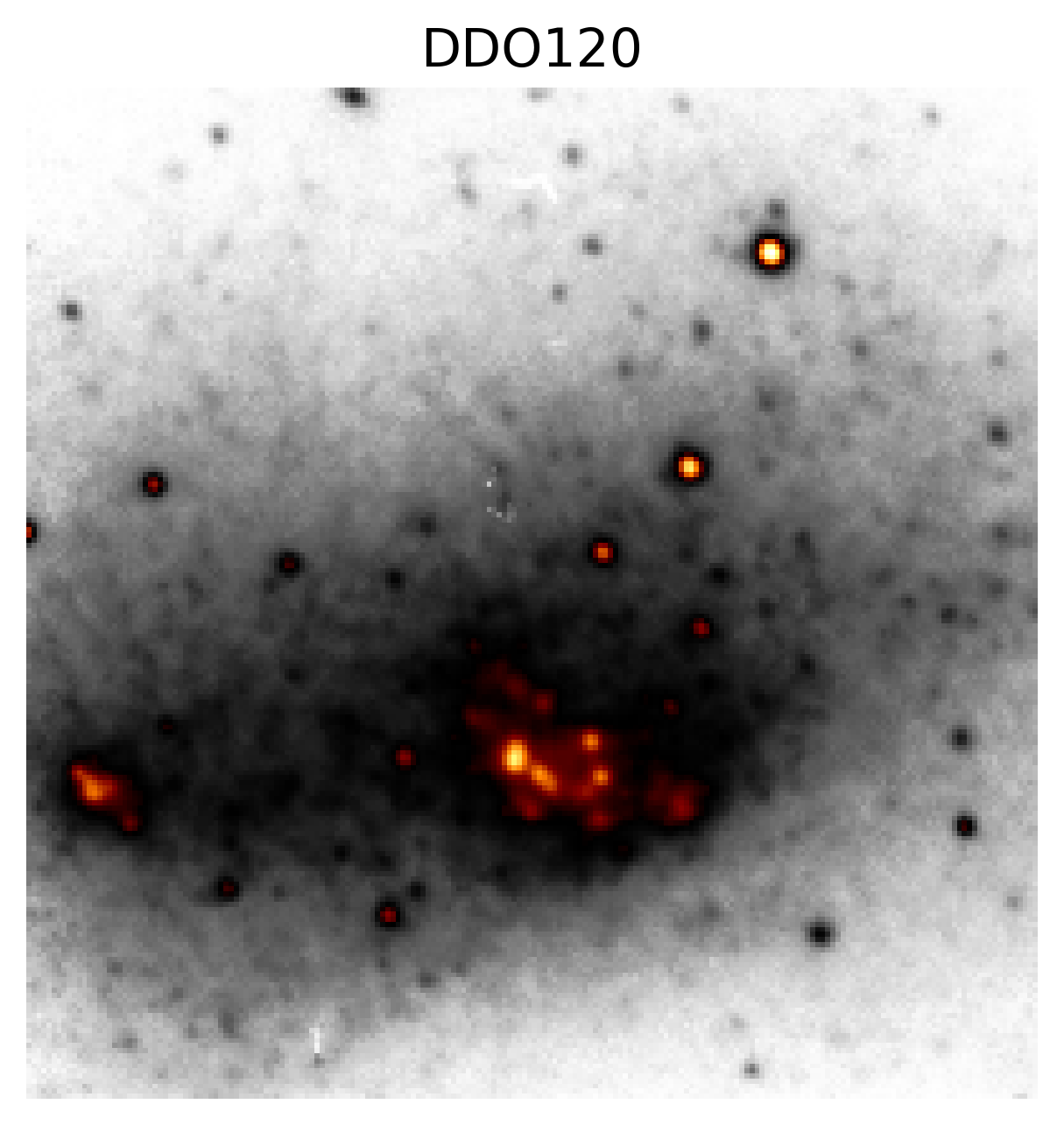}
	\includegraphics[width=0.163\linewidth]{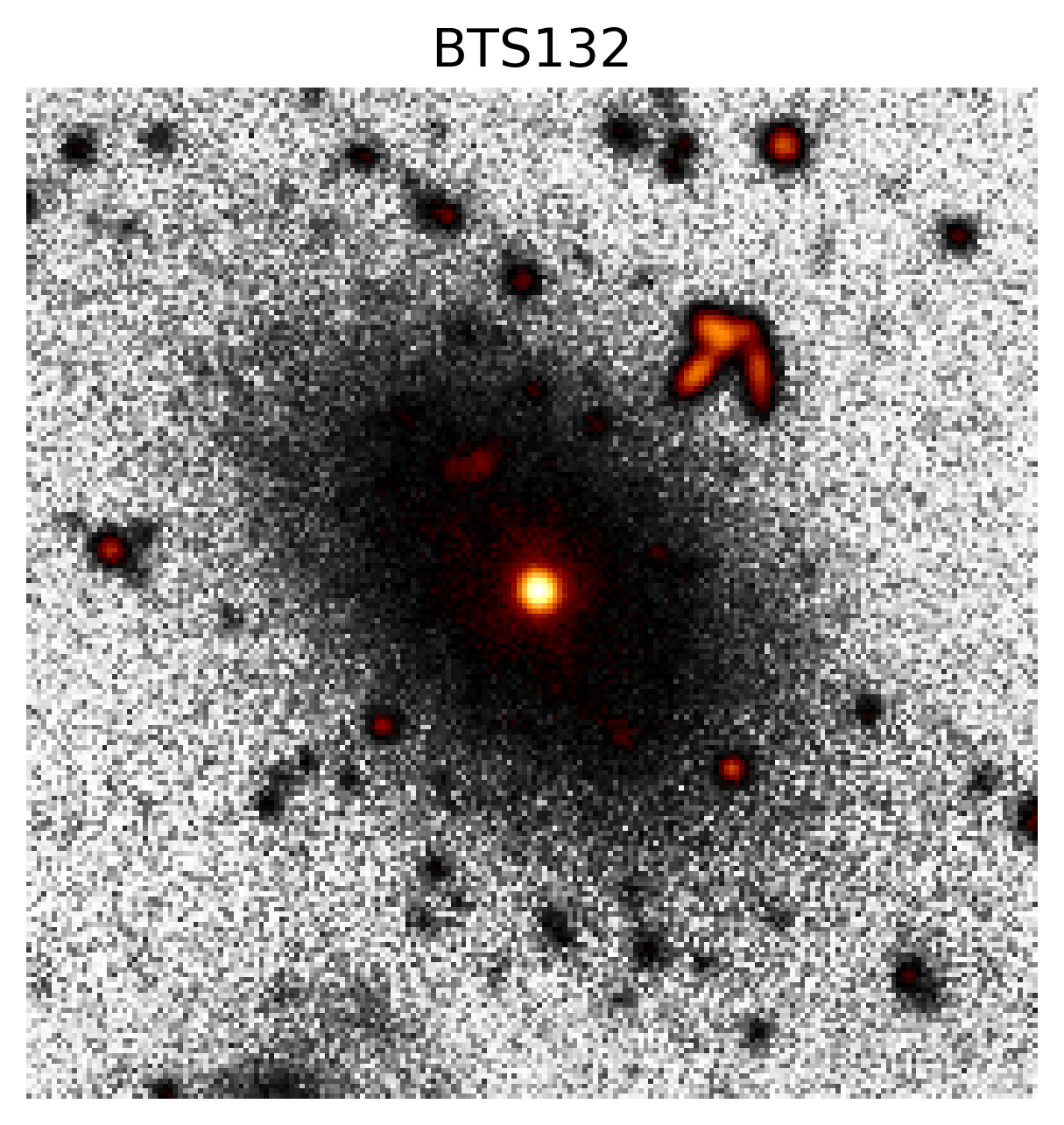}
	\includegraphics[width=0.163\linewidth]{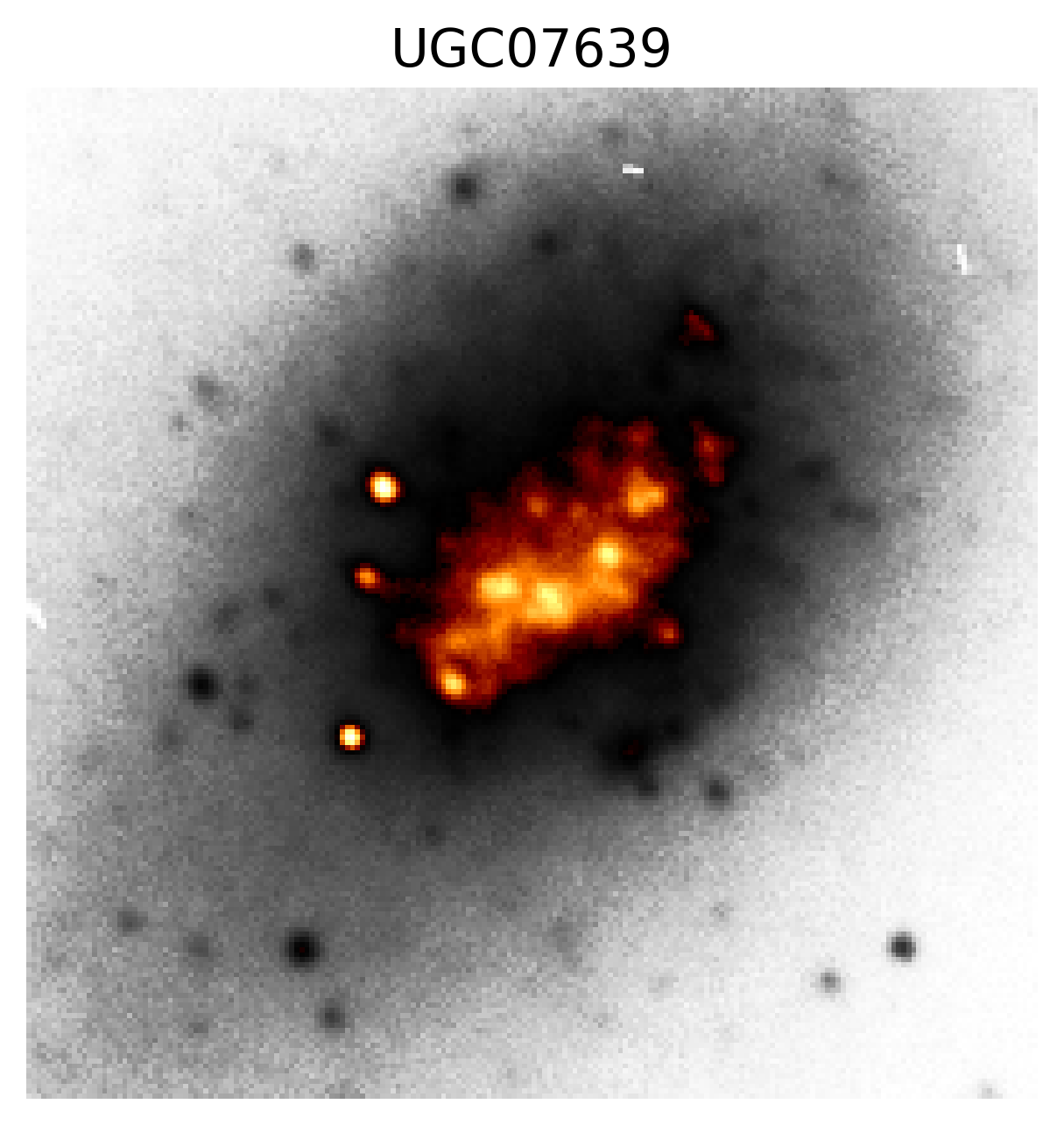}\\
\vspace{0.5cm}
    \includegraphics[width=0.163\linewidth]{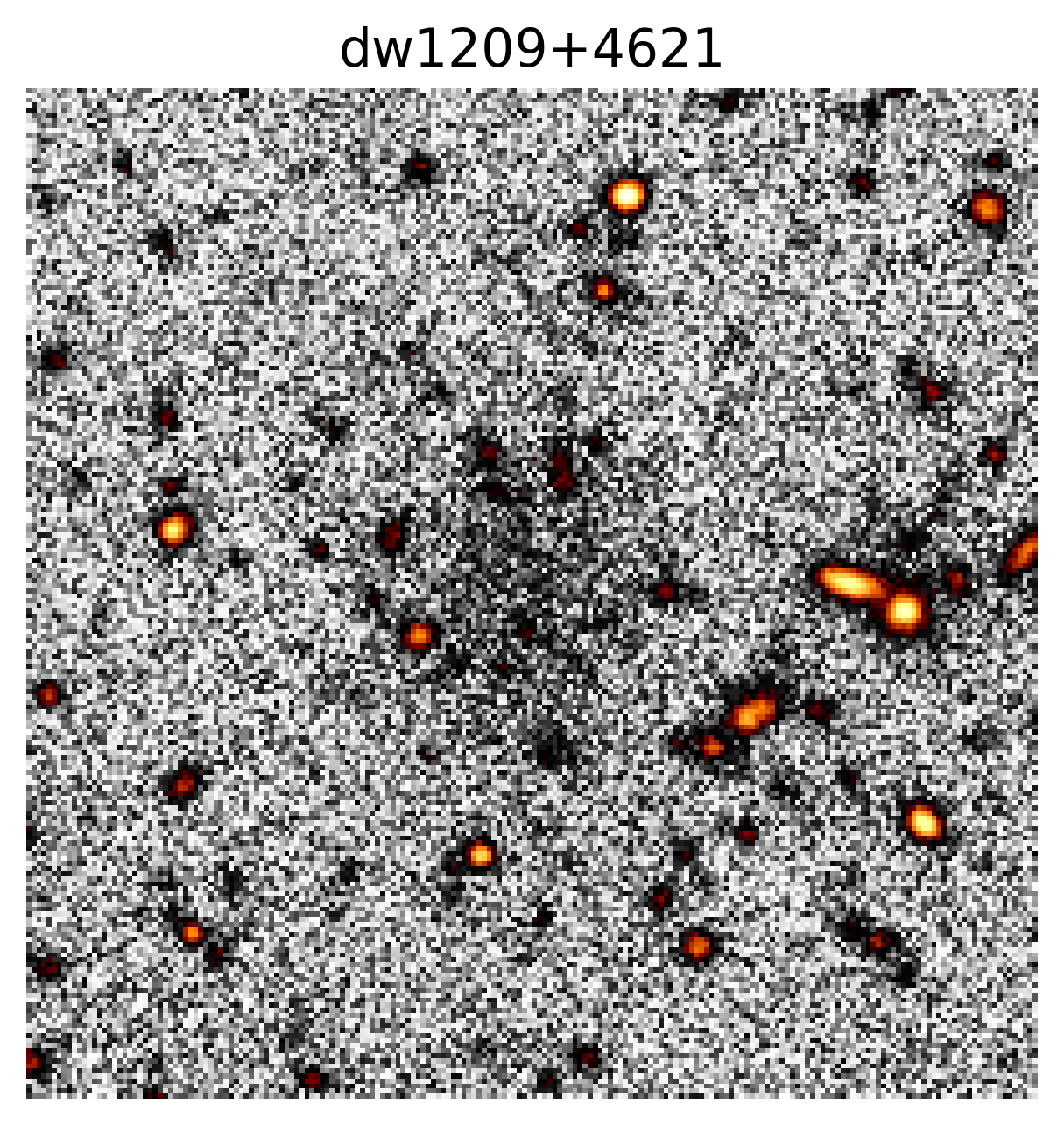}
    \includegraphics[width=0.163\linewidth]{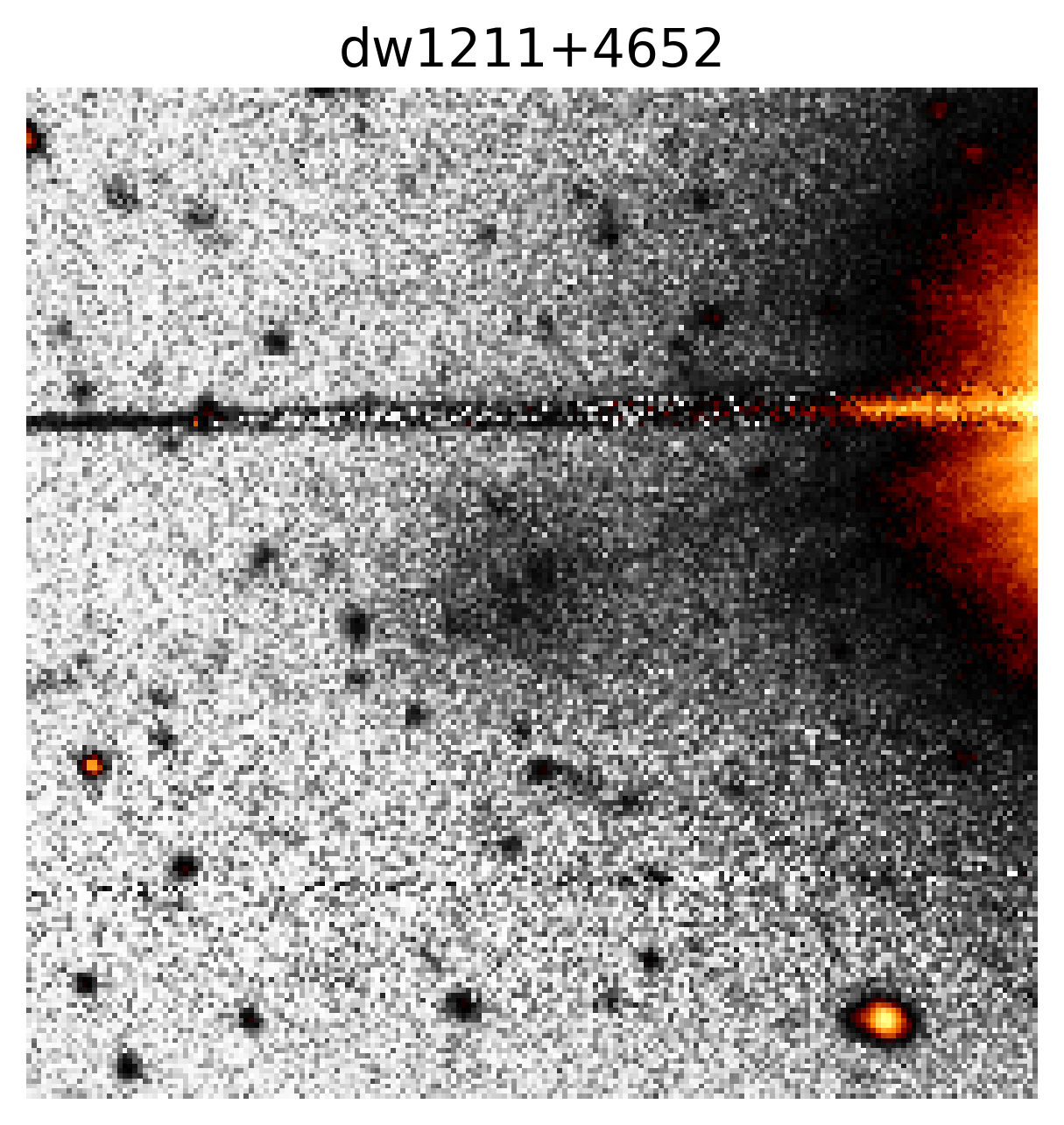}
	\includegraphics[width=0.163\linewidth]{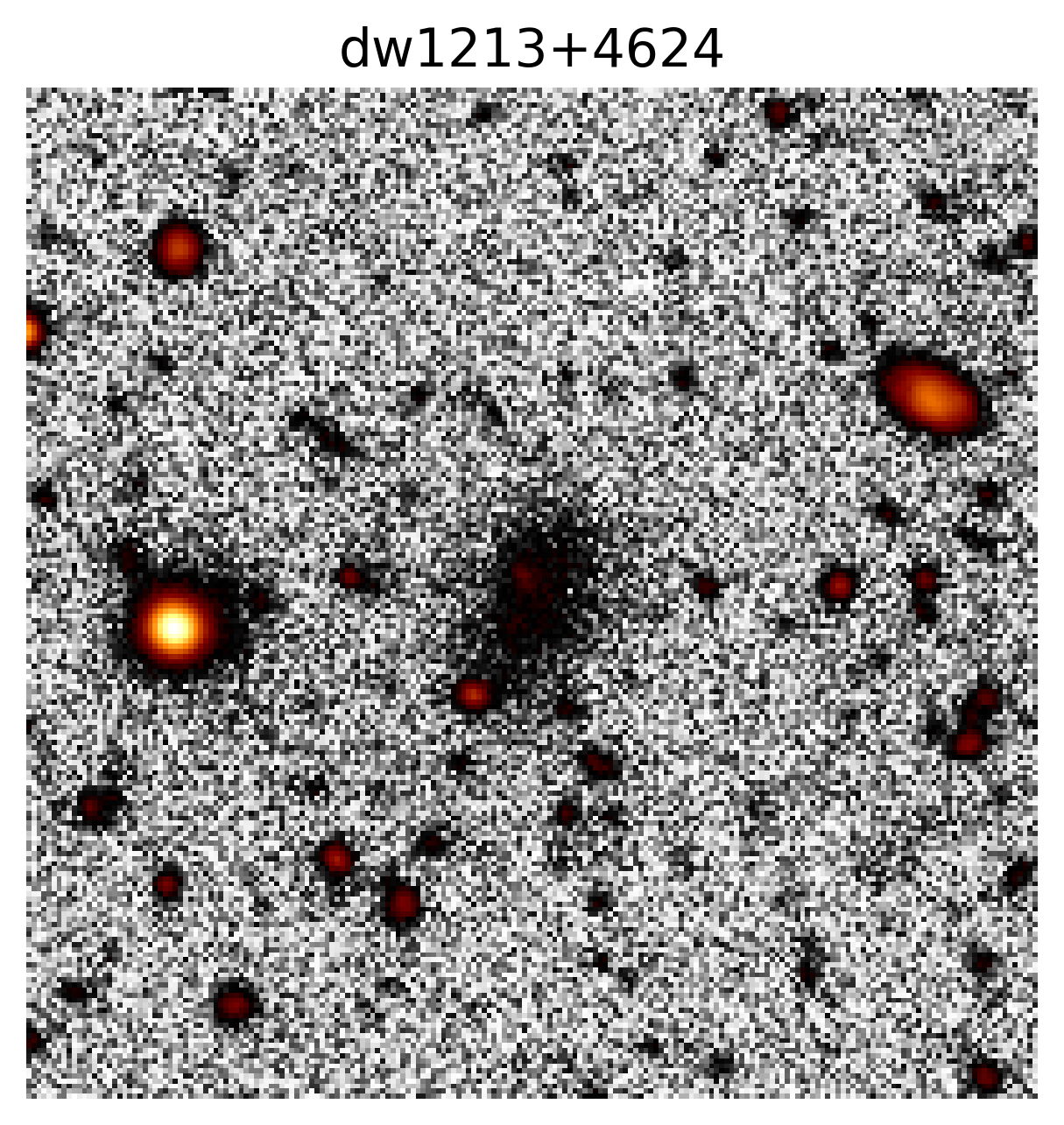}
	\includegraphics[width=0.163\linewidth]{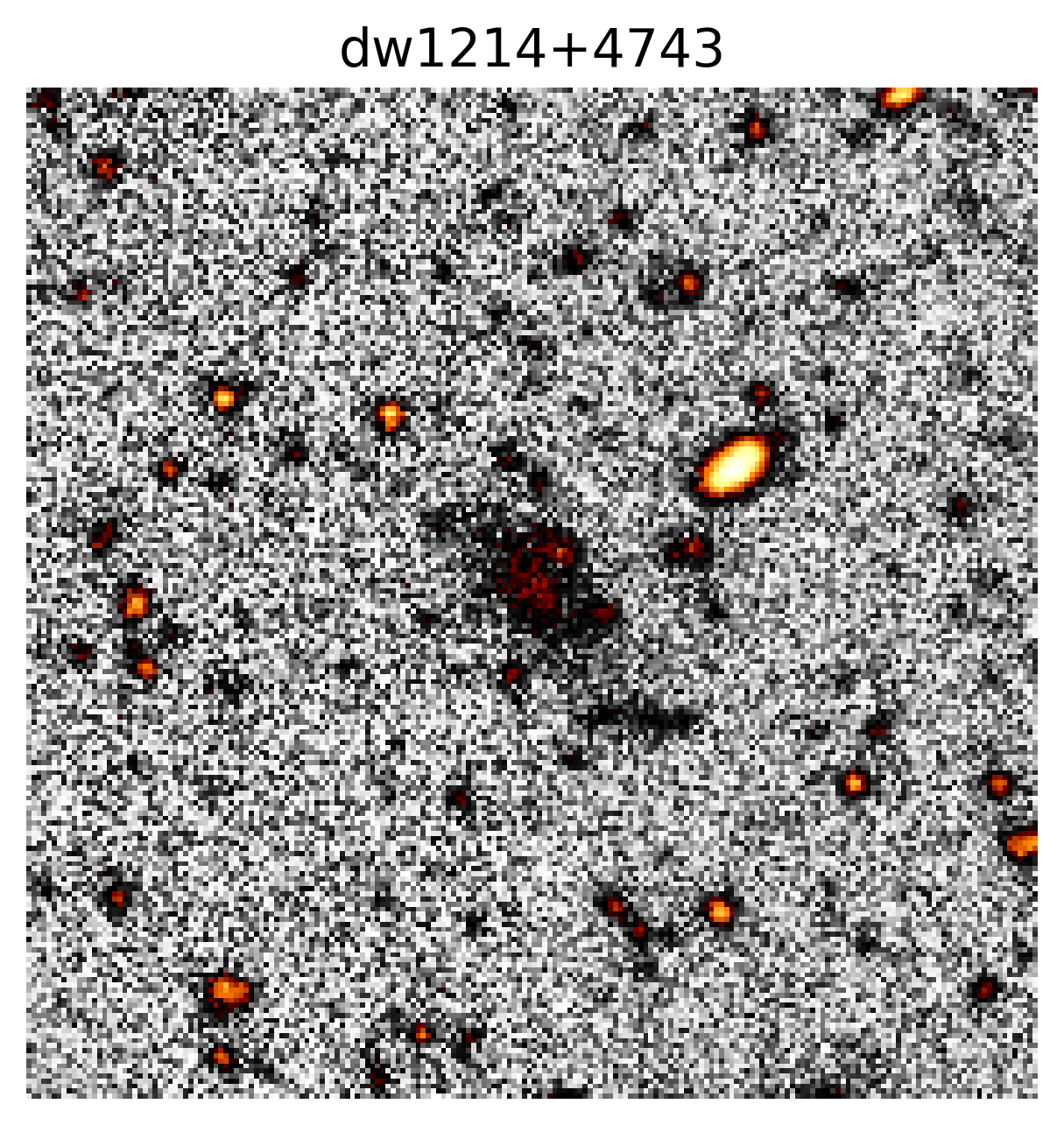}
	\includegraphics[width=0.163\linewidth]{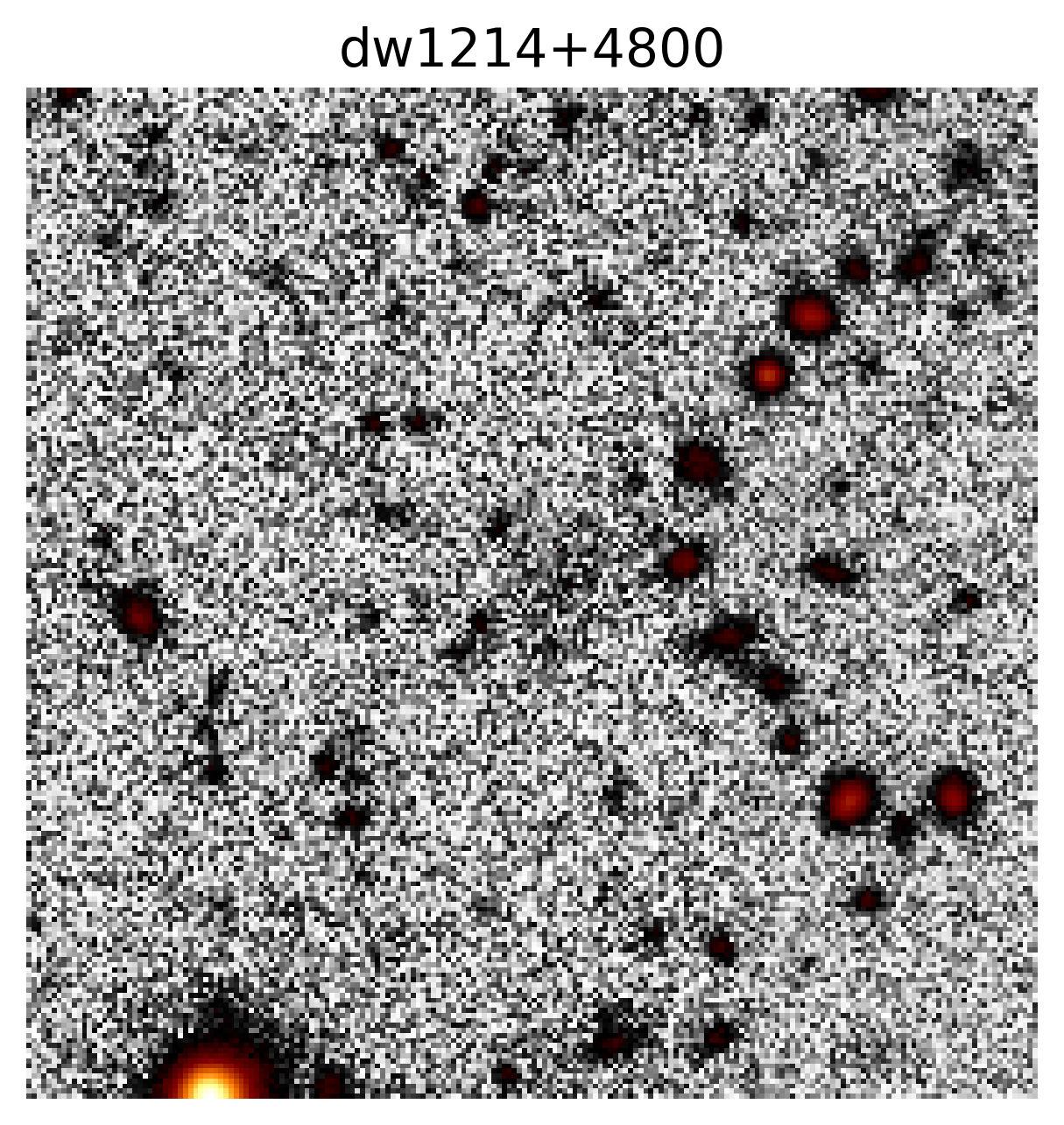}
	\includegraphics[width=0.163\linewidth]{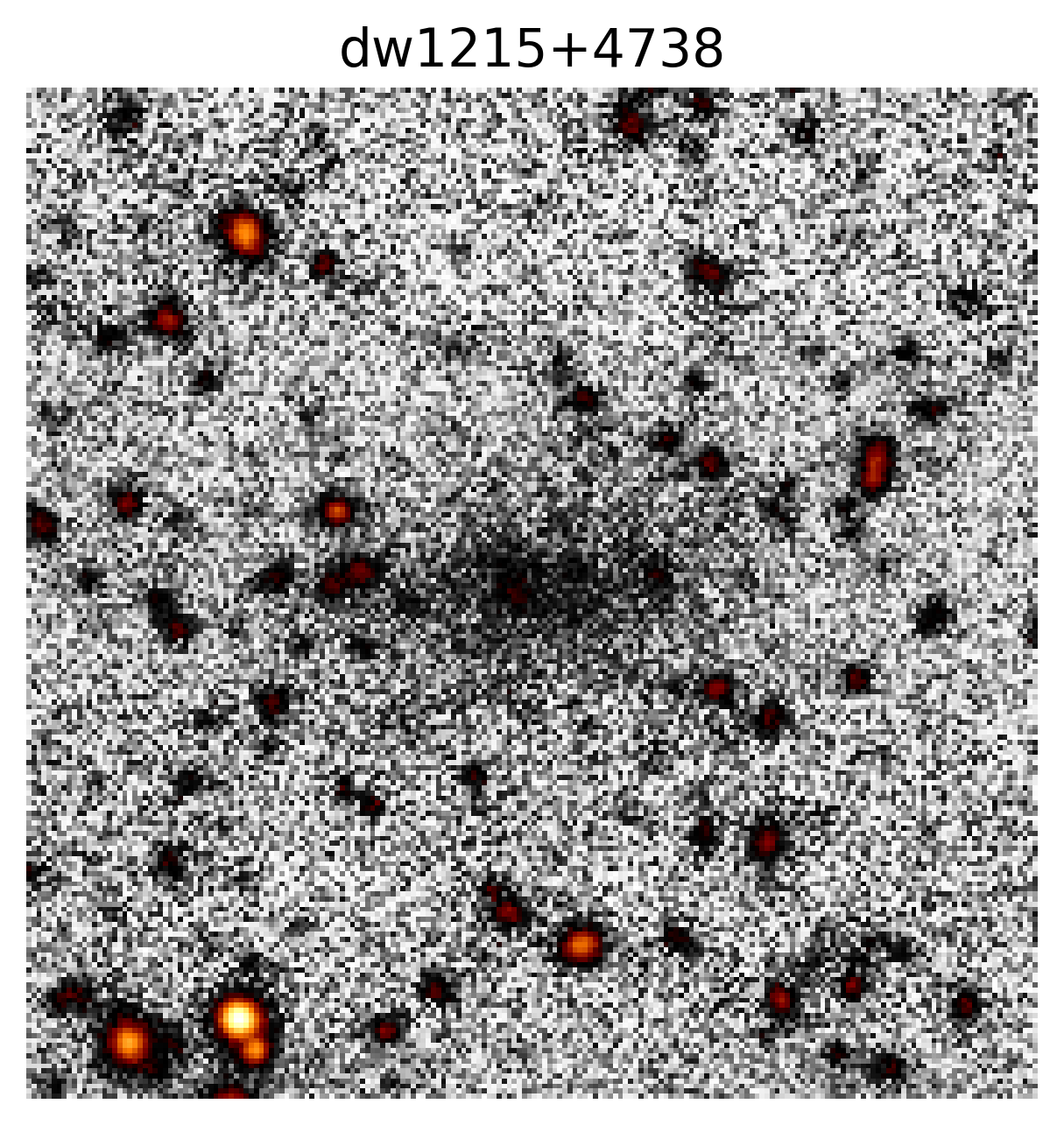}\\
	\includegraphics[width=0.163\linewidth]{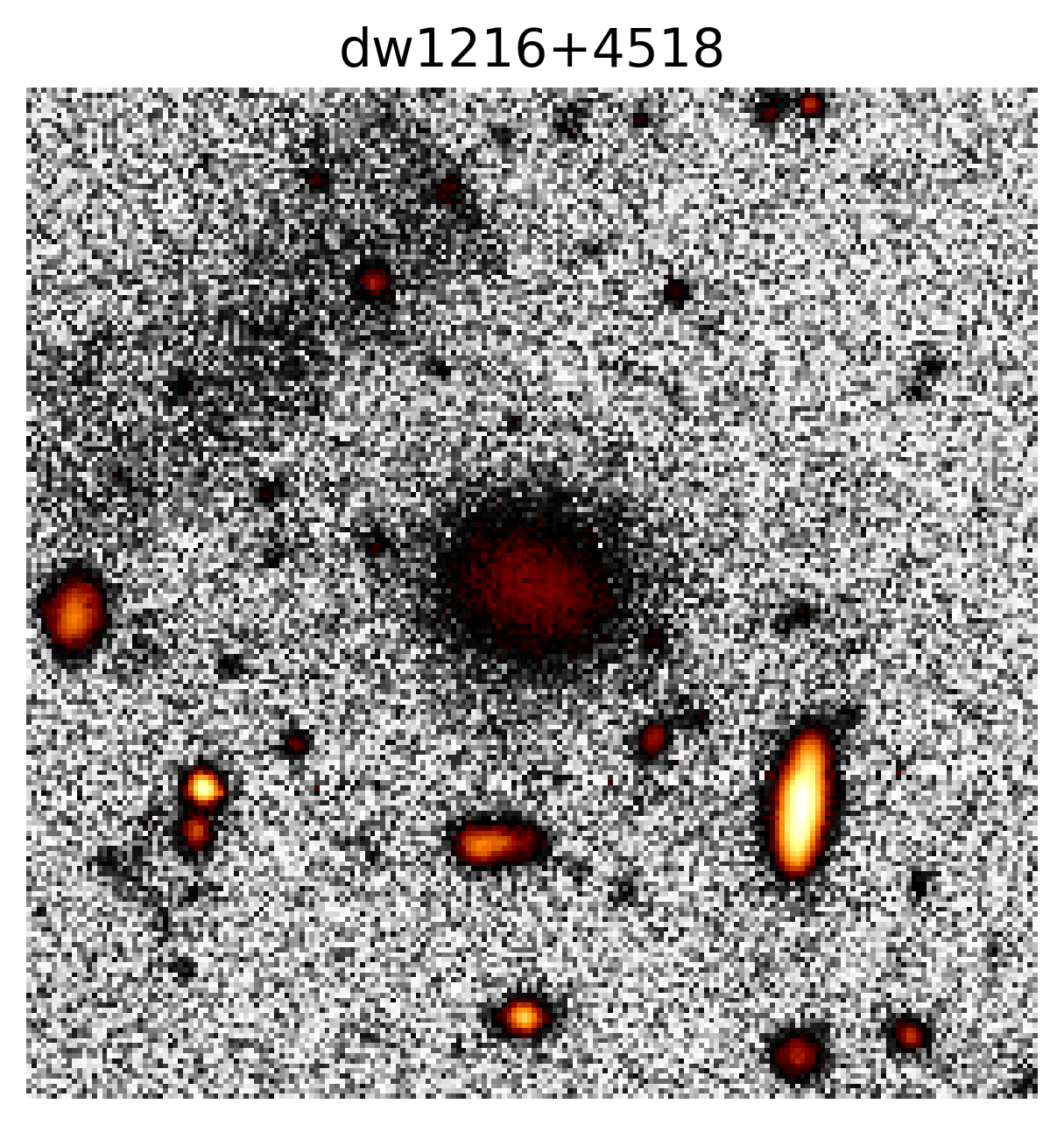}
	\includegraphics[width=0.163\linewidth]{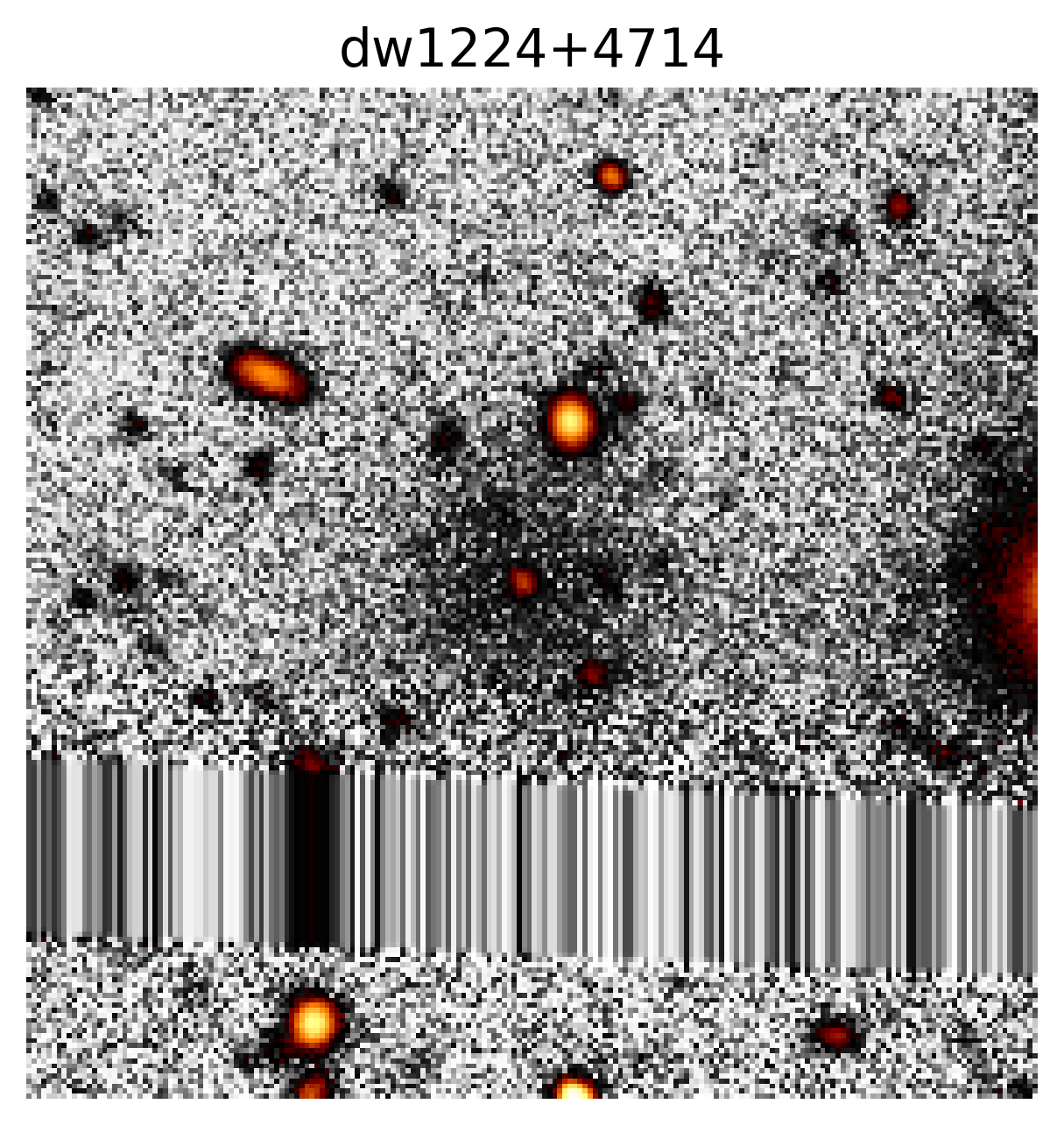}
	\includegraphics[width=0.163\linewidth]{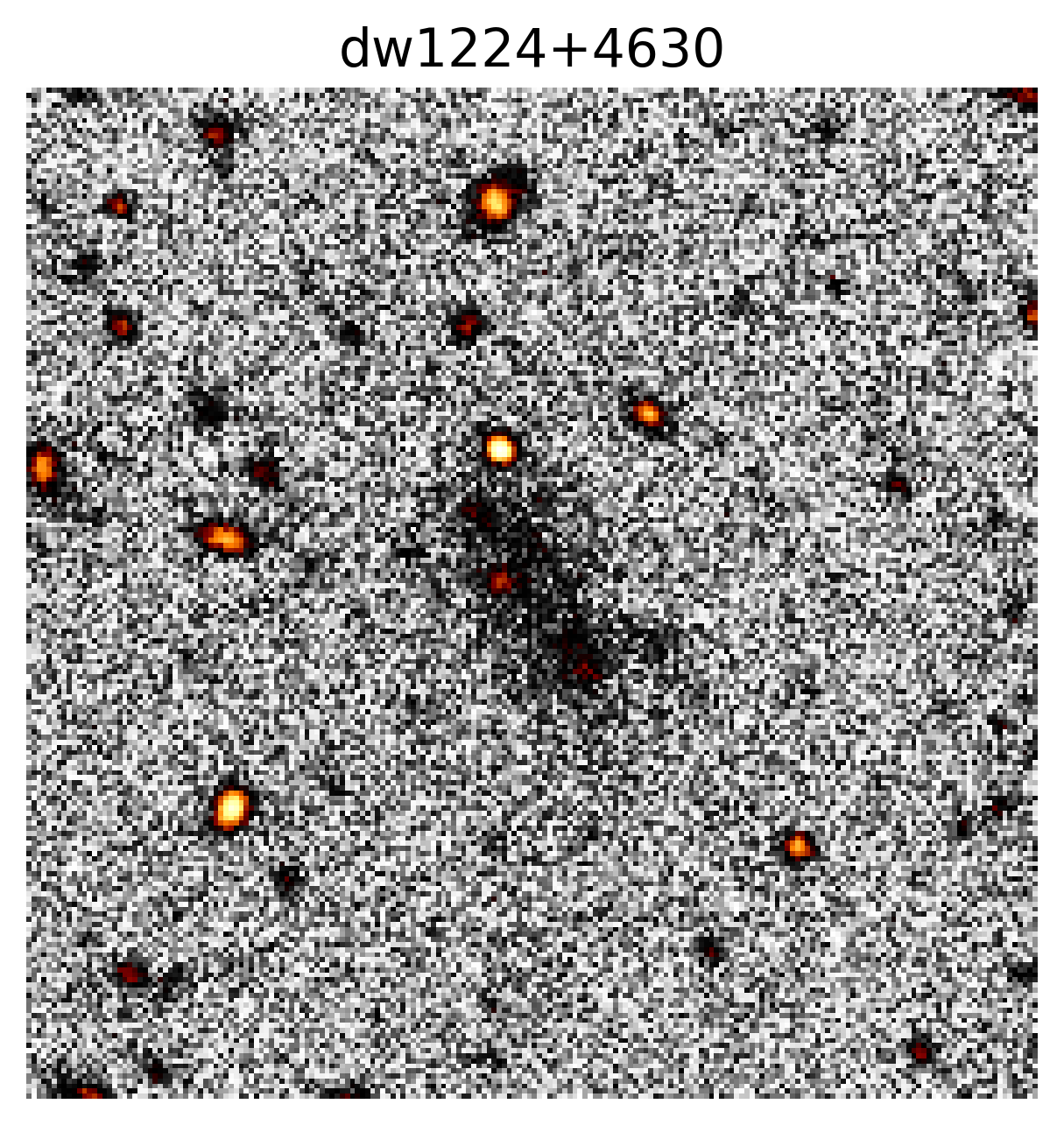}
    \includegraphics[width=0.163\linewidth]{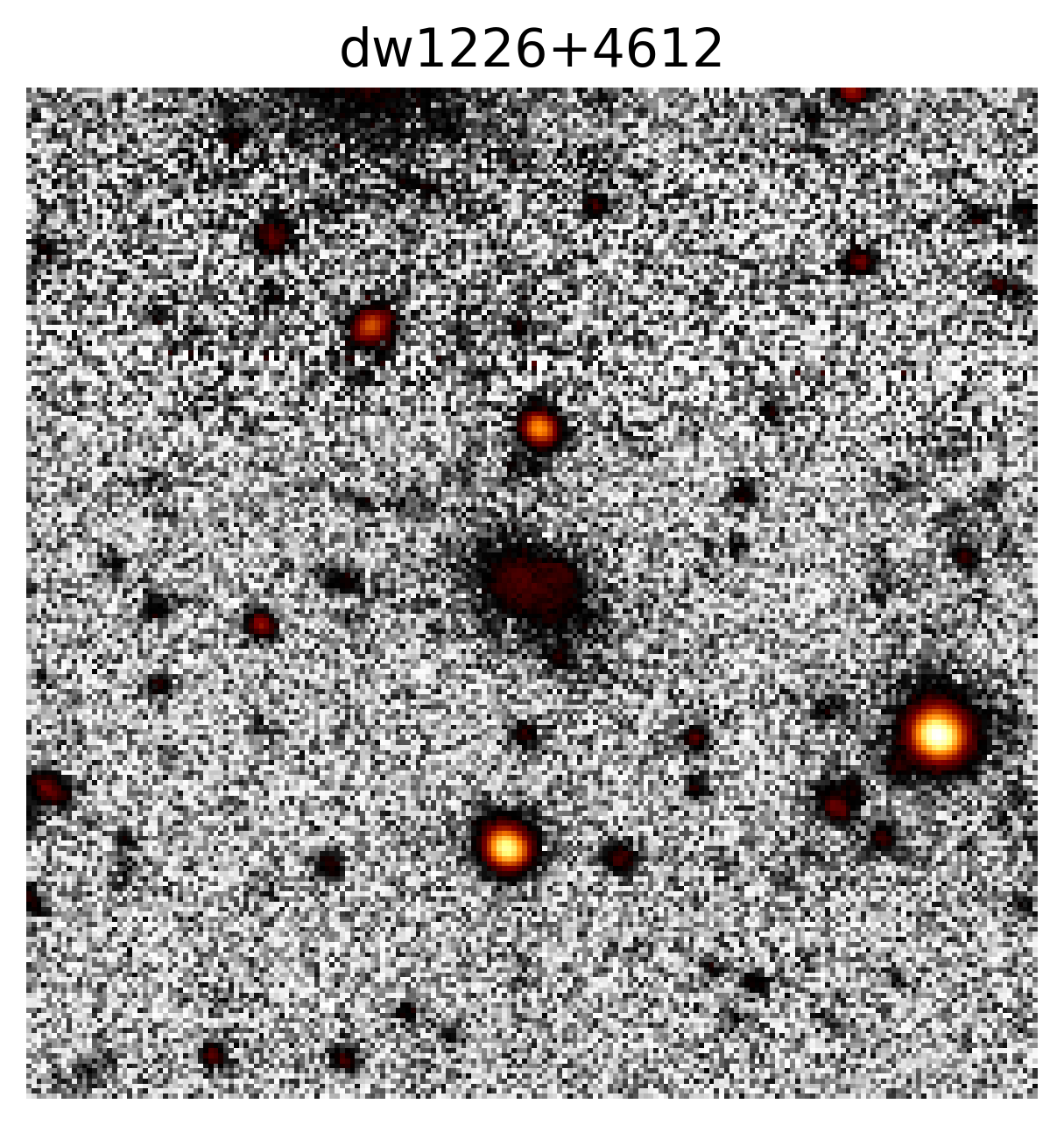}
    \includegraphics[width=0.163\linewidth]{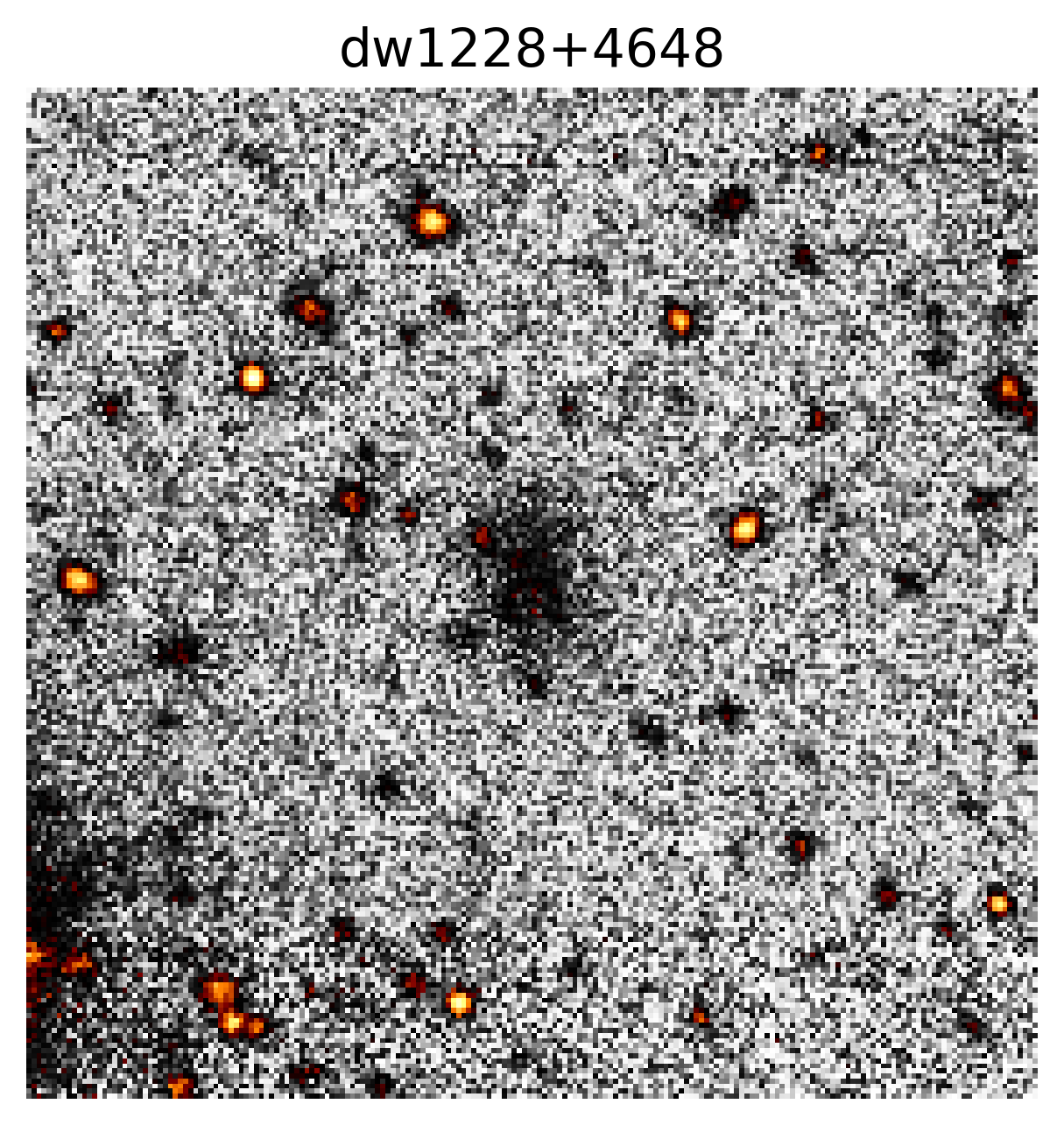}
	\caption{Cutouts of the confirmed  dwarf galaxies (top two rows) and dwarf candidates (bottom three two) of the M\,106 group.}
	\label{fig:M106_known}
\end{figure*}

\begin{table*}[h!]
\caption{The known dwarfs ($^*$) and new dwarf galaxy candidates around NGC\,3521.}             
\centering                          
\begin{tabular}{lcccrcc}        
\hline\hline                 
       Name &          RA  &          Dec &  $g$ &   $r_{eff}$ &  $\mu_0$ &  
        $\mu_{eff}$ \\ 
        &         (J2000.0) &          (J2000.0) &  (mag) &   (arcsec) &  (mag/arcsec$^2$) &  
       (mag/arcsec$^2$) \\ 
\hline      \\[-2mm]                  
 dw1101-0108 & 11:01:12.10 & -01:08:51.10 &  21.2 $\pm$     0.4 &  5.9 $\pm$   1.2 &  26.8 $\pm$     0.1 &  27.0 $\pm$     0.1 \\
 dw1102-0019 & 11:02:40.42 & -00:19:09.38 &  23.4 $\pm$     1.6 &  3.1 $\pm$   2.4 &  25.1 $\pm$     0.5 &  27.8 $\pm$     0.9 \\
 Dw1104+0004$^*$ & 11:04:38.52 & +00:04:54.08 &  20.1 $\pm$     0.1 &  5.4 $\pm$   0.3 &  25.2 $\pm$     0.0 &  25.8 $\pm$     0.1 \\
      KKSG20$^*$ & 11:04:40.23 & +00:03:29.27 &  16.4 $\pm$     0.1 & 16.8 $\pm$   1.0 &  23.4 $\pm$     0.0 &  24.5 $\pm$     0.1 \\
 Dw1104+0005$^*$ & 11:04:41.11 & +00:05:37.84 &  20.0 $\pm$     0.2 &  9.4 $\pm$   1.0 &  26.0 $\pm$     0.1 &  26.9 $\pm$     0.1 \\
    UGC06145$^*$ & 11:05:34.59 & -01:51:46.18 &  15.2 $\pm$     0.1 & 24.1 $\pm$   1.2 &  23.6 $\pm$     0.0 &  24.1 $\pm$     0.1 \\
      KKSG22$^*$ & 11:06:07.67 & -01:26:52.45 &  17.2 $\pm$     0.1 & 15.2 $\pm$   0.8 &  24.3 $\pm$     0.0 &  25.1 $\pm$     0.1 \\
 Dw1106-0052$^*$ & 11:06:25.95 & -00:52:46.10 &  19.8 $\pm$     0.2 &  5.9 $\pm$   0.5 &  25.4 $\pm$     0.0 &  25.6 $\pm$     0.0 \\
 dw1107-0038 & 11:07:06.20 & -00:38:01.17 &  20.6 $\pm$     0.3 &  5.9 $\pm$   1.1 &  26.1 $\pm$     0.0 &  26.5 $\pm$     0.1 \\
NGC3521dwTBG$^*$ & 11:07:13.30 & -00:11:12.34 &  17.4 $\pm$     0.2 & 21.3 $\pm$   1.9 &  25.5 $\pm$     0.1 &  26.1 $\pm$     0.1 \\
 dw1109-0016$^*$ & 11:09:18.17 & -00:16:42.27 &  17.9 $\pm$     0.1 & 15.7 $\pm$   1.2 &  25.3 $\pm$     0.1 &  25.8 $\pm$     0.1 \\
 dw1110+0037$^*$ & 11:10:29.53 & +00:36:59.50 &  16.6 $\pm$     0.2 & 13.8 $\pm$   1.2 &  23.2 $\pm$     0.1 &  24.3 $\pm$     0.1 \\
 Dw1111+0049$^*$ & 11:11:23.72 & +00:49:53.93 &  18.8 $\pm$     0.1 & 15.4 $\pm$   0.8 &  26.2 $\pm$     0.0 &  26.7 $\pm$     0.0 \\
 dw1113-0053 & 11:13:25.94 & -00:53:57.89 &  20.9 $\pm$     0.4 &  7.2 $\pm$   1.6 &  26.7 $\pm$     0.1 &  27.1 $\pm$     0.2 \\
\hline
\end{tabular}
\label{tab:group_NGC3521}
\end{table*}

\begin{figure*}[h!]
\raggedright
	\includegraphics[width=0.163\linewidth]{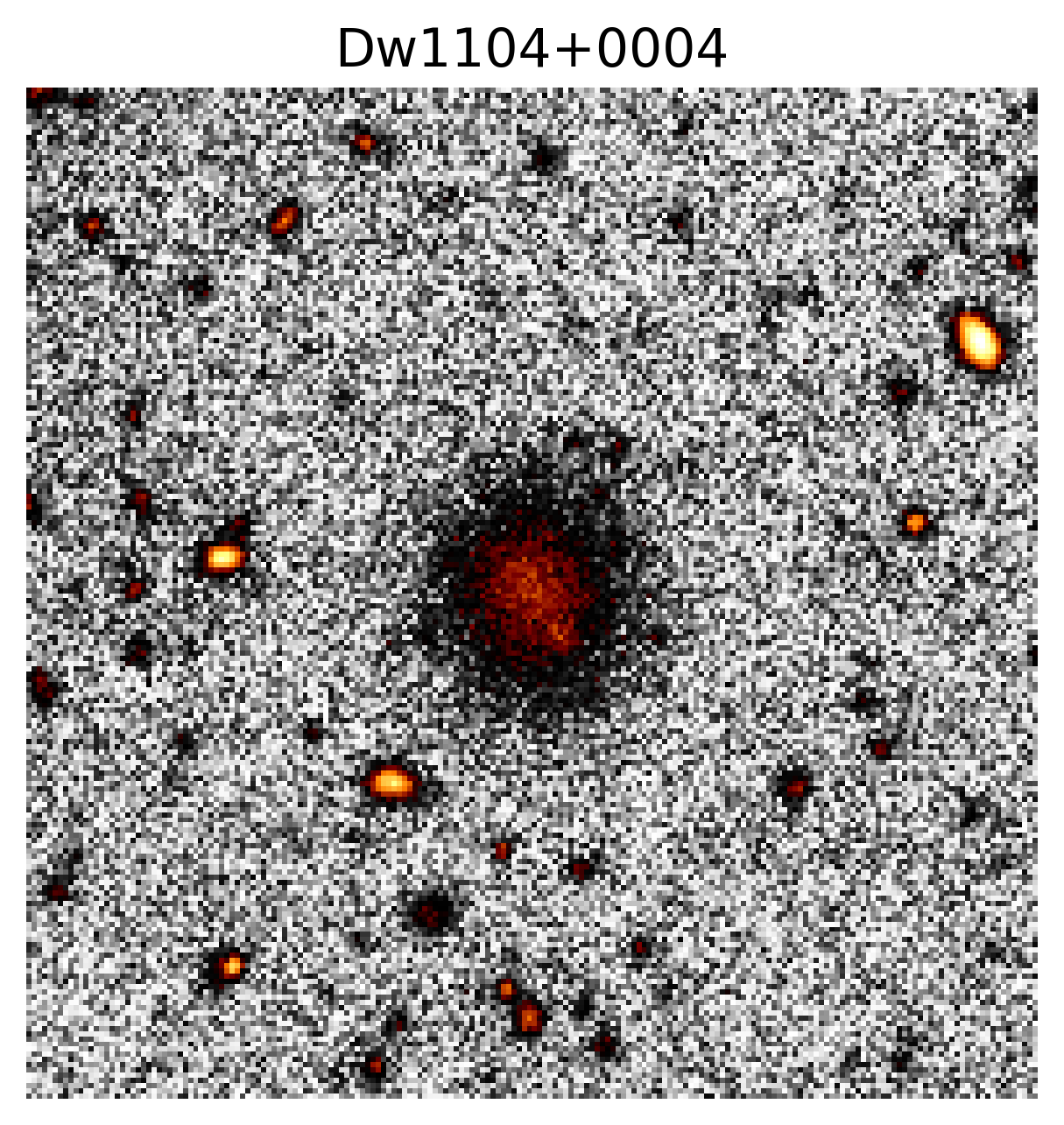}
    \includegraphics[width=0.163\linewidth]{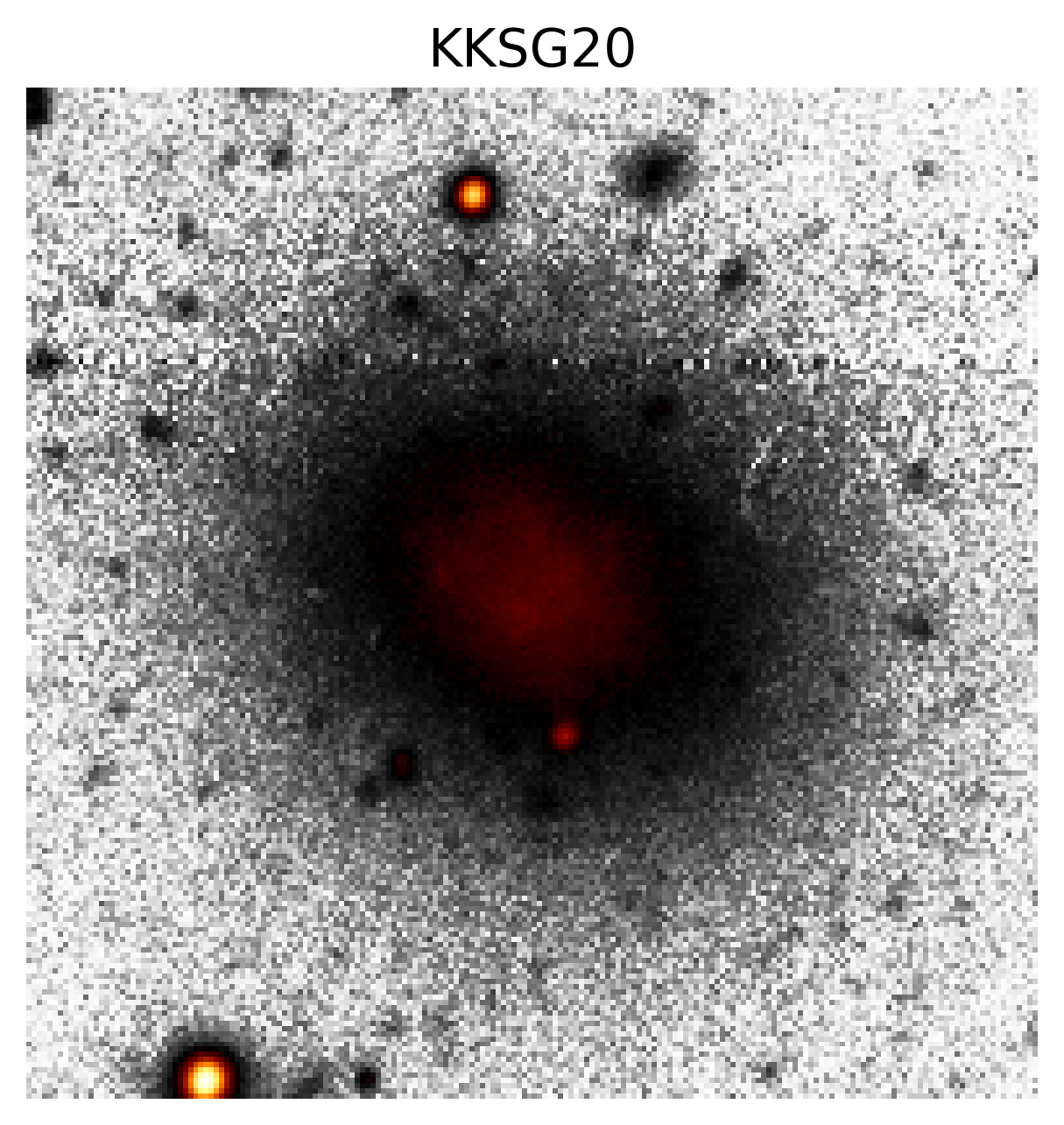}
	\includegraphics[width=0.163\linewidth]{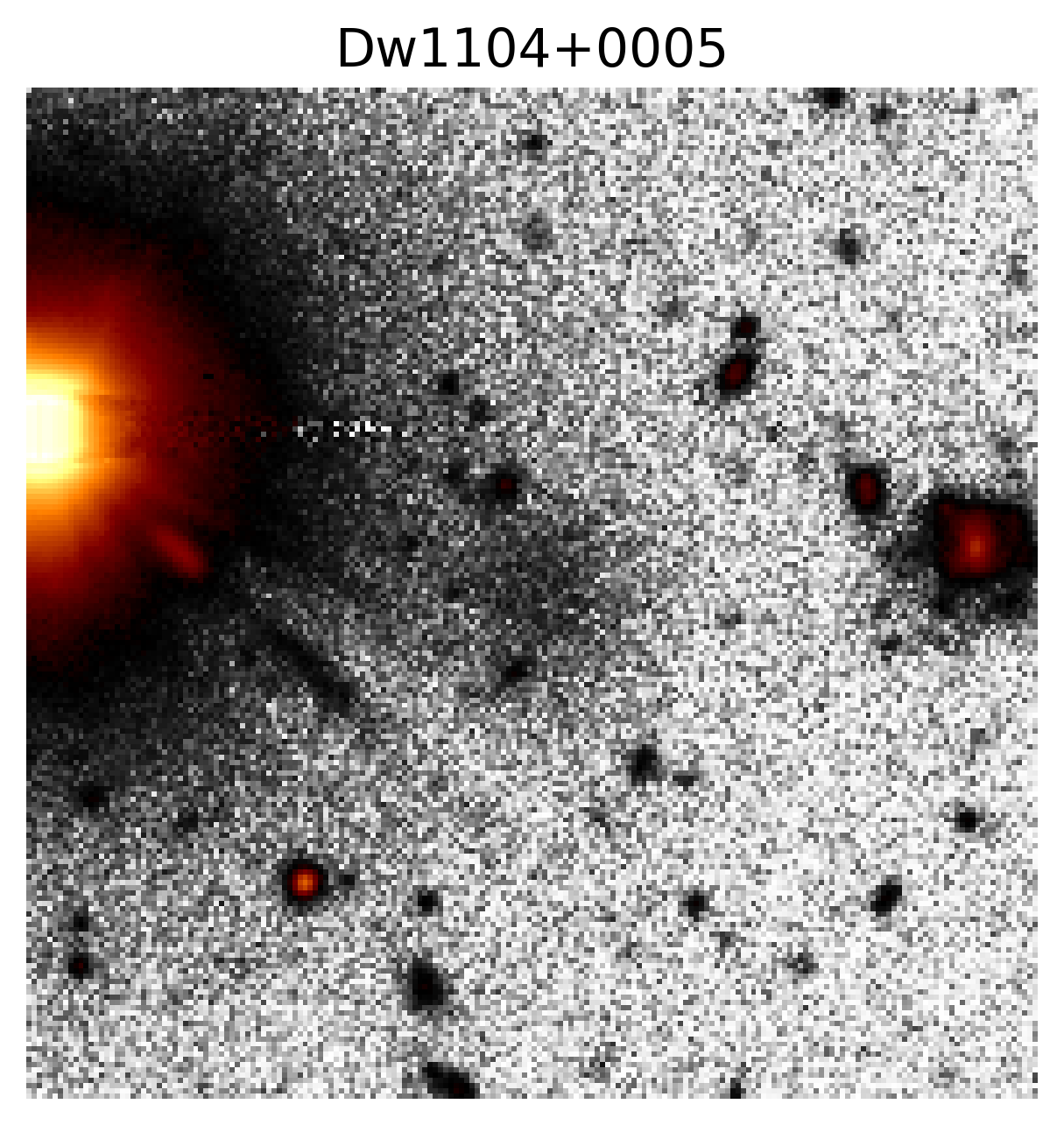}
	\includegraphics[width=0.163\linewidth]{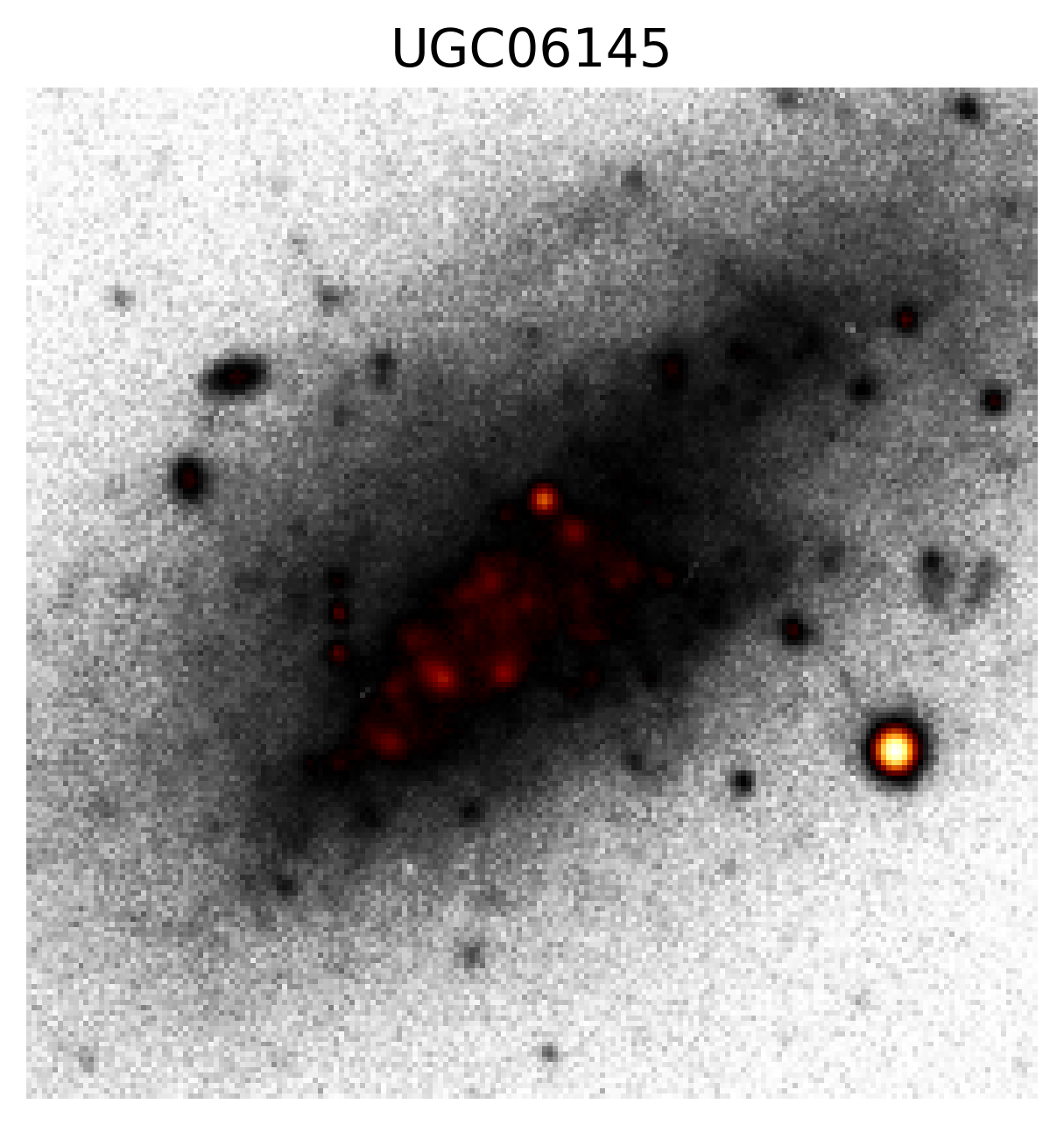}
	\includegraphics[width=0.163\linewidth]{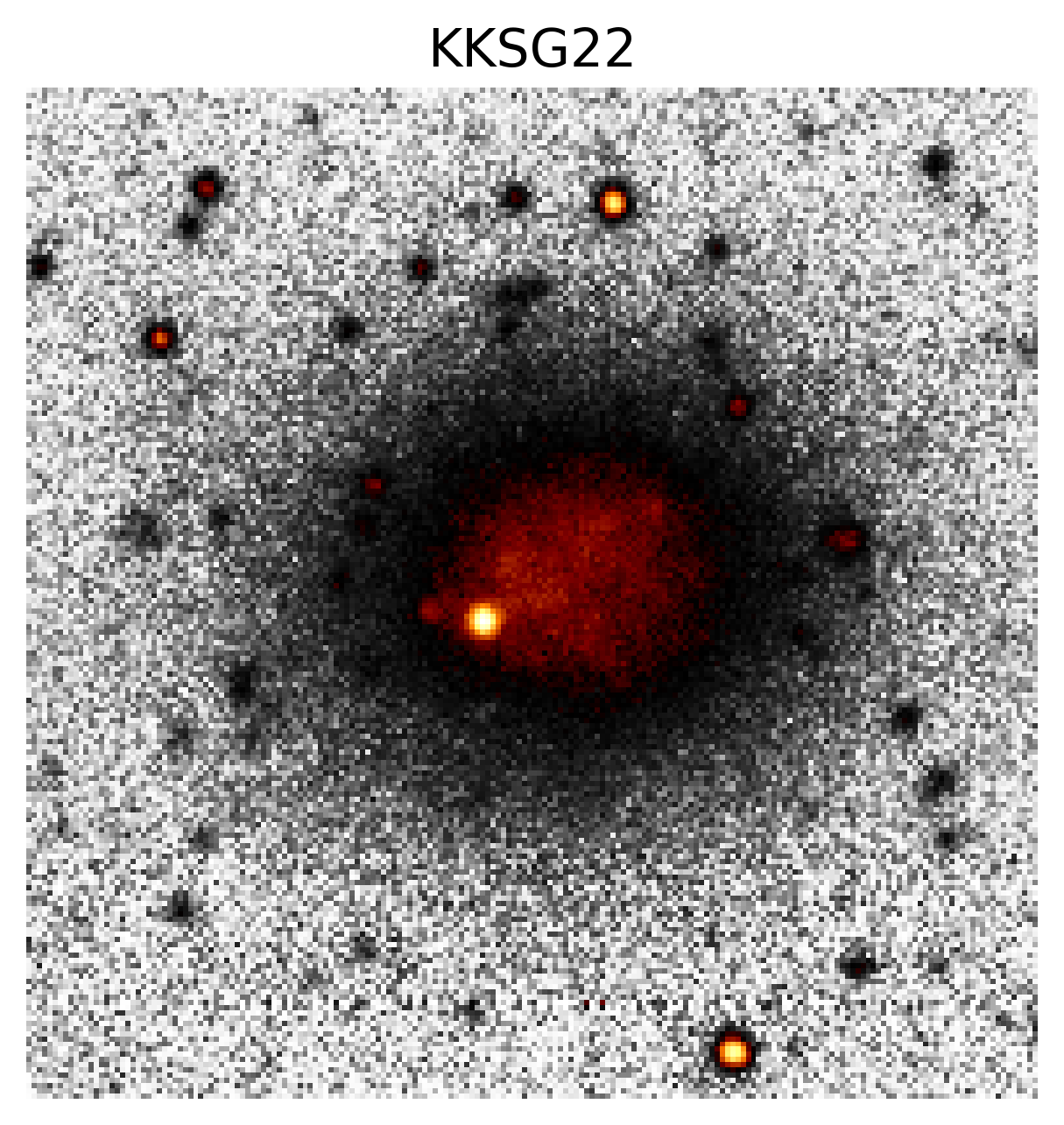}
	\includegraphics[width=0.163\linewidth]{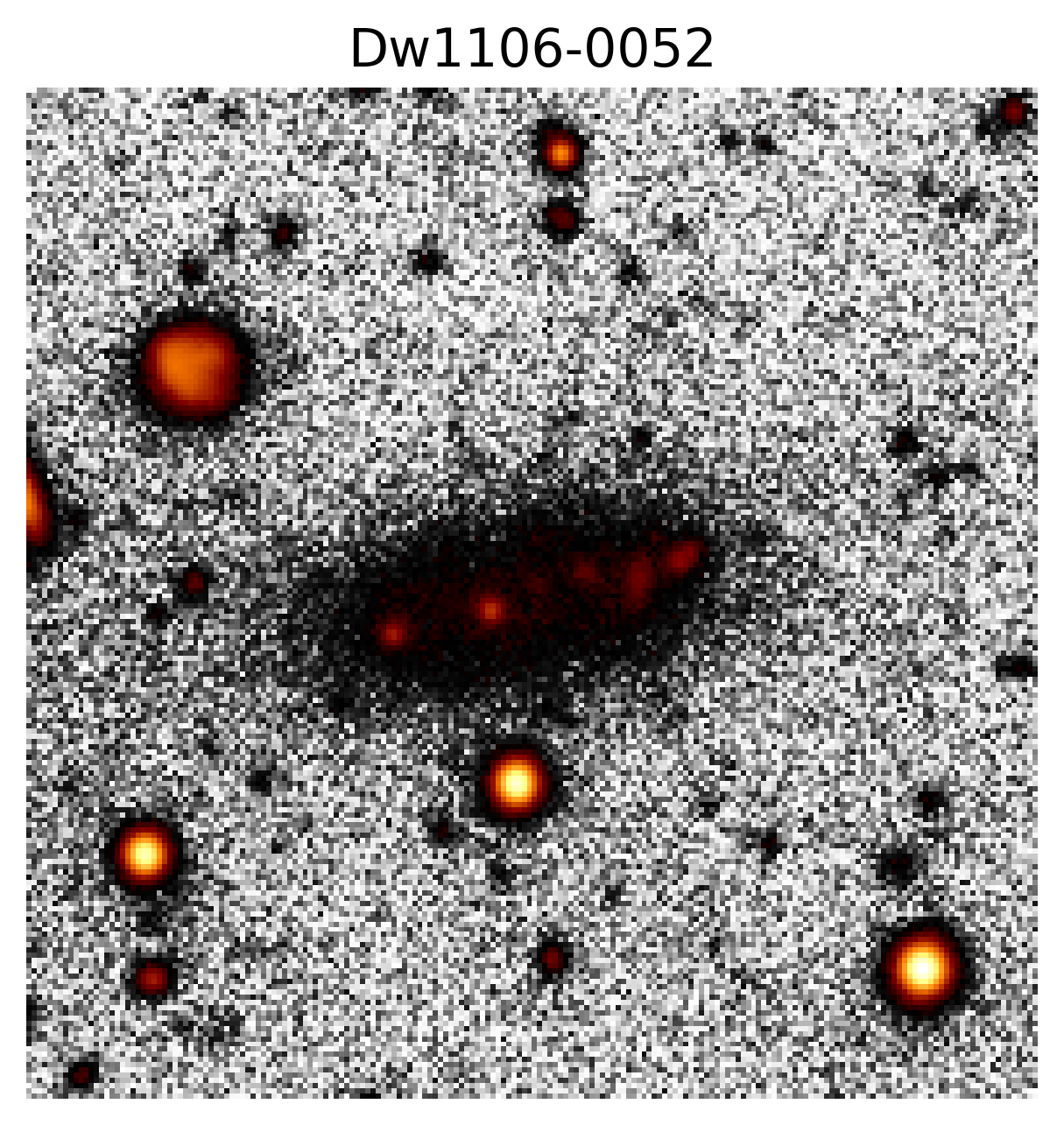}\\
	\includegraphics[width=0.163\linewidth]{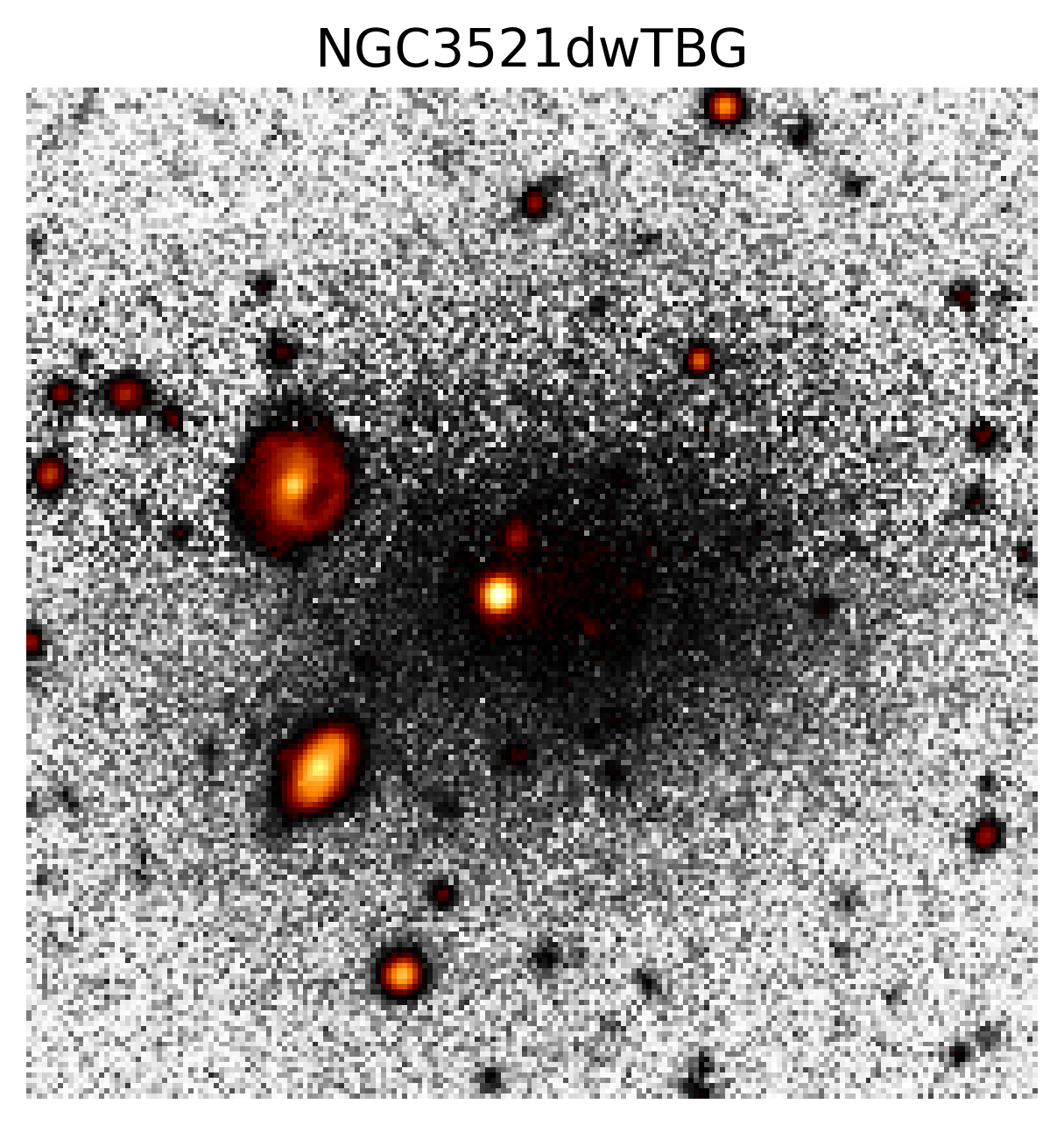}
    \includegraphics[width=0.163\linewidth]{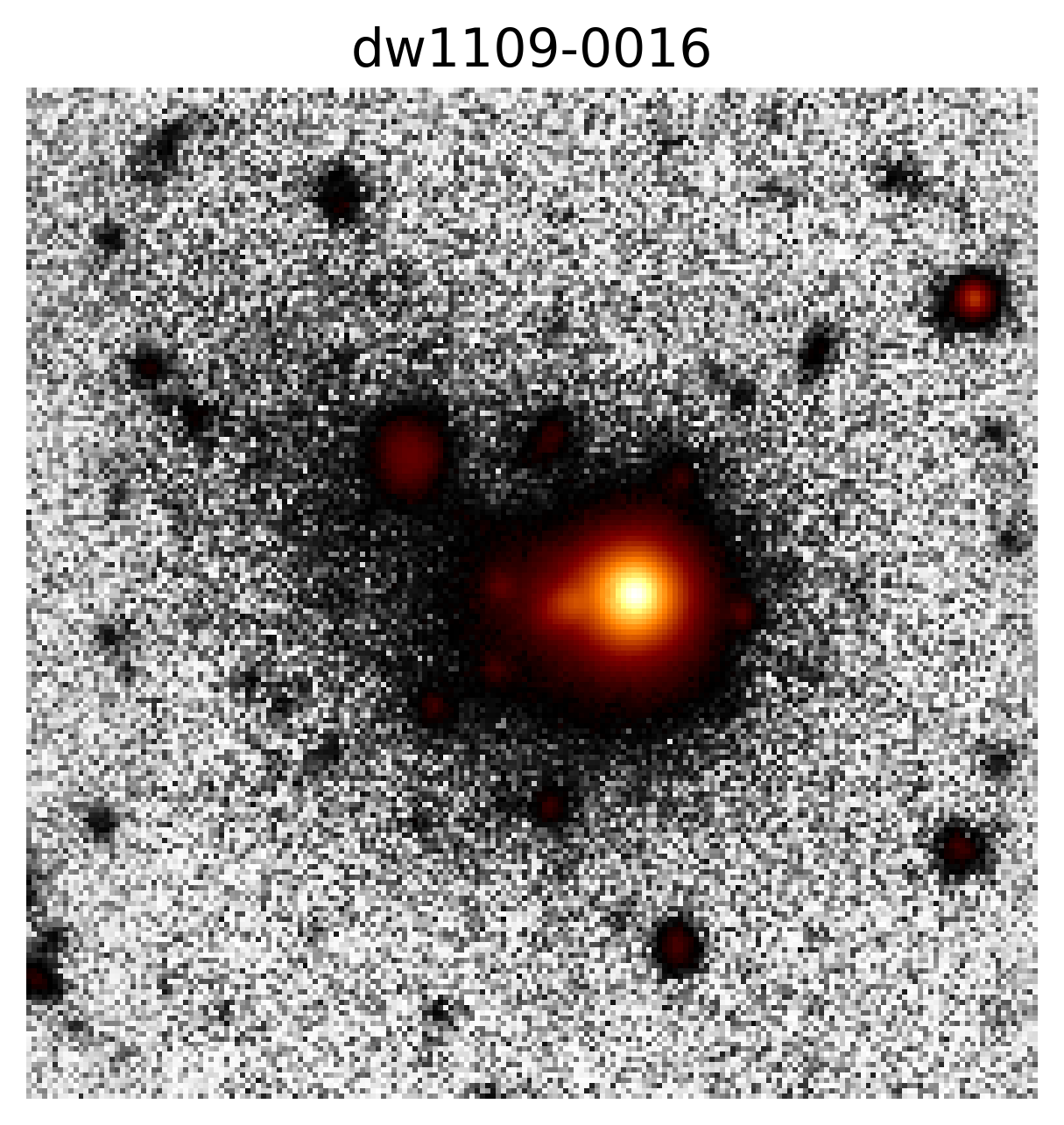}
    \includegraphics[width=0.163\linewidth]{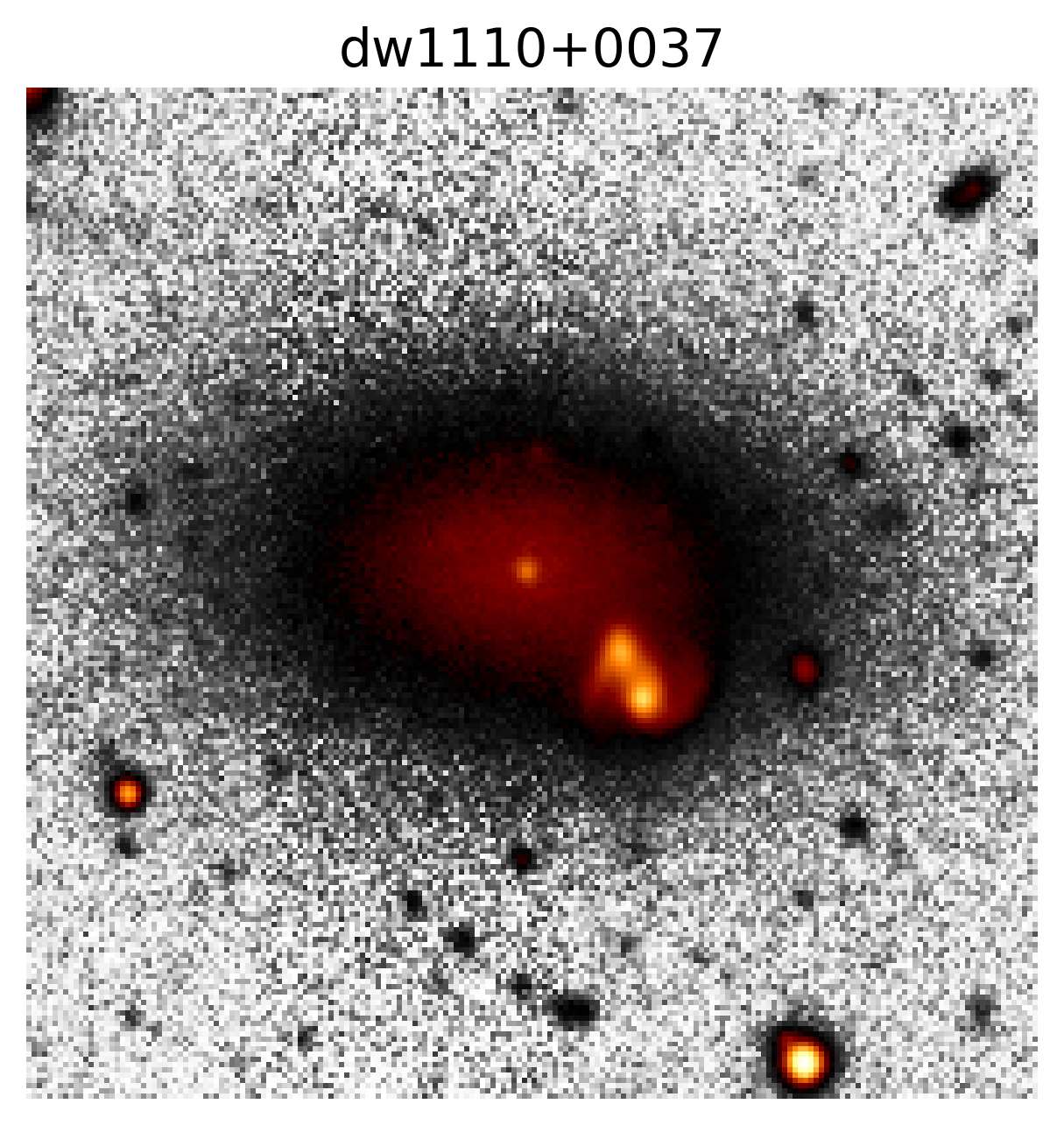}
    \includegraphics[width=0.163\linewidth]{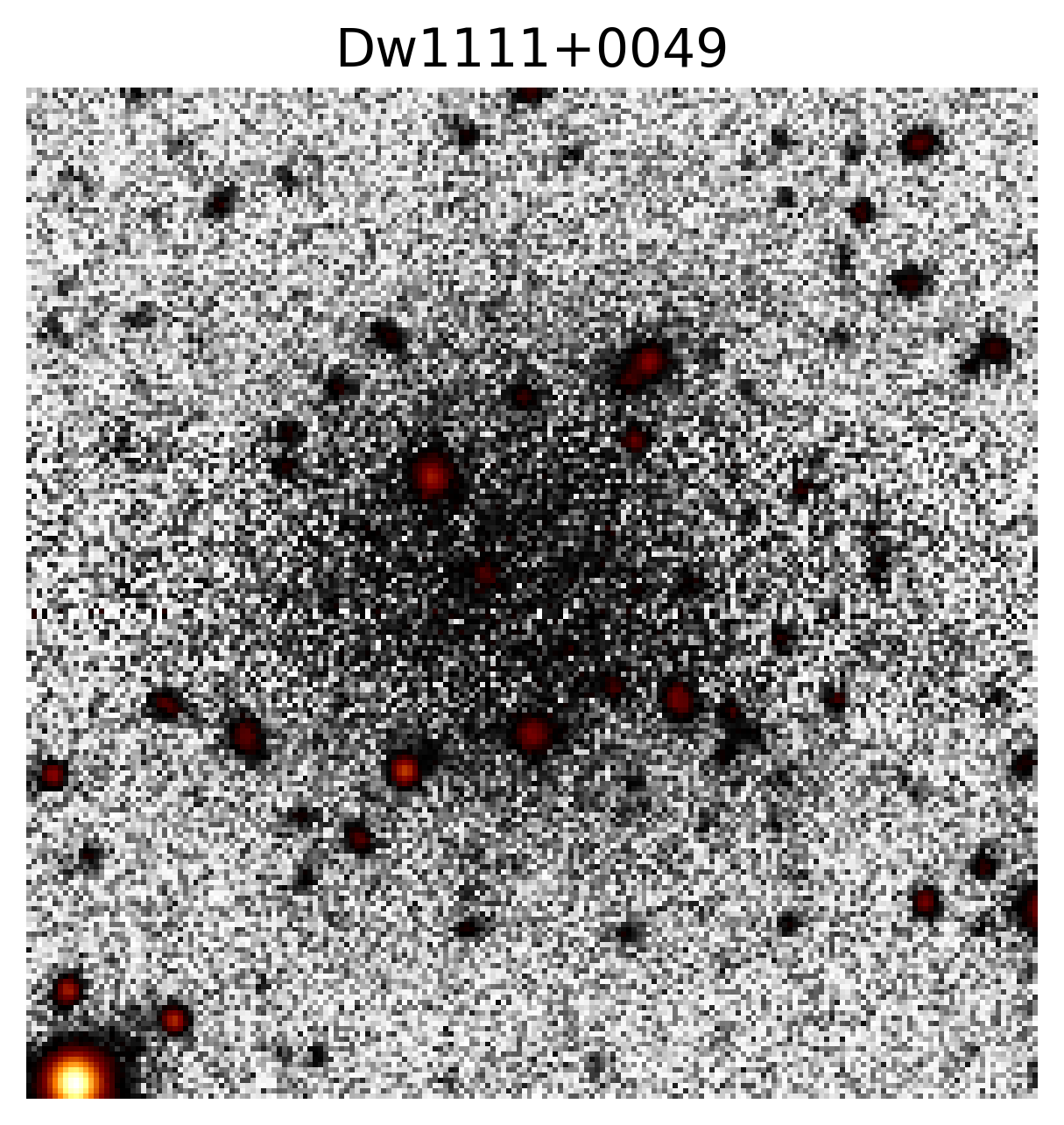}
\\
        \vspace{0.5cm}
    \includegraphics[width=0.163\linewidth]{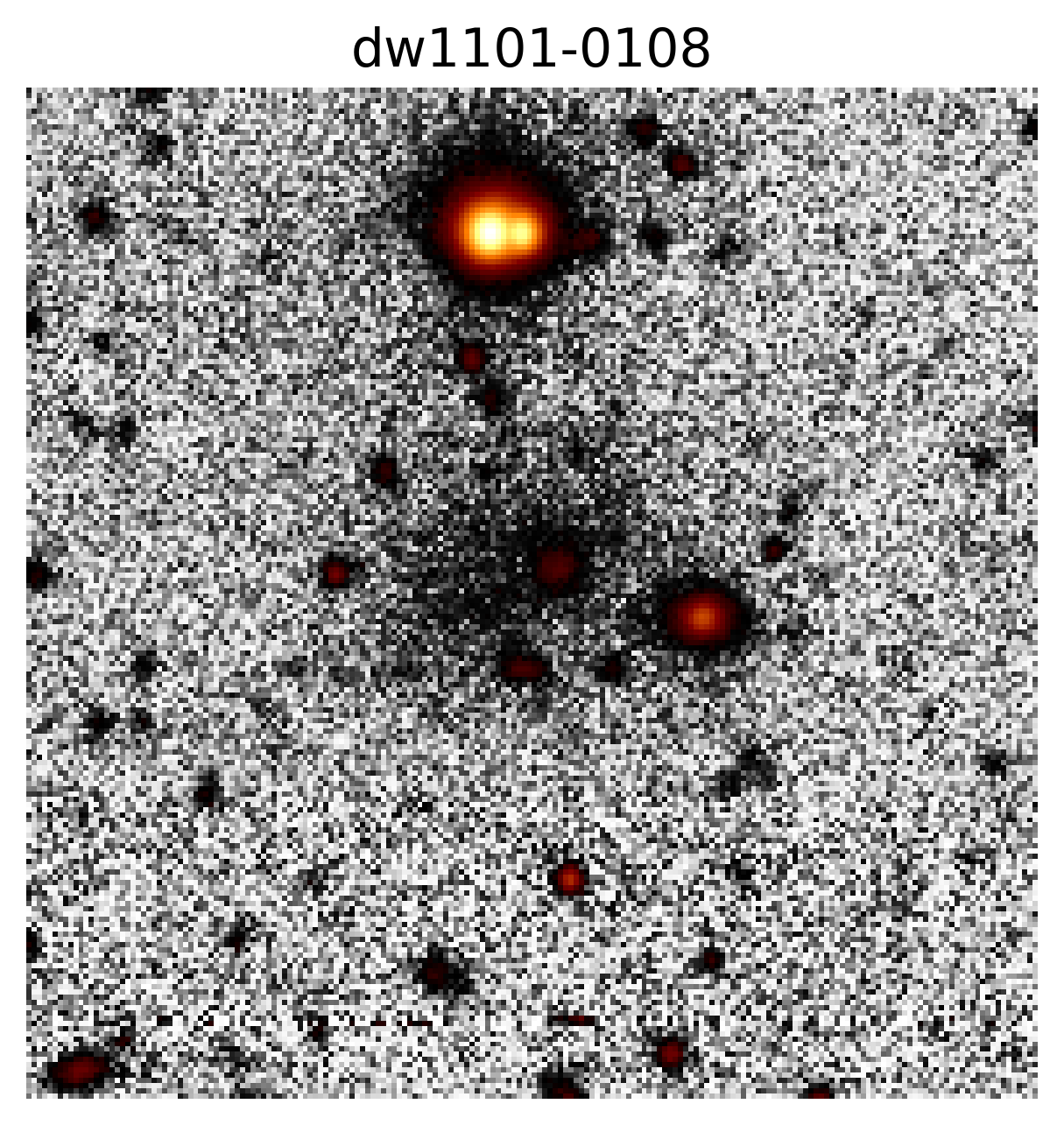}
    \includegraphics[width=0.163\linewidth]{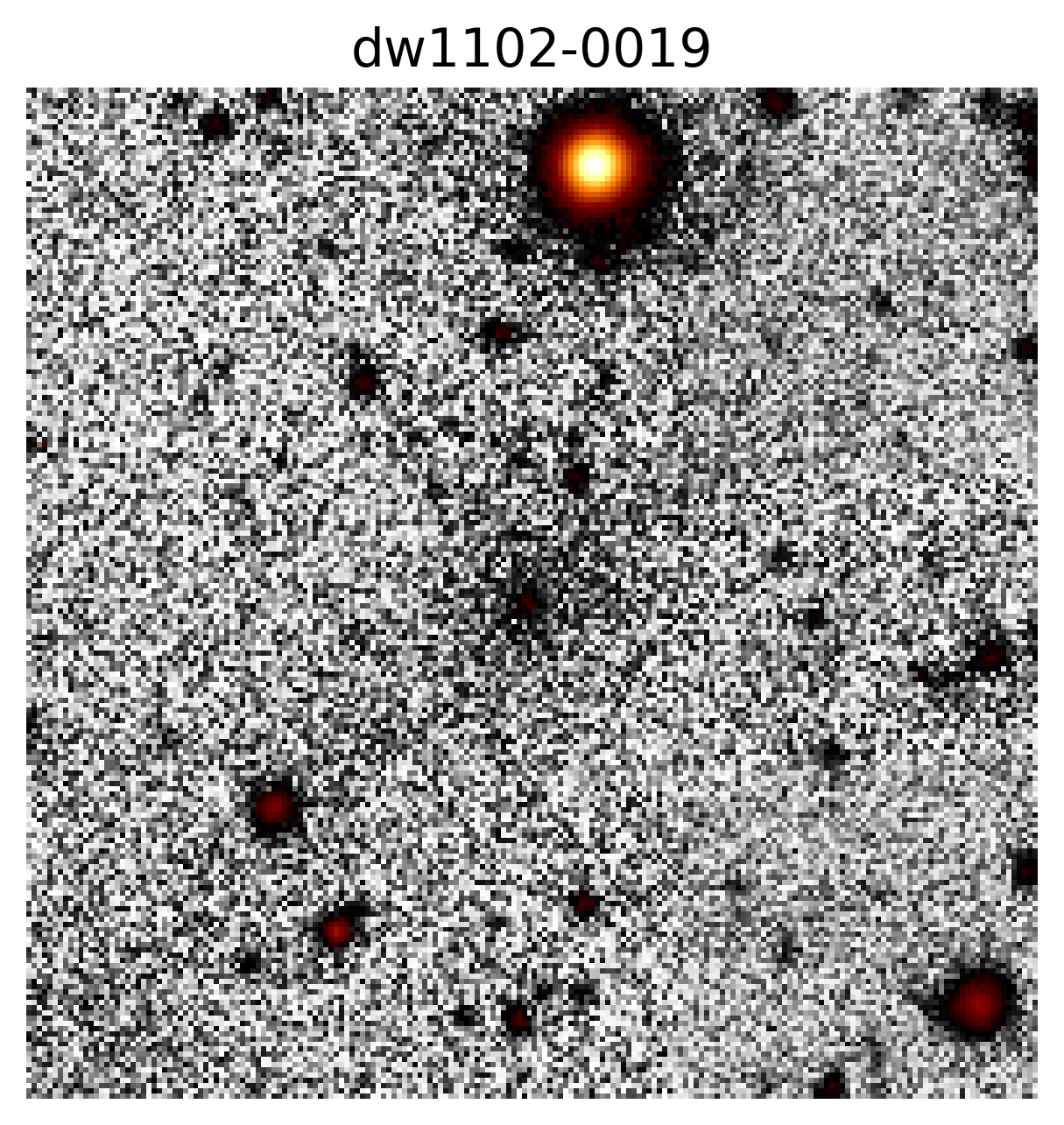}
    \includegraphics[width=0.163\linewidth]{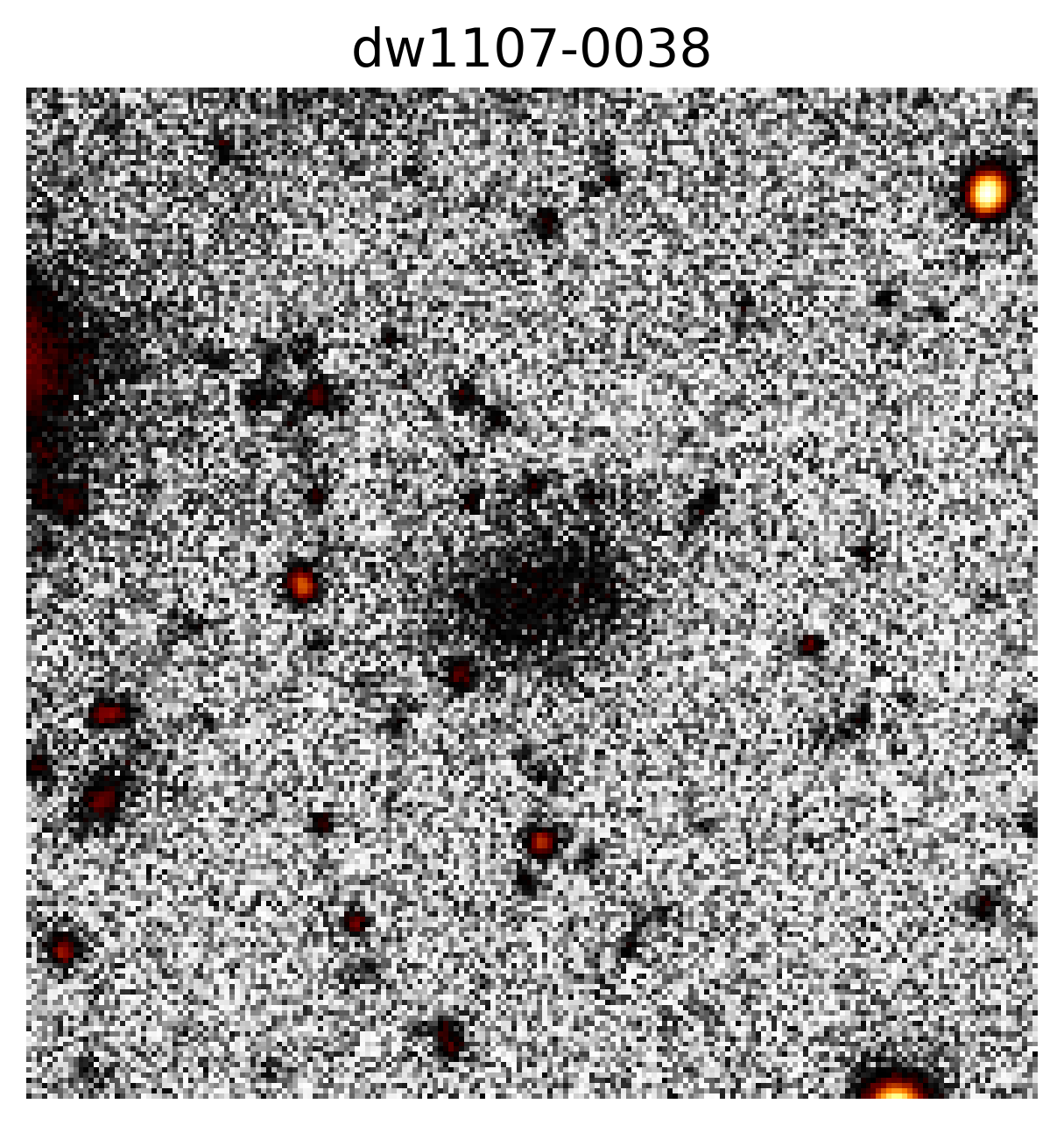}
    \includegraphics[width=0.163\linewidth]{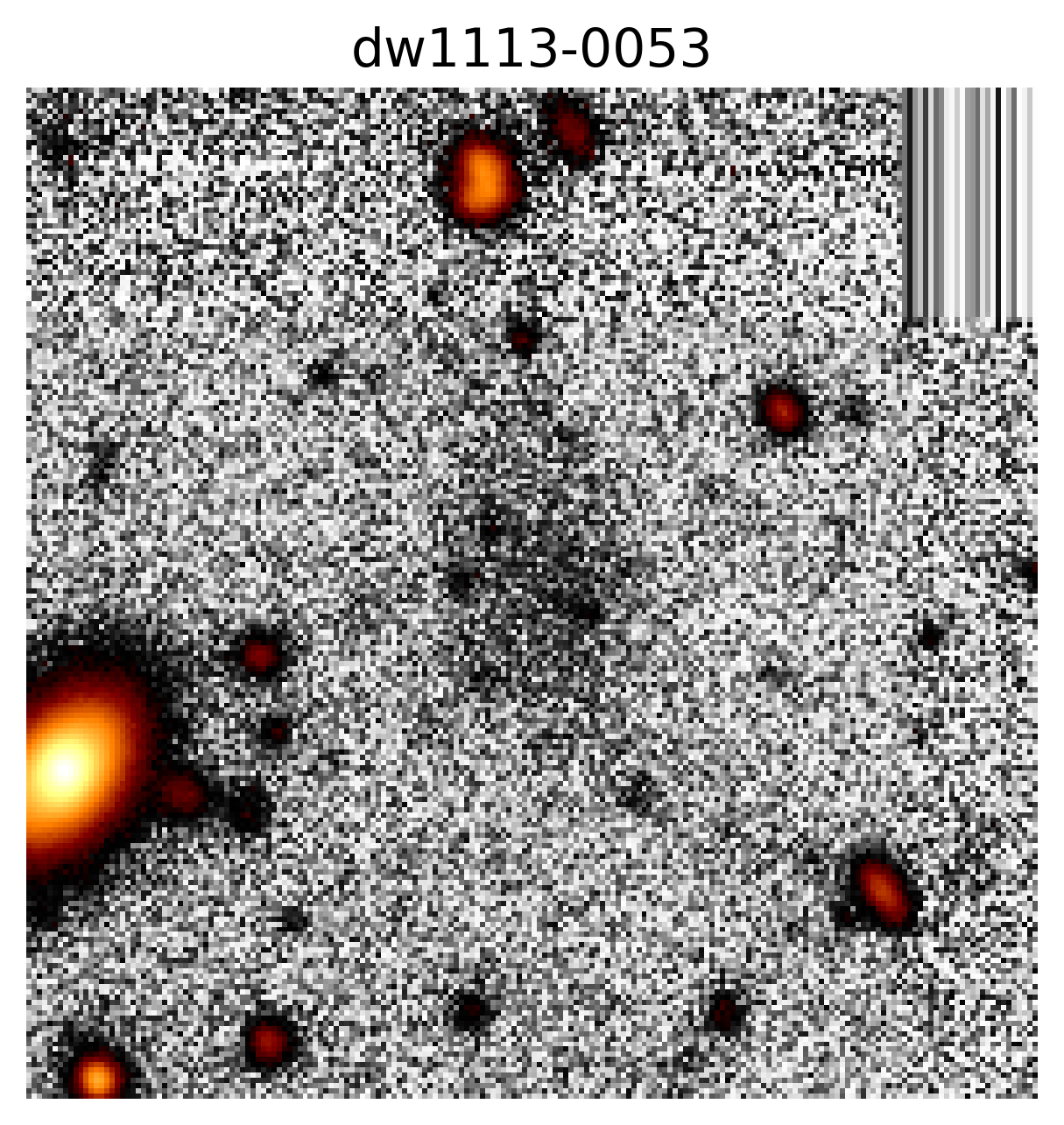}
	\caption{Cutouts of the confirmed  dwarf galaxies (top two rows) and dwarf candidates (bottom row) of the NGC\,3521 group.}
	\label{fig:NGC3521_known}
\end{figure*}

\newpage
\mbox{~}
\newpage

\begin{table*}[h!]
\caption{The known dwarfs ($^*$) and new dwarf galaxy candidates around UGCA127.}             
\centering                          
\begin{tabular}{lcccrcc}        
\hline\hline                 
       Name &          RA  &          Dec &  $g$ &   $r_{eff}$ &  $\mu_0$ &  
        $\mu_{eff}$ \\ 
                &         (J2000.0) &          (J2000.0) &  (mag) &   (arcsec) &  (mag/arcsec$^2$) &  
       (mag/arcsec$^2$) \\ 
\hline      \\[-2mm]                  
dw0618-0825 & 06:18:45.52 & -08:25:32.86 &  19.3 $\pm$     0.1 &  7.3 $\pm$   0.5 &  25.1 $\pm$     0.1 &  25.6 $\pm$     0.1 \\
dw0619-0947 & 06:19:40.26 & -09:47:48.34 &  19.5 $\pm$     0.4 &  9.1 $\pm$   1.9 &  25.7 $\pm$     0.1 &  26.3 $\pm$     0.2 \\
dw0619-0822 & 06:19:39.27 & -08:22:33.16 &  19.5 $\pm$     0.2 &  5.9 $\pm$   0.7 &  23.8 $\pm$     0.1 &  25.4 $\pm$     0.2 \\
dw0620-0931 & 06:20:28.57 & -09:31:58.63 &  20.7 $\pm$     0.6 &  5.8 $\pm$   1.7 &  25.4 $\pm$     0.2 &  26.5 $\pm$     0.3 \\
 UGCA127sat$^*$ & 06:20:54.66 & -08:39:02.40 &  16.9 $\pm$     0.1 & 13.9 $\pm$   0.8 &  23.6 $\pm$     0.0 &  24.6 $\pm$     0.1 \\
dw0621-0836 & 06:21:15.95 & -08:36:55.02 &  21.6 $\pm$     0.8 &  5.5 $\pm$   2.4 &  26.2 $\pm$     0.2 &  27.2 $\pm$     0.5 \\
WHIB0619-07$^*$ & 06:22:13.05 & -07:50:36.66 &  15.2 $\pm$     0.6 & 28.4 $\pm$   7.5 &  22.6 $\pm$     0.2 &  24.4 $\pm$     0.3 \\
dw0623-1002 & 06:23:23.76 & -10:02:32.92 &  20.1 $\pm$     0.4 &  7.4 $\pm$   1.7 &  25.7 $\pm$     0.2 &  26.4 $\pm$     0.2 \\
\hline
\end{tabular}
\label{tab:group_UGCA127}
\end{table*}

\begin{figure*}[h!]
\raggedright
	\includegraphics[width=0.163\linewidth]{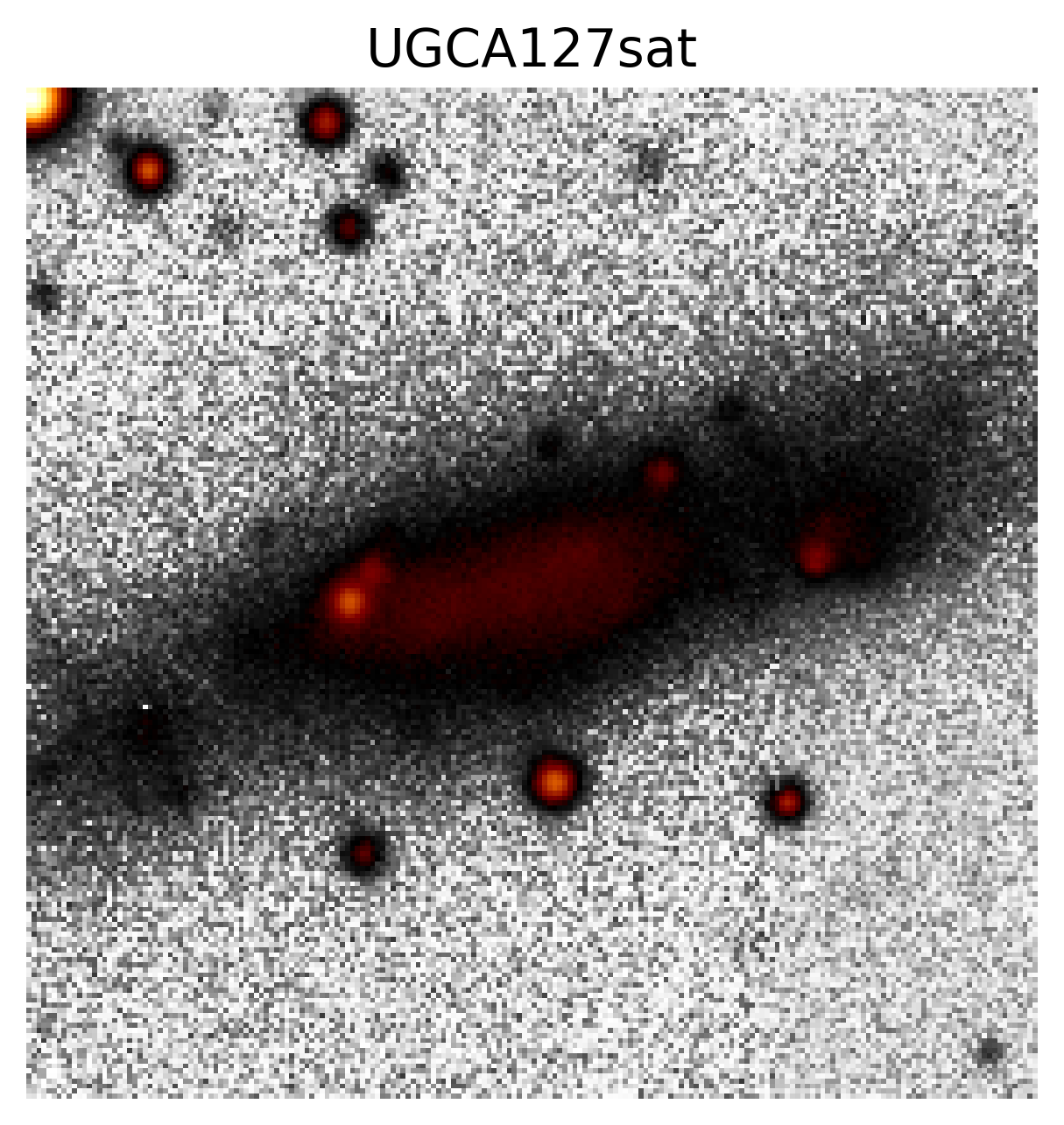}
    \includegraphics[width=0.163\linewidth]{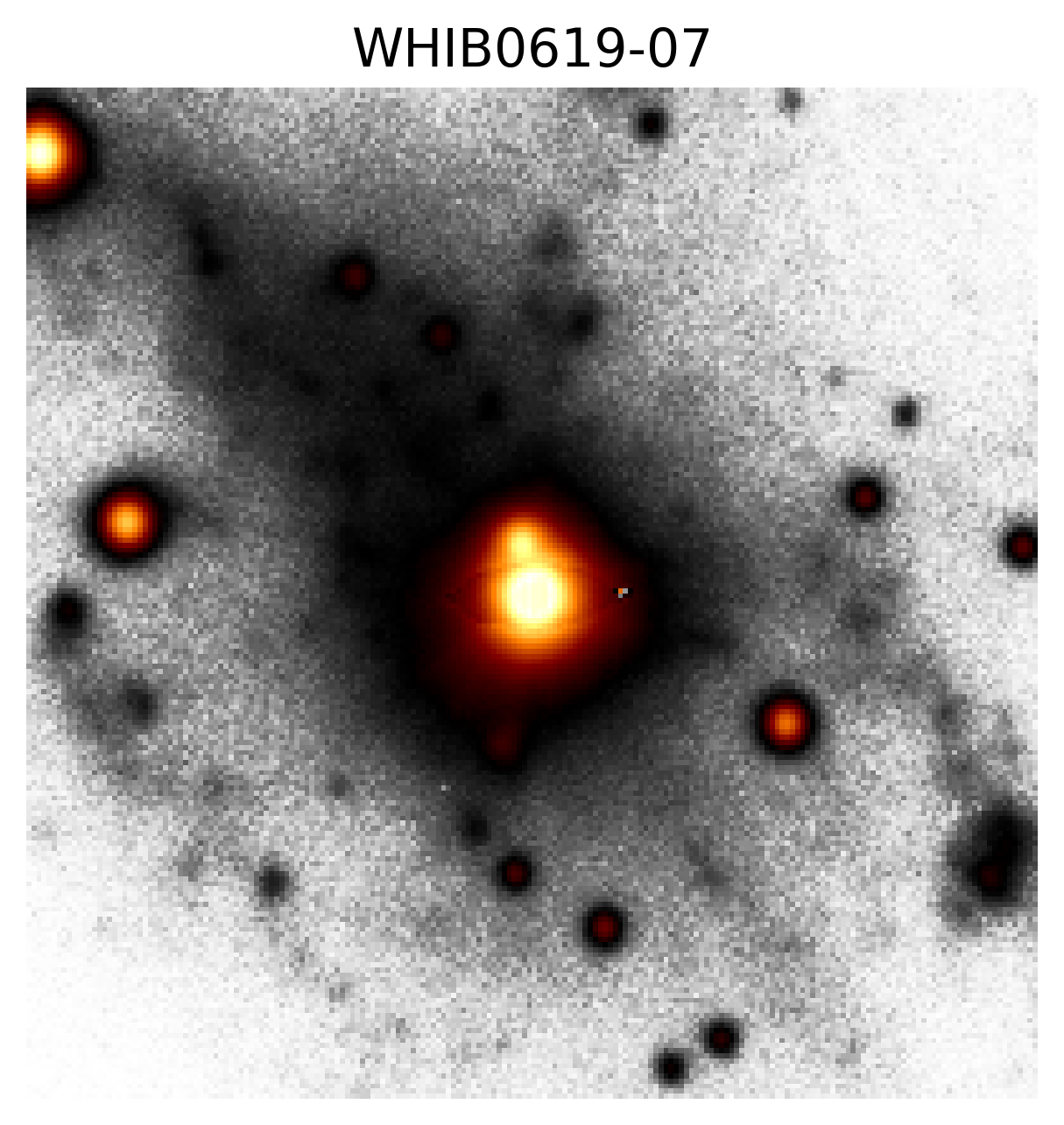}\\
        \vspace{0.5cm}
    \includegraphics[width=0.163\linewidth]{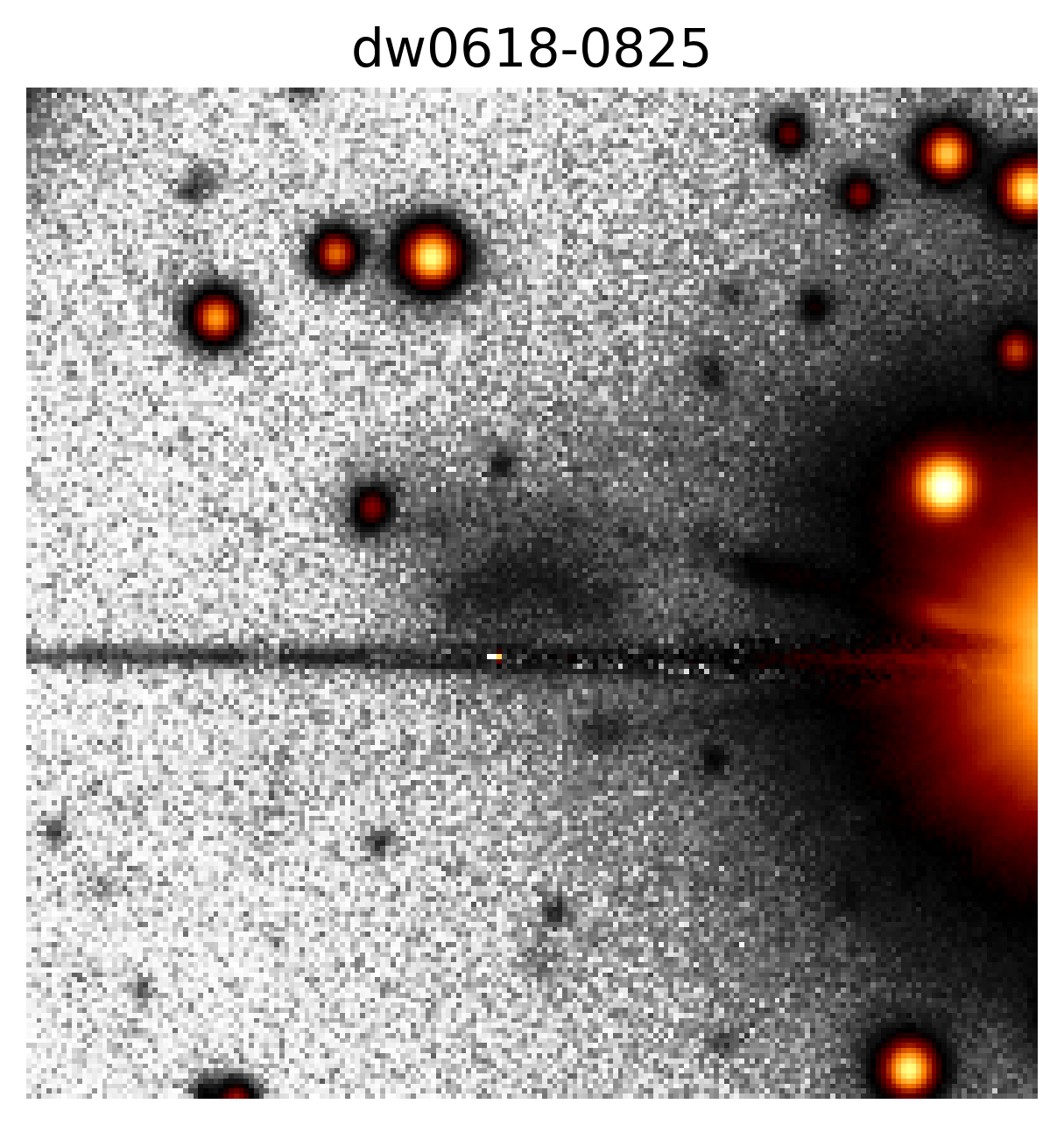}
    \includegraphics[width=0.163\linewidth]{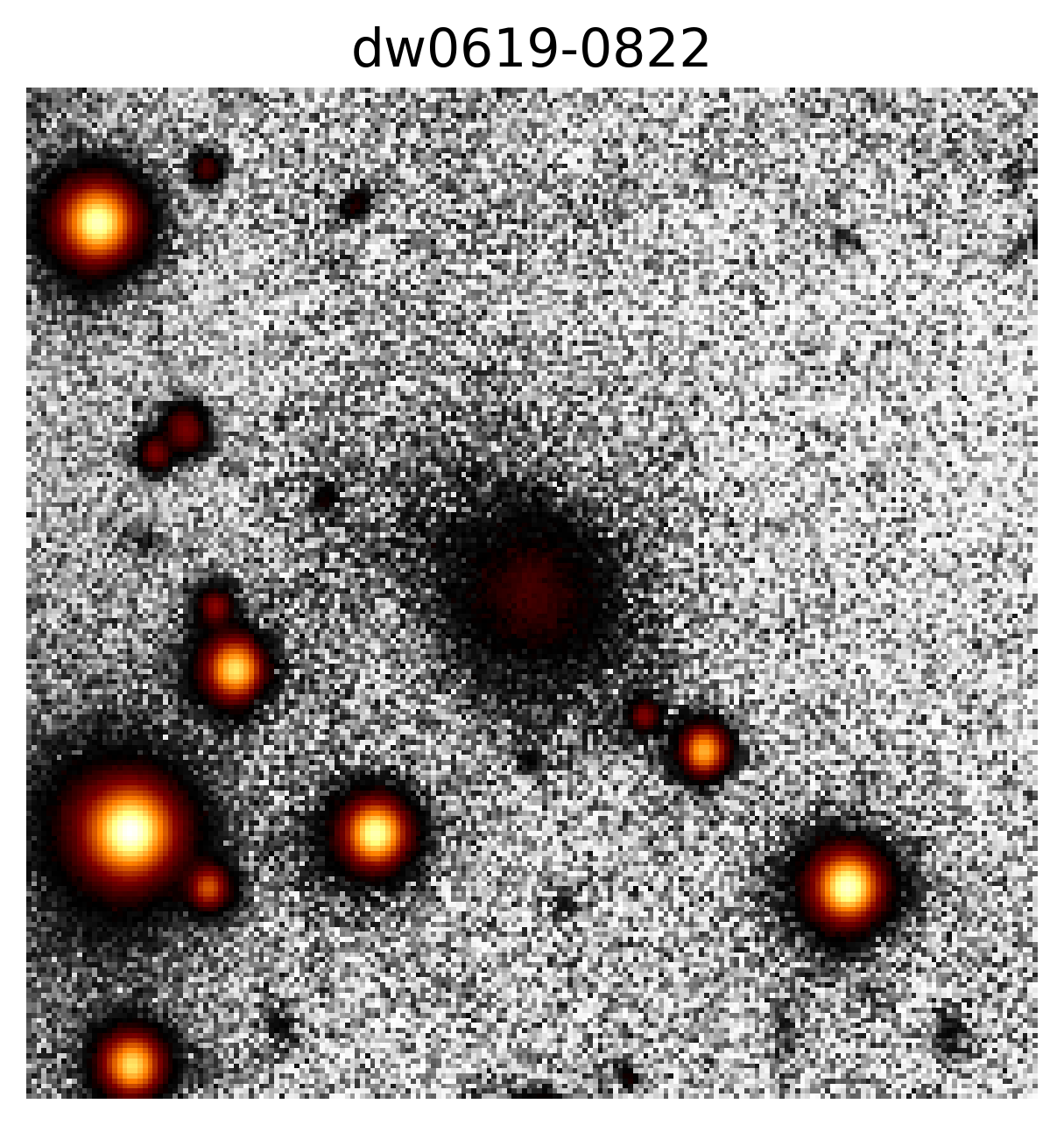}
    \includegraphics[width=0.163\linewidth]{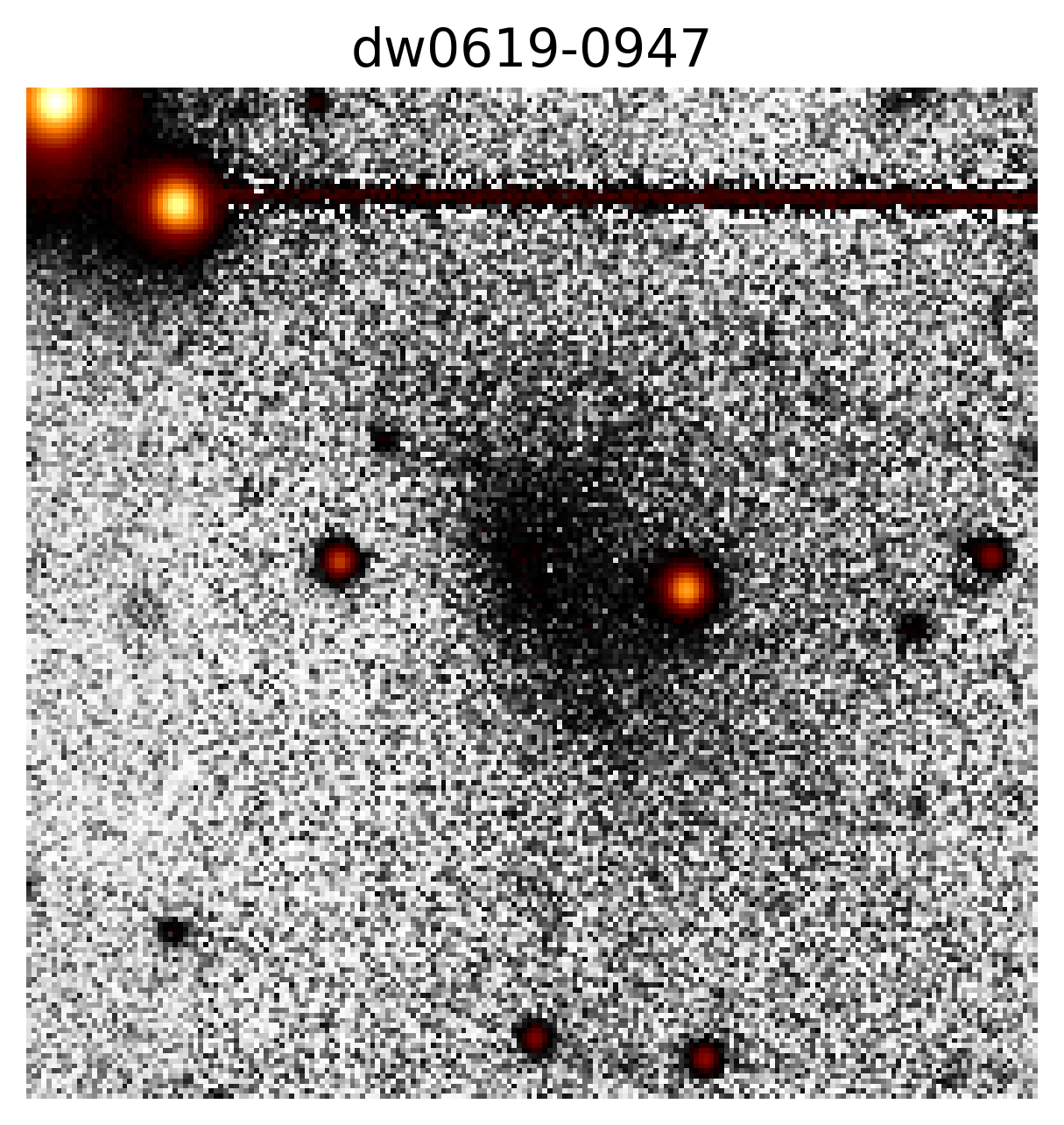}
    \includegraphics[width=0.163\linewidth]{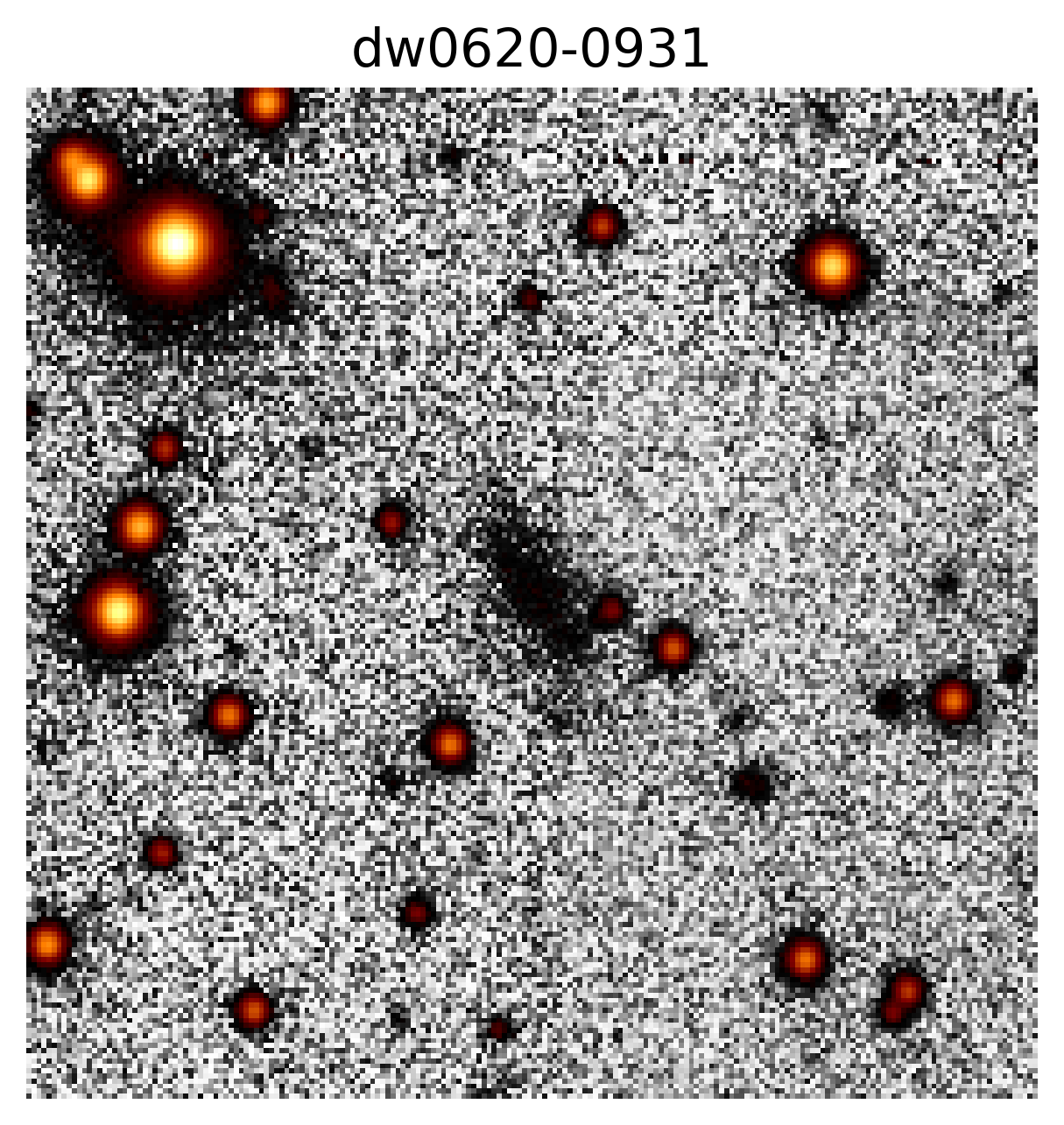}
    \includegraphics[width=0.163\linewidth]{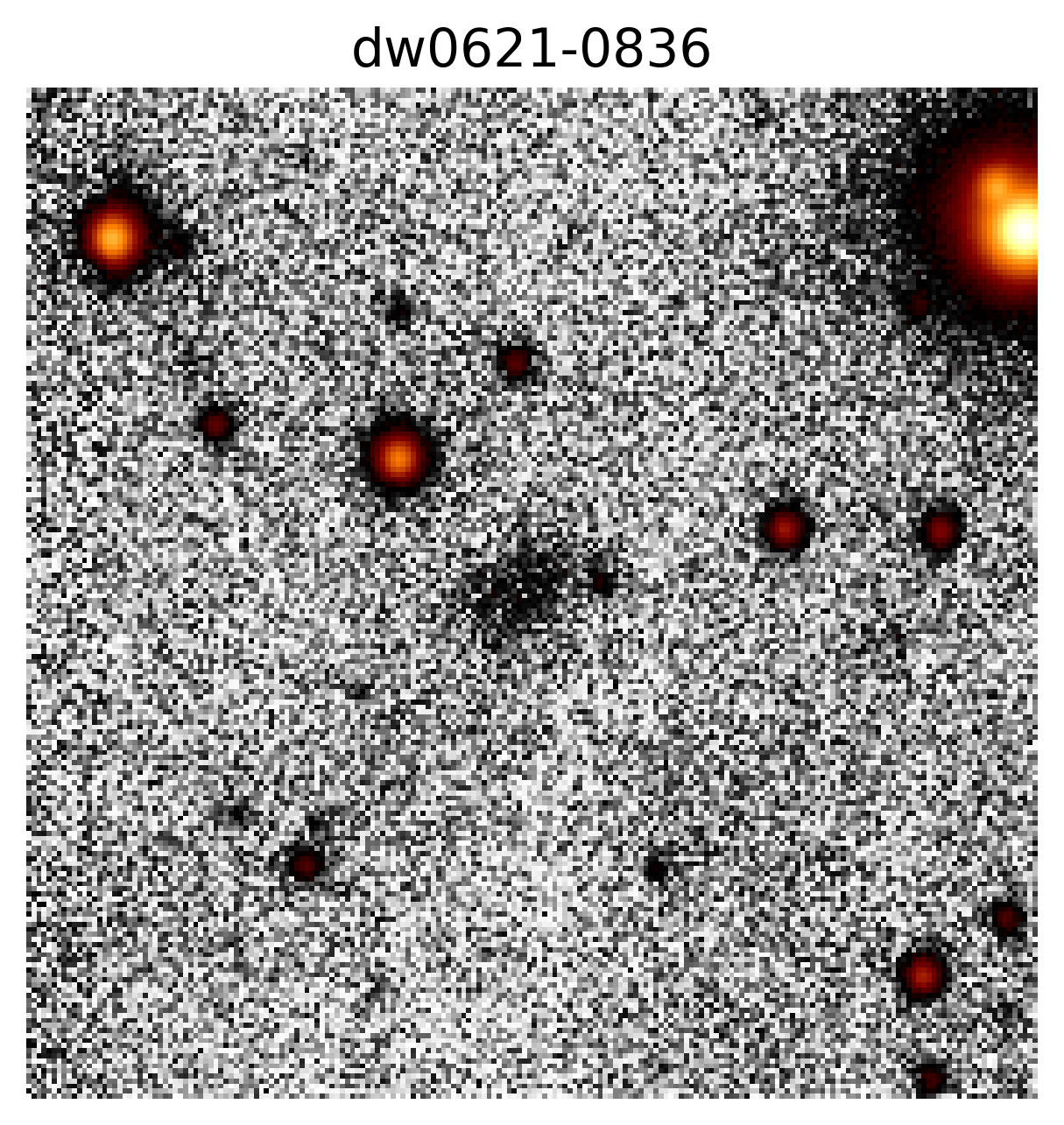}
    \includegraphics[width=0.163\linewidth]{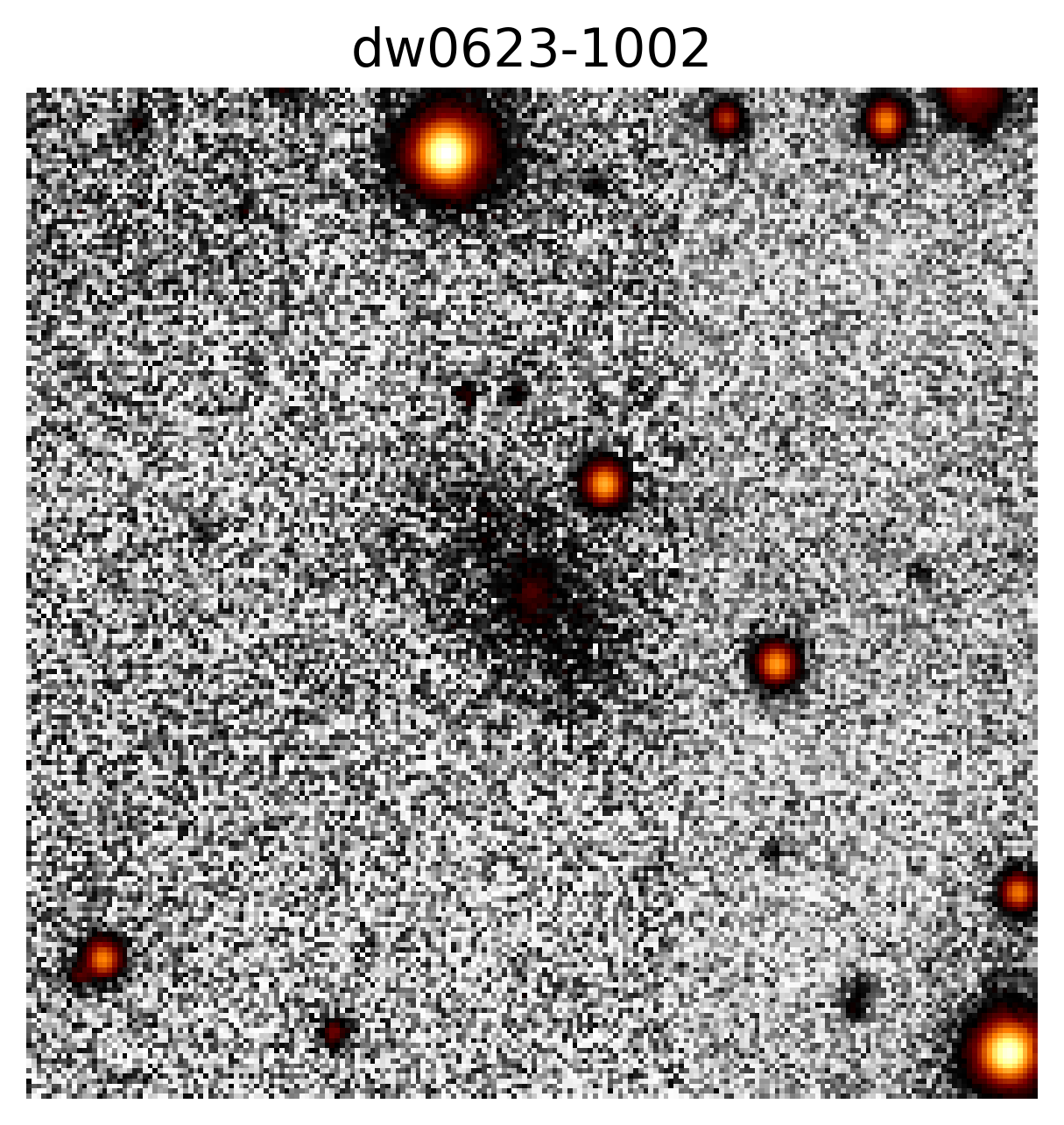}
	\caption{Cutouts of the confirmed  dwarf galaxies (top row) and dwarf candidates (bottom row) of the UGCA127 group.}
	\label{fig:UGCA127_known}
\end{figure*}

\end{document}